\documentclass[12pt]{article}
\usepackage{amsfonts}
\usepackage{setspace}
\usepackage{bm}
\usepackage{graphicx}
\usepackage{array}
\usepackage{color}
\usepackage{multirow}
\usepackage{mathrsfs}
\usepackage{tabularx,tabulary}
\usepackage{booktabs}
\usepackage{float}
\usepackage{bbold}
\usepackage{verbatim}
\usepackage{dsfont}
\usepackage{xcolor,colortbl}
\usepackage[sort]{natbib}
\usepackage[T1]{fontenc}
\usepackage{eurosym}
\usepackage{titling}
\usepackage{lmodern}
\usepackage[normalem]{ulem}
\usepackage[fontsize=11.2bp]{fontsize}
\usepackage{bbm}
\usepackage{hyperref} % Pour les liens internes
\DeclareGraphicsExtensions{.jpg,.eps,.pdf} % Formats images
\usepackage{enumerate}
\usepackage{amsmath,amssymb,amsfonts,amsthm} % Typical maths resource packages
\newtheorem{theorem}{Theorem}

\newtheorem{lemma}[theorem]{Lemma}

\newtheorem{proposition}[theorem]{Proposition}

\DeclareMathAlphabet{\mathpzc}{OT1}{pzc}{m}{it}

\theoremstyle{remark}

\definecolor{bleu}{cmyk}{1,0.4,0,0}
\definecolor{bleu2}{cmyk}{0.75,0.55,0,0}
\definecolor{orange}{cmyk}{0,0.56,0.88,0}

\newcommand{\E}{\mathbb{E}}

\newcommand{\PP}{\mathbb{P}}

\newcolumntype{a}{>{\columncolor{Gray}}c}

\definecolor{Gray}{gray}{0.95}
\definecolor{White}{gray}{1.00}
\definecolor{bblue}{rgb}{0.2,0.5,0.85}
\definecolor{dblue}{rgb}{0.05,0.2,0.4}
\definecolor{orange}{rgb}{0.89,0.55,0.2}
\definecolor{lightgray}{gray}{0.9}

\usepackage{xcolor}
\hypersetup{
citebordercolor=white,
filebordercolor=red,
linkbordercolor=blue
}
\hypersetup{
colorlinks,
citecolor=blue,
linkcolor=red,
urlcolor=blue}

\usepackage{palatino}
\fontfamily{ppl}

\definecolor{green}{rgb}{0.15,0.6,0.3}

\usepackage{xcolor}
\hypersetup{
citebordercolor=white,
filebordercolor=red,
linkbordercolor=green
}
\hypersetup{
colorlinks,
citecolor=green,
linkcolor=red,
urlcolor=green}

\begin{document}

%----------------------------------------------------------------------------------------------------------------%

%\title{Compounded adjustments in empirical studies}
\setlength{\droptitle}{-10mm}
\title{Forking paths in financial economics}
\author{Guillaume Coqueret\thanks{EMLYON Business School, 23 avenue Guy de Collongue, 69130 Ecully, FRANCE. coqueret@em-lyon.com. I thank Jonathan Lewellen, David Sraer, Olivier Scaillet, Fabio Trojani, St\'ephane Guerrier, Aur\'elien Baillon, Hossein Kazemi, Tony Guida, Gaetan Bakalli and Jean-Yves Gnabo for their constructive feedback. I am also grateful for comments from particpants from the 2022 AFFI conference, as well as the Financial Data Professional Institute, the Geneva Finance Research Institute and the Research Center for Statistics (UNIGE)  seminars. Lastly, I am indebted to Andrew Y. Chen for providing data for published asset pricing anomalies.  \vspace{3mm}} }
\maketitle
%\vspace{-6mm}

\begin{abstract}
%We propose a theoretical framework that characterizes the diversity of outcomes in empirical studies, depending on the nature and number of design choices that researchers make. We outline several ways to exploit the multiplicity of outcomes. Our ideas are illustrated in two studies. The first is an exercise of equity premium prediction with nine decision layers. We find that each additional degree of freedom in the protocol expands the average range of $t$-statistics by \textit{at least} 30\%. The second study pertains to portfolio sorts and shows that resorting to forking paths instead of bootstrapping in multiple testing greatly raises the bar in order to reach significant results. In the first case, at the 5\% confidence level, the threshold for statistics is 4.5, while in the second, it is at least 7.7, a bar much higher than those currently used in the literature.

We argue that spanning large numbers of degrees of freedom in empirical analysis allows better characterizations of effects and thus improves the trustworthiness of conclusions. Our ideas are illustrated in three studies: equity premium prediction, asset pricing anomalies and risk premia estimation. In the first, we find that each additional degree of freedom in the protocol expands the average range of $t$-statistics by \textit{at least} 30\%. In the second, we show that resorting to forking paths instead of bootstrapping in multiple testing raises the bar of significance for anomalies: at the 5\% confidence level, the threshold for bootstrapped statistics is 4.5, whereas with paths, it is at least 8.2, a bar much higher than those currently used in the literature. In our third application, we reveal the importance of particular steps in the estimation of premia. In addition, we use paths to corroborate prior findings in the three topics. We document heterogeneity in our ability to replicate prior studies: some conclusions seem robust, others do not align with the paths we were able to generate.

%We propose a theoretical framework that characterizes the diversity of outcomes in empirical studies, depending on the number of design choices that researchers make. Our ideas are illustrated in two studies. The first is an exercise of equity premium prediction with ten decision layers. We find that each additional degree of freedom in the protocol expands the average range of $t$-statistics by \textit{at least} 30\%. The second study pertains to portfolio sorts and shows that resorting to forking paths instead of bootstrapping in multiple testing raises the bar for significance: at the 5\% confidence level, the threshold for bootstrapped statistics is 4.5, whereas with paths, it is at least 8.2, a bar much higher than those currently used in the literature. %Such a conservative level also increases the odds of false negatives, a risk many money managers may be willing to take.  

\end{abstract}

% Keywords: p-hacking, robustness checks, asset pricing
% JEL: C12, C18, C51, G12

\vspace{-5mm}

%----------------------------------------------------------------------------------------------------------------%

%\clearpage
%\setstretch{1.5}

\section{Introduction}

``\small{\textit{Because the empirical economist must deal with nature in all her complexity, it is optimistic in the extreme to hope or believe that standard parametric economic models or probability models are sufficiently adequate to capture this complexity.'' - }\cite{white1996estimation}}

\vspace{3mm}

\subsection{Replication: a necessity and a challenge}

Empirical studies are built on many choices. Recently, five separate studies\footnote{See \cite{huntington2021influence}, \cite{breznau2022observing}, \cite{gould2023same}, \cite{huber2023competition} and the fincap study (exploited in \cite{menkveld2021standard} and \cite{perignon2022reproducibility}). Another enlightening exercise is provided by FivethirtyEight, in its \href{https://fivethirtyeight.com/features/science-isnt-broken/}{\textit{Science is not broken}} article, in the political science discipline.} have revealed that, given the same replication task, independent researchers can reach vastly different conclusions, including on trivial items such as sample sizes. Such divergences are the consequence of the large number of design options that are left to the appreciation of the empiricist. Examples of these choices are for instance listed in \cite{mitton2021methodological} for the field of corporate finance. 

In the best case scenario, minor shifts will lead to small adjustments, and solid conclusions will not be altered. Nevertheless, sometimes, articles may be retracted because the results are contradicted or simply not robust enough (\cite{rampini2021risk} and \cite{boissel2022dividend} are two recent occurrences).\footnote{Retractions are closely followed by a handful of researchers (\href{https://retractionwatch.com}{https://retractionwatch.com}) who have compiled and continuously maintain a database of scientific articles that have been retracted from their journals:
\href{http://retractiondatabase.org}{http://retractiondatabase.org}. 
See also \cite{shepperd2023analysis} for an analysis of retractions in computer science.} The replication of results has become a strong imperative in modern research,\footnote{We point to \cite{boylan2016reproducibility}, \cite{duvendack2017meant}, \cite{christensen2018transparency}, \cite{mueller2019replication}, \cite{vilhuber2020reproducibility}, \cite{colliard2021economics}, \cite{hofman2021expanding}, \cite{cortina2022we}, \cite{peters2022economists} and \cite{vu2022can} for discussions on replication and reasons that explain why it may be hard. Recent initiatives are the \href{https://i4replication.org/}{Institute for Replication} in economics and the \href{https://openmkt.org/}{Open Marketing platform}. One of the first attempts of the former pointed towards inflated conclusions (\cite{kjelsrud2023cooperative}).} so that data and code sharing policies have for instance been enforced in most major journals in finance and economics.\footnote{In \textbf{finance}, we can list: Journal of Finance, Review of Financial Studies, Journal of Financial Economics, and Review of Finance. In \textbf{economics}, there are at least the following: Econometrica, journals of the American Economic Association (including the American Economic Review), Quarterly Journal of Economics, Review of Economic Studies. In parallel, news services have emerged that propose framework to ease reproductibility, e.g., \href{https://nuvolos.cloud}{nuvolos} or \href{https://www.cascad.tech/}{cascad}.} When results can only be partly replicated (e.g., as in \cite{hou2020replicating} in asset pricing or \cite{kapoor2022leakage} in applied machine learning), debates can be fostered to determine if initial conclusions remain valid. 

The sensitivity of results to design choices is well-known in the research community, which is why many papers incorporate robustness checks in which analyses are replicated with small alterations in the original empirical protocol. %This can sometimes be computationally costly (see \cite{catherine2023robustness}), so that these checks can be limited or require ingeniosity. 
For instance, researchers can reproduce their results on sub-samples, or with alternative estimators, or when testing different values in parametric methods and techniques. They can also evaluate sensitivity measures of estimated parameters (\cite{leamer1985sensitivity}, \cite{jorgensen2023sensitivity}). %

Given the publication bias towards positive results,\footnote{This phenomenon, also known as the \textit{file drawer problem}, has been widely documented in many fields, especially \textbf{psychology} (\cite{rosenthal1979file}, \cite{open2015estimating}, \cite{stroebe2019can}), \textbf{medicine} (\cite{dickersin1987publication}, \cite{begg1988publication}, \cite{olson2002publication}, \cite{barnett2019examination} and \cite{van2019publication} to cite a few references), \textbf{economics} (\cite{leamer1983reporting}, \cite{de1992all}, \cite{stanley2005beyond}, \cite{doucouliagos2009publication}, \cite{doucouliagos2013all}, \cite{brodeur2016star}, \cite{camerer2016evaluating}, \cite{ioannidis2017power}, \cite{brodeur2020methods} and \cite{kasy2021forking}), \textbf{finance} (\cite{lo1990data}, \cite{harvey2017presidential}, \cite{morey2018documentation} and \cite{harvey2021uncovering}), and \textbf{accounting} (\cite{chang2021p}). In a recent experimental study encompassing 500 researchers, \cite{chopra2022null} conclude: ``\textit{we document that studies with a null result are perceived to be less likely to be published, of lower quality, and of lower importance than studies with statistically significant results even when holding constant all other study features}''. Moreover, \cite{serra2021nonreplicable} find that papers that are hard or impossible to replicate are more cited than more transparent studies. Other related issues are bias in research (\cite{fanelli2017meta}) and conventionality, whereby mainstream studies with expected outcomes have a higher probability of being published in top journals (\cite{dai2022dissemination}).} authors can be, knowingly or not, incentivized to produce low $p$-values in order to demonstrate that their results are \textit{significant}, and hence, worthy of publication.\footnote{The topic of $p$-hacking is now widely documented in many fields since the seminal work of \cite{sterling1959publication}, and we point to a few articles on the matter, among many others: \cite{head2015extent}, \cite{christensen2018transparency} and \cite{brodeur2022we}.} Equipped with a large palette of adjustments (i.e., leeway) in the research design, researchers may be tempted to mostly report those that confirm their priors or theoretical predictions. Indeed, positive results are easier to push forward and increase the odds of acceptance by reviewers and journal editors. From a scientific standpoint, this is suboptimal, because failure to reject the null is often informative (\cite{abadie2020statistical}), as it signals to other researchers where \textit{not} to look. That being said, \cite{blanco2020publication} have shown that editors have the power to change this habit in the refereeing community.

The troubles of false discoveries are also potent and problematic outside academia. In the money management industry, quantitative researchers compete to discover profitable trading strategies. Unfortunately, many of the latter which perform well in-sample end up disappointing out-of-sample in practice (see \cite{bailey2014pseudo}, \cite{deprado201810}, \cite{chen2021zeroing}). This has spurred a debate on whether there is a crisis of reproducibility in the field.\footnote{See \cite{bailey2021finance}, \cite{chen2021most}, \cite{harvey2021uncovering}, \cite{harvey2021skeptical}, \cite{chen2022publication} and \cite{chen2022peer}. This debate is also ongoing in the medical sciences (see \cite{ioannidis2005most} and \cite{leek2017most}, as well as the discussion in \cite{fanelli2018opinion})} As a solution, many scientists argue in favor of redefining the notion of \textit{statistical significance} (\cite{harvey2017presidential}, \cite{benjamin2018redefine}), or even propose abandoning it purely and simply,\footnote{Among many others, we can point to: \cite{carver1978case}, \cite{ziliak2008cult}, \cite{woolston2015psychology},   \cite{amrhein2019scientists}, \cite{mcshane2019abandon}, \cite{wasserstein2019moving}. A more nuanced take is given in \cite{imbens2021statistical}. } mainly for two reasons. First, because people sometimes confuse $p$-values with the probability of the null being true, given the data. And second, because if $p$-values are no longer the ultimate yardstick of scientific discovery, researchers will be less inclined to tilt their empirical protocols so as to produce the sought output (i.e., $p$-values below some chosen threshold, usually 1\% or 5\%). Without $p$-values, no more $p$-hacking.

The research community has also proposed other avenues to tackle these issues, e.g., by identification and detection methods for false discoveries and $p$-hacking (\cite{simonsohn2014p, simonsohn2014p2}, \cite{elliott2021detecting}), and even solutions via model averaging (\cite{moral2015model}, \cite{steel2020model}), extreme bound analysis (\cite{leamer1983reporting}, \cite{granger1990reasonable}), correction measures (\cite{andrews2019identification}), shrinkage (\cite{van2021significance}), robust aggregation (\cite{rytchkov2020information}),  noise dissemination (\cite{echenique2021screening}), new critical values (\cite{mccloskey2023critical}), specification curve analysis (\cite{simonsohn2020specification}), or Bayesian publication decisions (\cite{frankel2022findings}).

In the present paper, similarly to \cite{fabozzi2018being}, we argue that one solution, albeit a costly one, is to report the outcomes of a large number of forking paths in empirical studies. Large scale experiments are for example also advised in \cite{milkman2021megastudies} in the field of behavioral science. This is likely to dramatically increase the transparency of the research process and to strengthen the robustness of findings. In a similar spirit, \cite{coker2021theory} propose to report ``\textit{p-hacking intervals}'', which correspond to the total range of possible outcomes (e.g., coefficients, $t$-statistics, $R^2$, etc.) that are obtained when spanning sets of hyper-parameters in supervised learning models.

\subsection{Beyond replication: confirmation}

Being able to replicate findings is an important and necessary step, but it does not ensure the generalization of a result to  different empirical settings. When a result is published, its long-term validity will depend on whether it is \textit{confirmed} subsequently by studies which will seek to check if it holds for other data sources, in other geographical zones, or over alternative periods. If it does, then the result can be \textit{generalized} and sees its practical reach extended. \cite{perignon2022reproducibility} put forward two types of reanalyses: reproductions and replications. Reproduction occurs when the dataset is exactly the same as the original study, but the code can be different. Replication signals more leeway: the code and method can be different, or the data and code can also change. When the data, code and method change, the authors still talk of replication, but in the present paper, we will put forward the notion of \textit{confirmation}.  

Let us exemplify our argument with a well-documented phenomenon, the (cross-sectional) momentum in international markets. The idea is attributed to \cite{jegadeesh1993returns} but has since then been corroborated by many studies. The dimensions along which momentum has been confirmed are numerous, including geographical (\cite{rouwenhorst1998international}, \cite{chan2000profitability}, \cite{griffin2003momentum}, \cite{bhojraj2006macromomentum}), chronological (\cite{smith2022have}), industrial (\cite{moskowitz1999industries}), and across asset classes (\cite{asness2013value}) - to cite but a few. This accumulation of evidence, in spite of periodic crashes (\cite{barroso2015momentum}, \cite{daniel2016momentum}), establishes momentum as a multiply verified pattern in financial economics. This leads us to propose in Table \ref{tab:class} a classification of empirical conclusions, from the least reliable to the most robust evidence. The present paper focuses on the third type for which one contribution is able to span a large number of cases. 

\begin{table}[!h]
    \centering
    \footnotesize
    \begin{tabular}{r | c | c | c | c}
    \midrule
        \textbf{Evidence type} & \textbf{Protocols} &  \textbf{Publications} & \textbf{Dimensions} & \textbf{Evidence strength}  \\ \midrule
         Anecdote & one & one & one & none \\
         Robustness checks & few & one & few & weak \\
         Internal paths & many & one & many & sustained \\
         External confirmation & many & many & one & sustained \\
         Exhaustive documentation & many & many & many & strong \\ \midrule
    \end{tabular} \vspace{-2mm}
    \caption{\textbf{Classification of empirical conclusions}. \small{The number of protocols (i.e., paths) can come either from several publications that treat the same topic, or, as in the present paper, from one publication which generates and reports many results internally. By ``\textit{few}'', we mean a dozen at most, whereas by ``\textit{many}'', we mean a hundred at least. }  \label{tab:class} }
\end{table}

The generation of multiple outputs, based on small variations of similar datasets, shares some similarities with re-sampling techniques, as well as data augmentation and bagging, all of which are sometimes used in machine learning. The premise is that a model which relies on a diversified set of sub-models will benefit from a \textit{wisdom of the crowds} effect, as long as each individual model is relevant (loosely speaking) and that correlations between models are not too high. The best situation is when diversification operates and outcomes of forking paths reveal complementary facets of the initial problem. These ideas have blossomed in the frequentist (\cite{hansen2007least}, \cite{zhang2015consistency}, \cite{zhang2019inference} and \cite{zhu2021kernel}) and Bayesian (\cite{draper1995assessment}, \cite{raftery1997bayesian}) circles. For instance, Bayesian averaging has recently been used in \cite{avramov2022integrating} to cope with model uncertainty. 

Nevertheless, we forcefully underline that model averaging and extreme bound analysis (EBA) are only special cases of forking paths. In most papers on model averaging and EBA, the data is fixed, and models are generated through alternative independent variable combinations. With forking paths, this is also allowed, but, more importantly, the ways to construct the initial sample can be discretionarily many, and estimators are not necessarily unique. Crucially, most results in model averaging consider that the number of models is finite, while the sample size increases to infinity. In the present paper, it is the opposite: sample sizes are arbitrarily small or large, and it is the number of paths that increases. 

In addition to improving the robustness of reported results, framing empirical work as successive mappings helps organize code more neatly into a well-structured research pipeline. As is shown in \cite{perignon2022reproducibility}, coding skills are linked to the reproducibility of empirical studies. In our framework, each mapping has its own module, which opportunely prevents potential errors in lengthy scripts written in one block. It also forces to reflect upon the computational cost of each step of the research project and how to optimize it.\footnote{Even if modern computers allow for the parallelization of tasks, the complexity of most pipelines outweigh the CPU (and GPU/TPU) capabilities of standard machines. This is likely to limit the exploration of potentially promising but untested questions and configurations.} Consequently, an exhaustive approach to the reporting of results compels the analyst to focus on the first order choices of the research process and to filter out the unnecessary artifices.

\subsection{Summary of contributions}

Beyond model averaging and inference, we propose several ways to operationalize forking paths. The first one seeks to determine which design choices have an impact on the distribution of outcomes. This is very important because if some modelling stages change the sign or significance of an effect, they must be transparently documented. 

The second application pertains to \textit{partial} replication, a task which we also call \textit{confirmation}. The purpose here is not to exactly reproduce a given study but rather to test the generalization ability of a published result in slightly different contexts. To evaluate if a prior study has reported a \textit{plausible} value, we compare it with the ones that we are able to produce with many paths. We devise an indicator, which we call the \textit{ease to confirm} (EtC) and that measures the extent to which the original effect is compatible with the outcomes from the paths which we spanned. In essence, it measures the degree to which the original published value lies inside or outside the distribution of path-generated effects.  

Lastly, a third use of paths relates to multiple testing (MT). Modern approaches to MT rely on bootstrapping. This assumes that the data available to the researcher (e.g., one path) is representative of the full distribution of the effect under scrutiny. Such an assumption may be excessively strong, which is why we propose to replace bootstrapped samples by forking paths, a method we refer to as \textbf{exhaustive multiple testing}. Because paths are more diverse, they are likely to produce more extreme outcomes and significance hurdles that are higher, compared to thresholds that originate from bootstrap-based MT techniques.

We illustrate our framework and operational recommendations with three empirical studies. The first one relies on the prediction of the equity premium, a well-documented research question in financial economics.\footnote{Forking paths are tackled via the notion of \textit{non-standard errors} for asset pricing anomalies in \cite{soebhag2022mind} and \cite{walter2022non}.} We consider a large number of ways to run the empirical protocol and report the distributions of test statistics. The latter allow us to determine which design choices alter the average of the statistics and are hence strong drivers thereof. For instance, switching from a simple OLS estimator to the \cite{amihud2004predictive} specification has very little impact on coefficients. However, subsampling over two different periods generates substantial differences in estimates. In this first study, we also compare our results with those of \cite{goyal2021comprehensive} and find that their figures are mostly plausible and have magnitudes that can easily be reproduced.

 % + [Contaminated models]

Our second study revolves around asset pricing anomalies. Its core focus is on multiple testing whereby we seek to determine if the seemingly strongest factor is indeed strong enough. We compare the traditional approach to the one based on paths. We find that the improved thresholds generated by paths are significantly more conservative, compared to those of one traditional bootstrap-based method. From an investment standpoint, this means that our method implies that genuine anomalies are scarcer than the literature previously reported. In \cite{harvey2016and}, the authors recommend to raise the significance threshold of $t$-statistics to 3.0. With standard bootstrap-based multiple testing, we recommend to push it to 4.5. If we rely on forking paths, the bar is set at least at 8.2, a level that few anomalies are able to pass. Unfortunately, a strong reduction of the risk of false positives comes with an increase of false negatives. In some situations, bad investments matter more than missed opportunities. 

The requirement that a \textit{genuine} factor must remain strong under many specifications shares some intuition with the literature on invariance-based causality (see \cite{peters2016causal}, \cite{arjovsky2019invariant} and \cite{buhlmann2020invariance} for an overview). The premise therein is that in order to reveal a causal link between two variables, the relationship must hold in several sampling environments. If, for instance, each environment alters the distribution of $X$ while preserving the either correlation between $X$ and $Y$, or the conditional law of $Y$ knowing $X$, then inferential conclusions are stronger than if they rely on one environment only. If the effect remains invariant across multiple studies or datasets (i.e., it is \textit{replicable} outside its original sample), then it is robust, and possibly causal - this notion being out of the scope of the present paper.

Our second application revolves around \cite{fama1973risk} regressions and the estimation of risk premia. It reveals that while some choices do not matter much (winsorizing loadings after the first pass), others are more critical, especially the sampling of returns before the first pass. We compare our results with those of \cite{fama1973risk} and \cite{ang2020using}. We find that for the former, the reported figures are entirely reliable. For the latter, some market premia are realistic, but others are not.

The remainder of the paper is structured as follows. Section \ref{sec:lip} lays out a representation of research studies as compositions of operators. Therein, we also propose several ways to exploit the paths generated by these compositions, for instance by characterizing which mappings are significant for a given research output, or by means of model averaging. An important benefit from paths is that they allow to locate prior research results within the distribution from the paths: this allows to corroborate these prior findings - or not.
Our ideas are illustrated through three empirical studies which are mentioned throughout the paper and are located in Sections \ref{sec:premium}, \ref{sec:sorts} and \ref{sec:FM}. Finally, Section \ref{sec:conc} concludes. The appendix features some supplementary material.

\section{Theoretical groundwork}
\label{sec:lip}

This section comprises the analytical background of the paper which models the research process as compositions of operators. The foundations are laid in subsections \ref{sec:21} and \ref{sec:th}. The methods for the operationalization of the paths are presented in the remaining four subsections.

\subsection{Overarching framework and \textit{p}-hacking}
\label{sec:21}

\subsubsection{Notations}

We start with a few conventions on notation. Unless otherwise stated, the integer $N$ will always be the length of the vectors and the number of rows of matrices. Henceforth, lowercase bold letters $\mathbb{d}=\{\mathbb{d}_1,\dots,\mathbb{d}_N\}$ will denote vectors and uppercase bold letters matrices or tables. For the latter, we adopt the \textbf{tidy data} convention of \cite{wickham2014tidy}: rows are observations and columns are variables. Finally, we will sometimes (when there is little ambiguity) use the simplified notation $f(\mathbb{d})$ for the vector $[f(\mathbb{d}_1),\dots f(\mathbb{d}_N)]$. Moreover, for a matrix $\bm{M}$ or a vector $\bm{v}$, $\bm{M}'$ and $\bm{v}'$ will denote their transpose.

In addition, we will often compare two alternative inputs. Readers are accustomed to $\bm{X}$ and $\bm{y}$ for modelling purposes. To avoid any confusion, we work with the letters $\mathbb{D}$ and $\bm{D}$, which will stand for two versions of some data that is collected and then possibly transformed by the researcher. This choice of notation is disconcerting at first, but imperative because we will restrict the use of $\bm{X}$ and $\bm{y}$ letters to linear models later on.

We assume that the empirical part of research process starts with some input which we call $\mathbb{D}$ and can be thought of as the initial version of the data that is collected. The study is modelled as a sequence of operations $f_j$ that occur successively so that the reference research output $o_J$ (e.g., one $t$-statistic) is such that
\begin{equation}
o_J(\mathbb{D})=\left[\bigcirc_{j=J}^1 f_j\right](\mathbb{D})=f_J \circ f_{J-1} \circ \dots \circ f_1(\mathbb{D}),
\label{eq:seq}
\end{equation}
where $f_j:S_j\mapsto S_{j+1}$, with $S_1$ and $S_{J+1}$ encompassing the sets of feasible input $\mathbb{D}$ and output values, respectively. For simplicity, we can assume that $o_J$ is simply a real number, but it may be a more complex object, such as a vector (e.g., confidence interval) or a matrix. The index $J$ indicates that the output is the result of $J$ successive operations, which \cite{gelman2014statistical}, among others, refer to as \textit{forking paths}. Examples of such operations are provided in Appendix \ref{sec:exlip} and include for instance: missing data imputation or removal, winsorization, variable selection, variable scaling, subsampling, choice of estimator, etc. Here, the output depends on $J$ and the sequence of mappings $f_j$. Later on, for simplicity, we will index outputs with $p$, which will be the index of the corresponding path.

More precisely, we can write the output-generating process as
\begin{equation}
o_J(\mathbb{D}, \bm{P})=\left[\bigcirc_{j=J}^1 f_{j,\bm{p}_j}\right](\mathbb{D})=f_{J,\bm{p}_J}  \circ \dots \circ f_{1,\bm{p}_1}(\mathbb{D}),
\label{eq:seq2}
\end{equation}
where $\bm{P}$ encompasses the parameter sets $\bm{p}_j$ for all the mappings. These parameters may be fixed, or random, e.g., when sampling arbitrary thresholds for winsorization. The output $o_J$ is a random variable that depends on the realizations of the operators $f_j$ - and possibly on that of $\mathbb{D}$ if stochastic initial samples are allowed. In all generality, the realization of $f_j$ may depend on those of prior operators ($f_i$ for $i<j$). Plainly, the order of mappings may matter: it does not make sense to perform data imputation \textit{after} estimation.

%The formulation \eqref{eq:seq2}, because it relies on parameters at each layer, resembles feed-forward \textbf{neural networks} (NNs), though there are several notable differences between the two concepts. First, NNs are not traditionally viewed as random objects. In stochastic gradient descent, the randomness comes from the samples that are drawn, akin to $\mathbb{D}$, in our framework. Second, there is no supervision in Equation \eqref{eq:seq2}. The output is not known in advance (labelled), and we do not seek to optimize on the set $\bm{P}$ to obtain a desired value (or distribution) for $o_J$.\footnote{Going forth with this analogy, \textbf{data snooping} could be viewed as some form of supervised selection: each new forward pass in the network would only be retained if the outcome produces statistically significant results.} Lastly, in all generality, the operators are not differentiable, meaning that back-propagation is infeasible.

%Each path $p$ is determined by the

Henceforth, we assume that any mapping $f_j$ has $r_j$ deterministic options which the researcher must choose from, and which we write $\mathbb{f}_{j,r}$, for $r=1,\dots r_j$, where $r_j\ge 2$. For example, this can be alternative ways of handling missing data (deletion versus imputation), or the set of possible combinations of independent variables, in which case $r_j$ is the cardinal of this set (i.e., all permutations that are relevant for the study). This makes $P=\prod_{j=1}^Jr_j$ paths in total.

Paths are determined by the choice of their options for each layer. Thus, we define paths as $p:=\{\mathbb{f}_{j,r_{p,j}}\}_{1 \le j \le J}$, where $r_{p,j}$ is the option choice of path $p$ for layer $j$. To ease notation, we will sometimes write $\mathbb{f}_{j,r(p)}$ for the option of layer $j$ though which path $p$ passes.

A crucial facet of forking paths is their diversity, which we can measure via their proximity, or lack thereof. For each layer $j$, we assume we can define a distance function $d_j$ between all choices of the layer, e.g., between paths $p$ and $q$:
$d_j(p,q)=d_j(\mathbb{f}_{j,r(p)}, \mathbb{f}_{j,r(q)})$. We can then aggregate into a total distance between two paths
\begin{equation}
  d(p,q)=\sum_{j=1}^J\omega_jd_j(\mathbb{f}_{j,r(p)}, \mathbb{f}_{j,r(q)}),  
  \label{eq:distance}
\end{equation}
where $\omega_j$ specifies the relative importance of layer $j$. A simple and explicit form for $d(p,q)$ will be used in Section \ref{sec:th}.

%In most empirical studies, the baseline result correspond to one path and the robustness checks are minor deviations from it. For instance, these deviations can be obtained by altering decisions layers one at a time. Often, baseline protocols follow common practice and standard conventions from the literature. Sensitivity analyses help persuade the audience that small tweaks to the representative protocol do not alter the qualitative conclusions reached by the researchers.

%Naturally, all paths have to be spanned once to obtain the full distribution of outcomes. It is then possible to correct this distribution if some paths are more likely or more representative than others, by simply giving them more weight, e.g., by repetition.

\subsubsection{Divergence of outputs}

One interesting question pertains to the sensitivity of the output $o_J$ to a change in initial input $\mathbb{D}$. In order to derive theoretical results, we must impose some conditions on the mappings $f_j$ and we choose to work with Lipschitz smoothness. More precisely, we assume that for $\mathbb{D}, \bm{D} \in S_j$, there exists some constant $c_j>0$ such that
\begin{equation}
\|f_j(\mathbb{D}) - f_j(\bm{D}) \| \le c_j \| \mathbb{D}-\bm{D} \|,
\label{eq:lip}
\end{equation}
for some norms which are implicitly defined on $S_{j+1}$ and $S_j$. For simplicity, we do not precisely define these norms, except in Appendix \ref{sec:exlip} in which we explicit some Lipschitz constants for a few specific operators. In all generality, the object $\mathbb{D}$ can comprise several data types, categorical features notably (ordinal or nominal). Handling distances with such features is complex, though not impossible (see \cite{boriah2008similarity}), but for the sake of simplicity, the expos\'e will essentially assume that $\mathbb{D}$ and $\bm{D}$ are matrices of real numbers.  

Composing two operators yields
\begin{align*}
    \|f_{j+1}\circ f_j(\mathbb{D}) - f_{j+1}\circ f_j(\bm{D}) \|& \le c_{j+1} \|f_j(\mathbb{D}) - f_j(\bm{D} \| \\
    & \le c_j c_{j+1} \| \mathbb{D}-\bm{D} \|.
\end{align*}
Iterating this inequality leads to
\begin{equation}
    \left\| \left[\bigcirc_{j=J}^1 f_j\right](\mathbb{D})-\left[\bigcirc_{j=J}^1 f_j\right](\bm{D}) \right\| \le  \| \mathbb{D} - \bm{D} \| \prod_{j=1}^Jc_j.
\end{equation}

More generally, the accumulation of shifts may not start at the initial data sample $\mathbb{D}$, but at a later stage, say at $o_K$, after $K$ steps, for $K<J$. It is easy to prove the following lemma.

\begin{lemma}
If $o_J$ is given by Equation \eqref{eq:seq} and the mappings $f_j$ satisfy \eqref{eq:lip}, then for $1\le K <J$,
\begin{equation}
    \left\| \left[\bigcirc_{j=J}^{K+1} f_j\right](o_K(\mathbb{D}))-\left[\bigcirc_{j=J}^{K+1} f_j\right](o_K(\bm{D})) \right\| \le  \| \mathbb{D} - \bm{D} \| \prod_{j=K+1}^Jc_j.
\end{equation}
\label{lem:1}
\end{lemma}

It is obvious that the constants $c_j$ are the main drivers of the error bounds. In practice, a lower bound for the $c_j$ is often 1, meaning that their compounded effect can be sizeable, theoretically, especially if many $c_j$ are such that $c_j \gg 1$. Thus, adding steps in the design is likely to increase the amplitude of the difference between outcomes. If the latter are scalars and the norm is the max-norm, then we recover the width of the so-called $p$-\textit{hacking interval} introduced in \cite{coker2021theory}.

For illustration purposes, let us consider the sample mean, which is abundantly used, if only in summary statistics. If $f$ is the \textbf{sample mean} operation, we have, via H\"older's inequality in the last inequality,
\begin{align}
\|f(\mathbb{d})-f(\bm{d}) \|_p =   \left| \frac{1}{N} \sum_{n=1}^N \mathbb{d}_n - \frac{1}{N} \sum_{n=1}^N \mathpzc{d}_n   \right| \le \frac{1}{N}   \sum_{n=1}^N\left| \mathbb{d}_n -  \mathpzc{d}_n   \right| \le N^{-1/p} \| \mathbb{d} - \bm{d} \|_p, \label{eq:mean2}
\end{align}

so that in this case the Lipschitz constant is $N^{-1/p}$.

We list other important mappings along with upper bounds on their Lipschitz constants in Appendix \ref{sec:exlip}. Few are exactly sharp and in some cases, they even depend on the inputs, $\mathbb{d}$ and $\bm{d}$, or $\mathbb{D}$ and $\bm{D}$.

The upper bound for the range of outcomes provided in Lemma \ref{lem:1} is theoretical. Indeed, it is possible that conflicting effects in the mappings yield actual outcomes that are less dispersed. It is only after the paths have been spanned that we can evaluate the divergence, as in \cite{menkveld2021standard} for instance. This issue is postponed to Section \ref{sec:range}.

\subsubsection{$p$-hacking}

In practice, the paths are not necessarily spanned meticulously. Data snooping is a powerful generator of paths because it requires to reflect on the methodological choices underpinning the initial baseline. Let us now assume that paths and outcomes $o^{(p)}$ are indexed by some integer $p$ which represents the order in which the spanning has occurred: $o^{(1)}$ is the first output, etc.

Then, \textbf{simple \textit{p}-hacking} is simply the process of generating a sequence $o^{(p)}$ and stopping when the $p^{th}$ output is deemed satisfactory: it will be the one that is retained for publication. This usually requires two conditions. The first one is statistical significance, i.e., the outcome must be associated with a confidence level $1-\alpha$ that is high enough, usually 95\% or 99\%. The second condition is the sign of the outcome. Most of the time, the direction of the effect is important and it may occur that preliminary results be significant, but in the wrong direction. In which case, more paths need to be explored, unless the original hypothesis is validated, or revised, as in HARKing (\cite{kerr1998harking}, \cite{hollenbeck2017harking}). Note however that if $o^{(1)}$ is already statistically significant and in the desired direction, then $p$-hacking does not need to start.
\textbf{Robust \textit{p}-hacking} is more stringent. Here we do not require that only one path be fitting a narrative, but a neighborhood of paths. Thanks to this, it is possible to present deviations from the default path (robustness checks) that corroborate and strengthen its conclusions.

Finally, we mention the ultimate level of snooping, which we call \textbf{vicious \textit{p}-hacking}. This refers to the case when paths are explored until a sufficient amount of $p$-values can be extracted in such a way that they successfully pass a test that detects $p$-hacking, for instance those mentioned in \cite{elliott2021detecting}. Because such tests are very recent, it is highly unlikely that vicious $p$-hacking has already occurred.  Nevertheless, this notion is interesting because, given a vector of $p$-values generated by forking paths, we can try to evaluate the potential for vicious $p$-hacking.

Let us be more specific. Suppose we focus on the common case of one-sided $t$-tests, which we will cover in Section \ref{sec:sorts} for anomaly detection. In \cite{elliott2021detecting}, it is shown in Theorem 2 that, under mild technical assumptions and in the absence of $p$-hacking, the density of the $p$-values should be decreasing and convex. Because of their importance, discuss the relevance and applicability of $p$-hacking tests for forking paths in Appendix \ref{sec:tests}.

Let us assume that we have spanned many paths and obtained the corresponding $p$-values, which are distributed on the unit interval. We split the first half of this interval into $I$ sub-intervals of equal sizes and we write $n_i$ for the number of $p$-values inside these intervals. Simply put, the $n_i$ are simply the count values of the histogram.

Applied to sample values, the first condition for the absence of $p$-hacking is that $n_i>n_{i+1}$ for $i=1,\dots,\lfloor(I-1)/2\rfloor$. If it is satisfied, the second order condition requires that $n_i/n_{i+1}>n_{i+1}/n_{i+2}$. Given the actual values of $n_i$, it is possible to evaluate the number of these inequalities that are fulfilled. If some of them are not, we can also assess the extent to which they are violated. This yields a qualitative judgement of how much trafficking in the distribution is needed before it can pass a $p$-hacking detection test. This will be illustrated in Section \ref{sec:crosshack}.

%We consider a given original dataset, so that one given output is written $o_p=\bigcirc_{j=J}^1 f_{j,\bm{p}_j}$, and where each path $p$ can be seen as the sequence of mappings: $p=\{f_{1,p_1}, \dots, f_{J,p_J}\}$, where the parameters $p_j$ indicate the choices that have been made for $\mathbb{f}_{j,r}$. Since we can index the set of choices, we can write $p_j=\mathbb{f}_{j,i_j}$, where $i_j$ is the index of the chosen alternative for layer $j$.

\subsection{Paths as pseudo-environments}
\label{sec:th}

\subsubsection{Invariant effects}

In a large number of studies, the aim is to determine if a particular effect holds. For simplicity, let us assume that such effect can be summarized as a scalar, $b$, and that the null is $b=0$, so that the researcher seeks to reject it in order to obtain a publishable result. Importantly, the effect may not be constant and, in all generality, it can be viewed as a random variable that depends on the environment through which it is evaluated. This line of reasoning is in fact pregnant in several fields:
\begin{itemize}
\setlength\itemsep{-0.2em}
    \item For instance, in causal inference, it has been proven that if an effect remains invariant across several sampling environments, then this effect can be considered as causal in some sense (\cite{peters2016causal}, \cite{pfister2019invariant}). Causality is out of the scope of the present paper, but the analogy between paths and environments is at the core of our reasoning.
    \item Likewise, the idea that multiple models are useful to characterize variable importance is also popular in machine learning (\cite{fisher2019all}).
    \item Lastly, a final parallel can be drawn with the social sciences, for which the \textit{observer effect} may alter the gathering of data: the characteristic of a researcher can influence the behavior of the subjects being studied (\cite{monahan2010benefits}).
\end{itemize}

One important premise of the paper is that effects are indeed random, and that, most of the time, researchers seek to characterize the \textit{average effect}, $\E[b]$. For instance, in finance, it is well-known that risk premia depend on the economic environment. \cite{gagliardini2016time} have shown that traditional factor premia are strongly dependent on NBER cycles. Hence, premia will depend on the time-frame on which they are evaluated, and, more generally, on the data and estimator that is used. Consequently, to have a better understanding thereof, it is valuable to generate not only a representative average, but many plausible values that will serve to determine the distributions of the premia, taken as random variables.
%From an investment perspective, it is useful to know the long-run average return of a factor, but being able to time it is obviously even more valuable.

Said differently, there is considerable value in capturing effects in several environments. For instance, if estimated effects $\hat{b}_p$ are positive (and possibly statistically significant) across $P$ environments, there is arguably more evidence in favor of it than from a single point estimate. This is exactly the rationale of \cite{peters2016causal}, a seminal contribution on invariance-based inference. Therein, the authors argue that a set of causal predictors (parents in a directed acyclic graph) can be obtained by taking the intersection of variables which are significant across all available environments. %Given a set of predictors, the aim is to extract those that are consistently chosen as drivers of a separate variable of interest.  
%The present paper departs from this contribution in several ways. First, we do not assume we are given a potentially large set of predictors. Each predictor can be treated on a standalone basis, hence our framework is suited even for just one independent variable. Our aim is not to infer causality, but to offer a more accurate characterization of the distribution of effects. Another important departure is the assumption of independence across environments. With forking paths, this is not a reasonable hypothesis because two close paths are likely to produce similar (correlated) outcomes. Hence, we must propose another approach.
In the present paper, forking paths replace environments, and we refrain from invoking causal effects. Paths are simply used to shed light on multitudes of facets of effects and we exploit these facets to provide a more exhaustive characterization of effects.

\subsubsection{Stylized conditions for convergence}
\label{sec:condconv}

Henceforth, we \textit{want} (and sometimes need) to assume that estimated effects $\hat{b}_p$ are such that their distribution converges to the true distribution of $b$, as the number of paths, $P$, increases to infinity. Below, we outline conditions under which this may occur. Fundamentally, we are seeking a concentration inequality and such types of results are well-known when the underlying variables are i.i.d. The major issue here is that it is not realistic to assume that outcomes from paths are independent variables. Recently, there have been advances in concentration inequalities for correlated sequences.\footnote{We refer to either of Berry-Esseen results (\cite{bentkus1997berry}, \cite{jirak2016berry,jirak2022berry}), or to Dvoretzky–Kiefer–Wolfowitz inequalities (\cite{kontorovich2014uniform}, \cite{chen2017concentration}). \label{ft:refs} } However, the central result we rely upon is Theorem 1 of \cite{azriel2015empirical}, which we recall below. 

One way of representing our line of thought is to write the sought effect as $b=\bar{b}+\tilde{b}$, i.e., the sum of a constant $\bar{b}$, the mean effect, plus a zero mean Gaussian random term $\tilde{b}$ with variance $\sigma_b^2$. For the remainder of the section, we thus propose the following generic data generating process
\begin{equation}
    \label{eq:DGP}
    Y_p = X_p(\bar{b}+ \alpha_p \tilde{b}) + (1-\alpha_p)\epsilon_p, \quad p=1,\dots,P,
\end{equation}
where $p$ indexes the path (environment) and $\alpha_p$ is a deterministic perturbation intensity, which, in all generality can be path-specific. The error terms $\epsilon_p$ are assumed to be independent from $\tilde{b}$ and exogenous, i.e., such that $\E[\epsilon_p |X_p]=0$. All vectors $Y_p$, $X_p$ and $\epsilon_p$ have the same length. However, from one path to another, this length may vary, as different environments may entail changing sample sizes. For a given path $p$, the standard OLS estimator is then 
\begin{equation}
    \label{eq:ols}
    \hat{b}_p = (X_p'X_p)^{-1}X_p'Y_p=\bar{b}+\alpha_p \tilde{b}+(1-\alpha_p)e_p, \quad e_p=(X_p'X_p)^{-1}X_p'\epsilon_p.
\end{equation}

The shrinkage intensity $\alpha_p \in [0,1]$ implies that the estimator does not capture the entirety of the random component of $b$. Plainly, $e_p$ is a blurring term that depends on path $p$ and the shrinkage form is commonplace in contamination models (e.g., from \cite{huber1964robust} to \cite{chen2016general}), wherein the data imperfectly reflects reality. In short, the estimated effects capture the average, but they are perturbed by some independent Gaussian noise. Note that if the perturbation has variance exactly equal to $\sigma_e^2=\frac{1+\alpha_p}{1-\alpha_p}\sigma_b^2$, then $\sigma_b^2=\sigma_{\hat{b}_p}^2$ and the estimator $\hat{b}_p$ has the same distribution as $b$. This \textit{strong} assumption is required if we need $L^2$ convergence of the full distribution of the empirical $\hat{b}_p$. If we are only interested in the \textit{average} effect, the variance term matters much less, as long as it is finite. Finally, the dependence between two estimators from two different paths can be shown to be 
\begin{equation}
\label{eq:cor}
\mathbb{C}\text{or}(\hat{b}_p,\hat{b}_q)=\alpha_p\alpha_q+(1-\alpha_p)(1-\alpha_q)\mathbb{C}\text{or}(e_p,e_q),
\end{equation}
which depends on the shrinkage intensities and on the correlation between the paths' perturbations.

To theoretically show the benefits of forking paths, we then rely on a stylized special case of the above framework. We assume for simplicity and tractability that estimated effects across all paths $\hat{b}_1, \dots, \hat{b}_P$ follow standard $N(0,1)$ Gaussian variables with correlation matrix $\bm{\Sigma}_P$ that gathers all pairs of correlations defined in Equation \eqref{eq:cor}. We underline that general Gaussian variates can also be considered since they can be normalized (demeaned and scaled). Then, from \cite{azriel2015empirical}, there exists a constant $c>0$ such that

\begin{equation}
   \underset{x \in \mathbb{R}}{\sup} \, \mathbb{E}\left[ (\Phi(x)- \Phi_{b, P}(x))^2 \right] \le \frac{1}{4P}+ c \| \bm{\Sigma}_P \|_1,
    \label{eq:convergence}
\end{equation}
where $\Phi(x)$ is the Gaussian cdf of the true effect and $\Phi_{\hat{b}, P}(x)$ the empirical distribution of estimated effects $\hat{b}_p$, obtained by the generation of $P$ paths. We thus seek conditions under which $ \| \bm{\Sigma}_P \|_1$ will shrink to zero as $P\rightarrow \infty$. 

Therefore, it remains to characterize the correlation between path outcomes. As advocated earlier, it makes sense that it be driven by the proximity between paths. Hence we postulate that the correlations between paths $p$ and $q$ are such that $\rho(p,q)=\mathbb{C}\text{or}(\hat{b}_p,\hat{b}_q)=h(d(p,q))$, where $d(\cdot,\cdot)$ is the distance function defined in Equation \eqref{eq:distance} and $h$ is some strictly decreasing function from $\mathbb{R}_+$ to $[-1,1]$ such that $h(0)=1$. Because of Equation \eqref{eq:convergence}, we are interested in

\begin{equation}
 \|\bm{\Sigma}_P\|_1 = P^{-2} \sum_{1 \le p,q \le P} | \rho(p,q)|= P^{-1} + 2 P^{-2} \sum_{1 \le p<q \le P}  | h(d(p,q))|.
\end{equation}

Without further assumptions, it is impossible to provide additional information on the empirical distribution of the distances. We are bound to specify a more explicit form for $h \circ d$. To do so, we first posit a simple distance function which corresponds to the number of layer choices that differ between $p$ and $q$:
\begin{equation}
    d(p,q)= \#\{j, r_{p,j}\neq r_{q,j} \} \in \{0,1,\dots, J \},
    \label{eq:distass}
\end{equation}
where the operator $\#\{ A\}$ measures the cardinal of set $A$ and we recall that $r_{p,j}$ is the mapping option through with path $p$ passes for layer $j$. Because there are $J$ layers in the protocol, we consider that the number of non-identical mapping options can be a good proxy for the differences between the related outcomes. If two paths have only one non-common choice, then, their outcomes should be more correlated than if they have zero commonality. As we show in Appendix \ref{app:CLT}, this simplification allows to fully characterize the distribution of distances and to determine conditions for convergence as summarized in the following result.

\begin{proposition}
\label{prop:CLT}
Assuming $\rho(p,q)=\rho^{d(p,q)}$ with $d$ given by \eqref{eq:distass}, we have that $\|\bm{\Sigma}_P \|_1 \overset{J \rightarrow \infty}{\rightarrow}0$, and, by \eqref{eq:convergence}, $\Phi_{\hat{b},P}$ converges uniformly to $\Phi$ in $L^2$. 
\end{proposition}

One important feature of the proposition is that the number of paths has to grow to infinity via the number of mappings $J$, not via the number of mapping options $r_j$. As we show in the proof in Appendix \ref{app:CLT}, if $J$ remains finite, the norm does not decrease to zero and convergence to the true cdf does not occur. In practice of course, $J$ will be finite but should be large, so that the norm of the matrix will be small, thereby implying small errors on the true cdf. 

Unfortunately, Proposition \ref{prop:CLT} remains a theoretical result, because there is no straightforward way to measure or estimate $\rho(p,q)$, which would allow us to evaluate the plausability of the proposition's assumption. Indeed, in our framework, we are endowed with only one outcome for each path, which makes correlation evaluation impossible. As \cite{anderson2004modelbook} put it bluntly: ``\textit{have no basis to estimate the across-model correlation}''. One way to bypass this hurdle would be to resort to a resampling of the original datasets that serve as input to the paths, $\mathbb{D}$. But this is out of the scope of the present paper. 

Another route toward convergence would be to assume that beyond a certain distance (e.g., $d(p,q)>7$), two paths are sufficiently dissimilar so that we can assume a zero correlation, or even independence. This relates to so-called $m$-dependent variables and their link to central limit theorems is investigated in \cite{hoeffding1948central}, \cite{diananda1955central} and \cite{orey1958central}. But this is not the road we follow here.

\subsection{The range of outcomes and its rate of increase}
\label{sec:range}

Another essential question which arises from paths, and underlined in Lemma \ref{lem:1}, is the speed at which extreme outcomes diverge as a function of $J$, the number of mappings (i.e., design choices). Practically, it is impossible to assess the impact of each layer on the dispersion of, say, $t$-statistics, because all mappings are chained and outcomes are only produced by the final layer.

We thus propose a tractable method to evaluate the impact of the richness of the protocol (the number of paths) on the range of outcomes. First, we set an integer $K \in [1,J-1]$ which will correspond to the number of mappings that are fixed. For each $K$, we write $c_l(K)$ for each combination (set) of fixed mappings; they are indexed by $l=1,\dots, {J \choose K}$. Each $c_l(K)$ is associated with $\prod_{k=1}^Kr_{k \in c_l(K)}$ fixed configurations and each configuration has $\prod_{k=1}^{J-K}r_{k \notin c_l(K)}$ possible paths. The rationale is that by fixing $K$ mappings, we leave $J-K$ as degrees of freedom in the protocol, as if there were in fact $J-K$ actual layers of design choices. Furthermore, testing all permutations allows to obtain a very rich collection of cases.

\begin{figure}[!h]
\begin{center}
\includegraphics[width=13.5cm]{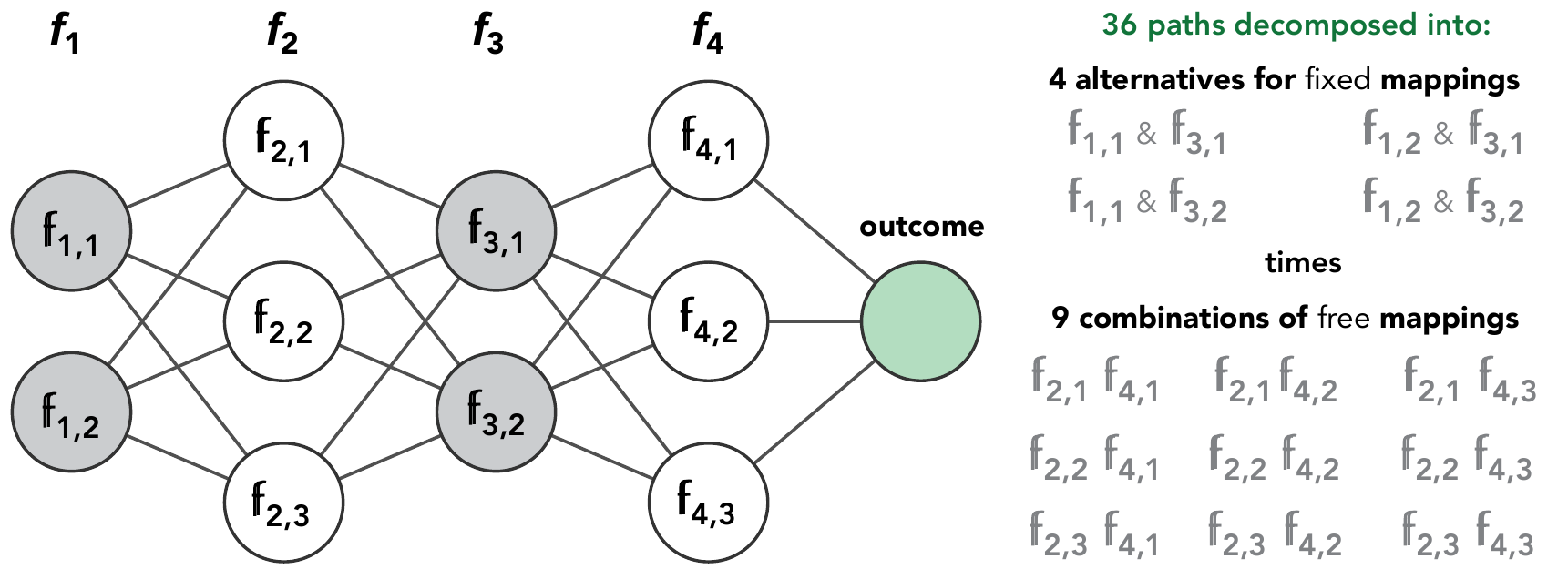}\vspace{-3mm}
\caption{\textbf{Spanning the paths: fixed versus free mappings}. \small The above representation features $J=4$ mappings with a number of options $r_1=r_3=2$ and $r_2=r_4=3$. The grey circles show the ($K=2$) \textbf{fixed mappings} while the white ones pertain to the \textbf{free} mappings.    }
\label{fig:paths}
\end{center}
\end{figure}

One such case is illustrated in Figure \ref{fig:paths} for $J=4$ and $K=2$. There are ${4 \choose 2}=6$ possible combinations of 2 mappings and we show one of them with the grey circles: we fix the first and third layers. Given the number of options, there are then 4 combinations for the fixed mappings, and 9 paths can be followed for these 4 combinations (when the first and third mappings are fixed to one of their possible alternatives).

For each $c_l(K)$ (i.e., one configuration like the one shown in Figure \ref{fig:paths}), it is possible to compute some statistics over all the outcomes generated by the free mappings. We write $S_{K,l}$ for the set of paths related to $c_l(K)$ and we are interested in the breadth of the outcomes related to the paths, which, following \cite{coker2021theory}, we call the range of \textit{hacking intervals}:
\begin{equation}
I_{K,l}=\underset{p \in S_{K,l}}{\max} \ o_p - \underset{p \in S_{K,l}}{\min} \ o_p.
\label{eq:hackint}
\end{equation}
Given these outputs, it is possible to calculate aggregate values, notably the average range of intervals, for each $K$:
\begin{equation}
    ARI(K)=\frac{1}{n_K}\sum_{l=1}^{n_K} I_{K,l},
    \label{eq:ari}
\end{equation}
where $n_K=\sum_{l=1}^{{J \choose K}}\prod_{k=1}^Kr_{k \in c_l(K)}$ is the total number of intervals for a fixed $K$. As $K$ increases, the number of free mapping decreases, and thus, the ARI is expected to decrease. In the spirit of Lipschitz constants mentioned in the previous section, we define the \textbf{growth rate of average intervals}, as the number of free mappings (i.e., $J-K$) increases:
\begin{equation}
\rho_{J-k} = \frac{ARI(J-k)}{ARI(J-k-1)}-1.
    \label{eq:rate}
\end{equation}
This rate is not supposed to be uniform with $k$. Indeed, in the early stages, the intervals are likely to expand rapidly, but, as $J-k$ increases, $\rho_{J-k}$ should shrink towards zero, so that additional design choices only impact the range of outcomes marginally. This will be confirmed in one of our empirical study in Section \ref{sec:expanding} below.

\subsection{Inference from paths}

Naturally, once the paths have been generated comes the question of the exploration of their outcomes. In this section, we revert back to the ubiquitous need for inference in empirical research when conclusions mostly depend on the statistical significance of a given effect. Below, we consider the question of the confidence intervals for the \textit{average} effect in Section \ref{sec:averaging}. To do so, we exploit a few ideas from the literature on model averaging. We refer to \cite{moral2015model}, \cite{zhang2019inference} and \cite{steel2020model} for surveys on the matter. In Section \ref{sec:cond_avg}, we outline our definition of \textit{conditional} averages.

\subsubsection{Model averaging}
\label{sec:averaging}

The expression for the estimated average effect is always written as
\begin{equation}
\hat{b}_* =\sum_{p=1}^Pw_p\hat{b}_p, \quad \text{with} \quad \sum_{p=1}^Pw_p=1,
\label{eq:coef}
\end{equation}
where the main issue is the determination of the weights $w_p$. This stage is very far from benign, as we will show and highlight in our empirical applications. When weights deviate from uniformity ($w_p=P^{-1}$), it means that paths are far from equal in generating trustworthy values for effects. 

We first consider a \textit{frequentist} average of the effects and follow \cite{buckland1997model} and \cite{anderson2004modelbook}. We define positive weights that rely on likelihood through information criteria:
\begin{equation}
w_p= \frac{e^{-\Delta_p/2}}{\sum_{k=1}^Pe^{-\Delta_k/2}}, \quad \Delta_p = AIC_p-\underset{p}{\min} \, AIC_p,
\label{eq:weights}
\end{equation}
where $AIC_p$ is the Akaike Information Criterion of model (i.e., path) $p$. For the estimation of the variance of the aggregate estimator, we follow Equation (1) in \cite{burnham2004multimodel} who make the conservative assumption of perfect correlation between estimators ($\mathbb{C}\text{or}(\hat{b}_p,\hat{b}_q)=1$):
\begin{equation}
\hat{\sigma}^2_*=\left(\sum_{p=1}^P w_p\sqrt{\hat{\sigma}_p^2 + (\hat{b}_*-\hat{b}_p)^2} \right)^2.
\label{eq:sigma}
\end{equation}
Assuming mild dependence conditions in order to be able to invoke the Central Limit Theorem to derive confidence intervals for $\bar{b}$ at the $\alpha$-level, we have: 
\begin{equation}
 \mathbb{P}\left[\bar{b} \in \left(\hat{b}_*-c_{\alpha/2}\hat{\sigma}_*/\sqrt{P} , \ \hat{b}_*+c_{\alpha/2}\hat{\sigma}_*/\sqrt{P} \right) \right] = 1-\alpha ,
    \label{eq:IC}
\end{equation}
where $c_\alpha$ is the quantile function of the standard normal law. Importantly, note that the speed of convergence, in contrast to most of the literature on model averaging for linear models,\footnote{We point for instance to \cite{zhang2019inference} and the references therein.} is not in the sample size, but in the \textbf{number of paths}. 

%In anticipation to our empirical results, we underline that the weights defined in Equation \eqref{eq:weights} will often be highly concentrated in a handful of models. The consequence is that standard deviations can be relatively small because the terms $(\hat{b}_*-\hat{b}_p)^2$ will be small whenever $w_p$ are non-negligible. 

\vspace{3mm}

For the sake of completeness, we also propose a weighting scheme from the perspective of \textbf{Bayesian} model averaging. We follow the standard nomenclature, as is for instance laid out in \cite{hoeting1999bayesian}. The quantity of interest is $b$, with posterior probability given the data $D$ equal to 
$$\PP[b | D]= \sum_{p=1}^P\PP[b|M_p,D]\PP[M_p|D],$$
where $\mathbb{M}=\{M_p,p=1,\dots,P\}$ is the set of models under consideration. In this paper, one model corresponds to one complete path. Notably, the above equation translates to the following conditional average and variance:
\begin{align}
    \E[b|D]&=\sum_{p=1}^P\hat{b}_p\PP[M_p|D] \label{eq:mean} \\
    \mathbb{V}[b|D]&=\sum_{p=1}^P \left(\mathbb{V}(b|M_p,D) + \hat{b}^2_p \right)\PP[M_p|D] - (\E[b|D])^2
    \label{eq:bayes_avg}
\end{align}
where $\hat{b}_p$ is the estimated effect from path $p$. The posterior model probabilities are given by
\begin{equation}
    \PP[M_p|D] = \left(\sum_{j=1}^P \frac{\PP[M_j]}{\PP[M_p]}\frac{l_D(M_j)}{l_D(M_p)} \right)^{-1},
    \label{eq:weights_2}
\end{equation}
with $l_D(M_j)$ being the marginal likelihood of model $j$. Because we are agnostic with respect to the relative importance of paths, we set the prior odds $\frac{\PP[M_j]}{\PP[M_p]}$ equal to one. Note that in this case, the posterior probabilities are then simply proportional to the likelihoods. The remaining Bayes factor is by far the most complex and we follow the recommendations of \cite{fernandez2001benchmark} (Equation (2.16), adapted for inhomogeneous sample sizes): 
\begin{equation}
\frac{l_D(M_j)}{l_D(M_p)}=\left(\frac{n_j}{n_j+1}\right)^{\frac{k_j}{2}}\left(\frac{n_p+1}{n_p}\right)^{\frac{k_p}{2}}\frac{\left(\frac{s_p+n_mv_p}{n_p+1} \right)^{(n_p-1)/2}}{\left(\frac{s_j+n_jv_j}{n_j+1} \right)^{(n_j-1)/2}},
\label{eq:odds}
\end{equation}
 where $n_j$ is the inverse of the number of observations used in model $j$ and $k_j$ is the number of predictors in this model, omitting the constant ($k_j=1$ in our case). Moreover, $n_jv_j$ is the sample variance of the dependent variable in model $j$. Finally, $s_j$ is the sum of squared residuals under model $j$. 

Averages of the form \eqref{eq:coef} along with confidence intervals \eqref{eq:IC} will be shown in Subsection \ref{sec:prem_baseline} both with frequentist \eqref{eq:weights} and Bayesian \eqref{eq:weights_2} weights.

\subsubsection{Conditional averages}
\label{sec:cond_avg}

Arguably one of the most important question with forking paths pertains to the impact of choices that researchers make on the distribution of the outcomes they generate. In Equation \eqref{eq:seq2}, the randomness in outcomes stems from the original sample $\mathbb{D}$, but also, and more importantly, from all modelling steps $f_j$ that constitute each path. One interesting extension pertains to the random variables $\hat{b}_p|\{f_j=\mathbb{f}\}$, which are the values of effects, conditional on the knowledge of one layer choice for mapping $j$, say, $\{f_j=\mathbb{f}\}$. Therefore, $\hat{b}_p|\{f_j=\mathbb{f}\}$ gathers all realizations of $\hat{b}_p$ such that the path $p$ ``\textit{passes}'' through the layer option $\{f_j=\mathbb{f}\}$. This notation will serve to determine if the operator $f_j$ has an important impact on the distribution of the outcome. 

%In practice, this is easily done by computing the same averages as those of the previous subsection, but on the subsets of results that correspond to particular paths. 

%For example, in our first study below, we find that that failing to correct an estimator for auto-correlation in residuals increases the tail of the distribution of $t$-statistics. This is expected, as in this case, the standard error is under-estimated and hence the statistic is over-estimated (in absolute terms). Hence the decision to resort to the OLS estimator or the HAC estimator will shift the distribution of statistics and potentially alter the conclusions. We show this by simply focusing of subgroups of paths, which is the avenue that we pursue in this subsection.

In order to test if one operator (i.e., layer) has an impact on the estimated outcomes, we define 
\begin{equation}
\hat{b}^{\{f_j=\mathbb{f}\}}_p \quad \text{and} \quad \hat{b}^{\{f_j=f\}}_{-p}
    \label{eq:cond_mean}
\end{equation} 
as the random variables $\hat{b}_p|\{f_j=\mathbb{f}\}$ and $\hat{b}_{-p}|\{f_j=f\}$, respectively, for two different mapping alternatives $\mathbb{f}\neq f$ at layer $j$. Note that we introduce a special notation with a negative index. The variable $\hat{b}^{\{f\}}_{-p}$ corresponds to the outcome of the path which is exactly the same as path $p$, except for layer $j$. In other words, the two effects defined in \eqref{eq:cond_mean} come from paths which are very close and have only one difference. This is depicted in Figure \ref{fig:cond_avg} below. The test layer ($f_j$) has two options (e.g., winsorizing or not) and we discriminate the paths based on whether they pass through one option (in \textcolor{bblue}{\textbf{dotted blue}}) or the other (in \textcolor{orange}{\textbf{orange}}). 

\begin{figure}[!h]
\begin{center}
\includegraphics[width=11.5cm]{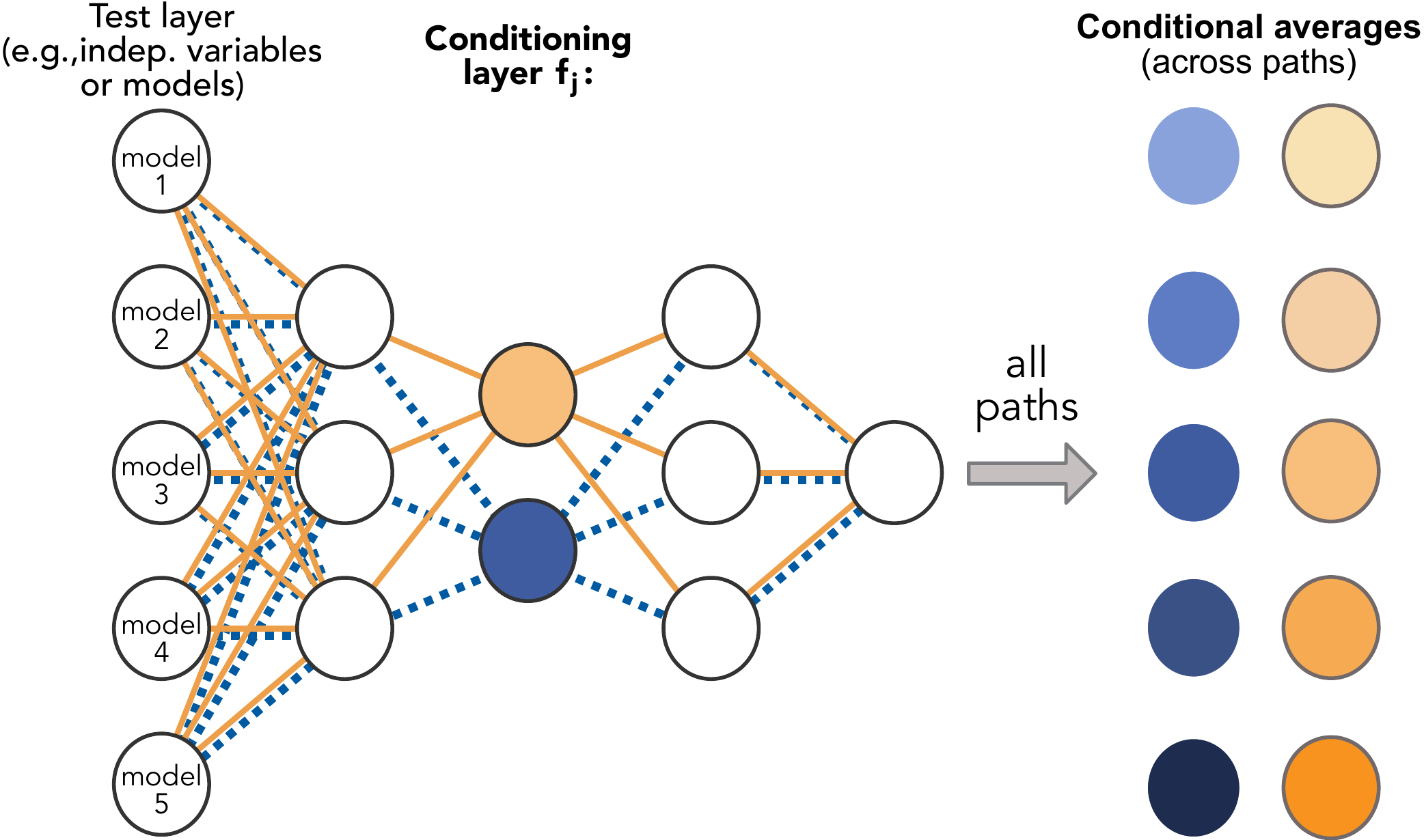}\vspace{-3mm}
\caption{\textbf{Spanning the paths: passing through alternative mapping options}. \small We depict the notion of average effects, conditional on some layer choices. For instance, the variables $\hat{b}^{\{f_j=\mathbb{f}\}}_p$ corresponds to the outcomes of all dotted \textcolor{bblue}{\textbf{blue paths}} while $\hat{b}^{\{f_j=f\}}_{-p}$ to the outcomes of all \textcolor{orange}{\textbf{orange paths}}.    }
\label{fig:cond_avg}
\end{center}
\end{figure}

Once the set of paths has been separated in two (or more, depending on the number of layer options), it is possible to test whether the distributions of $\hat{b}^{\{\mathbb{f}\}}_p$ and $\hat{b}^{\{f\}}_q$ are different. While tests on the full distribution (Kolmogorov-Smirnov) are available, we prefer to stick to simple mean tests. However, because estimation outputs are weighted prior to inference, we need to apply the corresponding weights before averaging ($w_p$ and $w_{-p}$). Thus, we can resort to a simple Student $t$-test to the series of outputs
\begin{equation}
    \Delta_p = \frac{2}{P}\left(w_p\hat{b}^{\{f_j=\mathbb{f}\}}_p -w_{-p} \hat{b}^{\{f_j=f\}}_{-p} \right) , 
    \label{eq:test}
\end{equation}
where the $2/P$ scaling is there to ensure that a simple mean of the above series over the $P/2$ paths will be exactly equal to the difference in weighted averages. This can easily be generalized to decision layers with more than two options. 

One canonical option for $f_j$ is the subsample. For instance, \cite{goyal2021comprehensive} report their results for the first and second halves of their samples. Their results underline the marked sensitivity of reported effects to the choice of the period. We will forcefully corroborate this phenomenon. The application of the above ideas are located in all of our empirical analyses, in Subsections \ref{sec:prem_cond_avg}, \ref{sec:sorts_stability},  and \ref{sec:FM_options}.

\subsection{Exhaustive multiple testing}

% critique of BS: \cite{young1994bootstrap}

\label{sec:multiple}

Not all mappings are equal. For instance, the choice of independent variable may very well be a strong modelling assumption. Consider the two alternative questions:
\begin{itemize}
\setlength\itemsep{-0.2em}
    \item Does variable $X$ predict variable $Y$?
    \item Can variable $Y$ be predicted?
\end{itemize}

In the first case, it the focus is clearly on the predictive ability of variable $X$, hence picking it as independent variable is not a design choice, it is an imperative. In the second option, choosing variable $X$ or $Z$, or $W$ is left to the appreciation of the researcher, and, in fact, it is conceivable to mine as much data as possible to find the few predictors that may indeed predict $Y$. Typically, in the debate on the predictability of the equity premium, dozens of variables have been proposed. If many are studied, statistical significance must be corrected for \textbf{multiple testing}.\footnote{Because we propose to generate series of outcomes, the ideas presented in the present paper are undoubtedly linked to this notion, a theme that goes back at least to \cite{bonferroni1936teoria}, and which is applied in serveral disciplines, including medicine (\cite{farcomeni2008review}), economics (\cite{viviano2021should}), finance (\cite{harvey2020false}, \cite{harvey2020evaluation}, \cite{giglio2021thousands}), generic model discrimination (\cite{hansen2011model}) and statistics more generally (\cite{fan2017estimation}, \cite{wang2017confounder} to cite but a few). In many cases, as in \cite{romano2005stepwise, romano2010balanced} or \cite{wilson2019harmonic}, the methods take as input series of test statistics (or $p$-values).}  Recently, several studies in finance have approached research questions by resorting to large scale tests in which many predictors are considered (\cite{yan2017fundamental}, \cite{chordia2020anomalies}, \cite{giglio2021thousands} and \cite{jensen2021there}).

\begin{figure}[!h]
\begin{center}
\includegraphics[width=13cm]{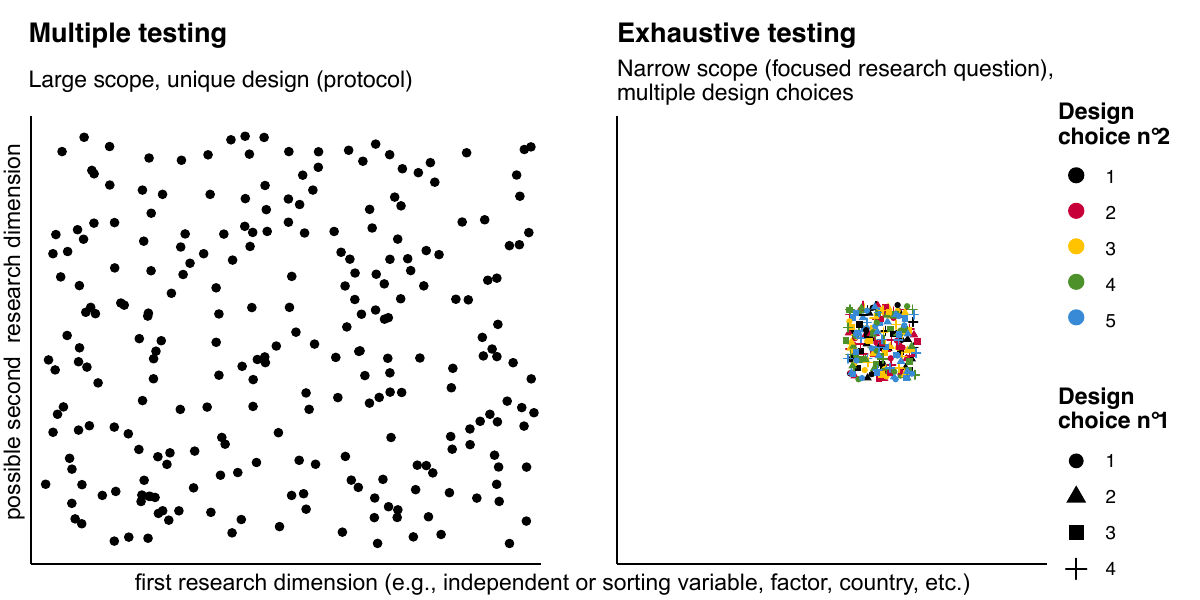}\vspace{-3mm}
\caption{Illustration of \textbf{model rich and protocol poor} (many hypothesis, one path) versus \textbf{model poor and protocol rich} studies (one hypothesis, many paths). }
\label{fig:scheme6}
\end{center}
\end{figure}

The approach we have advocated until now is slightly different. We have assumed that the researcher has a precise research question in mind, but that there are many different ways to answer it, thanks to small shifts, or tweaks, in the empirical protocol, exactly as in \cite{huntington2021influence} and \cite{menkveld2021standard}. % This is also the spirit of model averaging, or \textbf{multi-model inference} (\cite{burnham2004multimodel}). One common way of extracting multiple coefficients is to consider several combinations of control variables in regressions, as in \cite{zhang2019inference} and many references therein. One way to put it is to consider that exhaustive robustness checks must become the baseline result.
The difference with multiple testing is illustrated in Figure \ref{fig:scheme6}. In the left graph, the space of models is wide, and all hypotheses are tested with the same unique protocol (e.g., simple portfolio sorts, or linear models). In the right plot, the scope is narrower, and the number of hypothesis is small (e.g., just one), but the methods used to reach conclusions are heterogeneous. In financial economics, the first type can be found in \cite{jensen2021there}, in which the authors approach the topic of asset pricing anomalies via a large-scale study encompassing thousands of firms worldwide and testing hundreds of factors. The portfolios are all constructed using the same methodology. Two examples of the second type of studies are \cite{asness2013devil} and \cite{amenc2020intangible}, wherein the authors focus solely on the \textbf{value} anomaly, but propose alternative ways to construct value factors. Similar analyses have been carried out for the \textbf{momentum} anomaly (see \cite{novy2012momentum} and \cite{gong2015momentum}), and, more generally, to a broad scope of anomalies in \cite{soebhag2022mind} and \cite{walter2022non}.

In this subsection, we argue that it is in fact possible to combine the two, i.e., to enhance multiple testing with exhaustive protocols. The rationale is the following. Most state-of-the-art techniques used in multiple testing (see \cite{harvey2020evaluation}) rely on bootstrapping. Now, as is shown in \cite{romano2005stepwise} (Assumption 3.1), this requires that the sampling distribution that emanates from the data consistently estimates the true distribution of the effect. This is arguably a strong assumption which completely precludes sampling bias, which is why we propose an alternative procedure to generate more diverse samples.

Below, we formally recall the so-called \textit{bootstrap reality check} (BRC) of \cite{white2000reality}, as it is exposed in \cite{harvey2020evaluation}. It will serve as benchmark, and the method can effortlessly be extended to the StepM procedure of \cite{romano2005stepwise}. We are given $T \times N$ observations $x_{t,n}$, where $T$ stands for the sample size and $N$ is the number of tests. These observations are bootstrapped $B$ times to yield a $B\times T \times N$ tensor $x_{t,n}^{(b)}$. Here, $b$ is the index of the bootstrapped sample. Bootstrapped statistics are defined as
\begin{equation}
    t_n^{(b)}=\sqrt{T} \ \frac{\mu_n^{(b)}-\mu_n}{\sigma_n^{(b)}},
    \label{eq:boot}
\end{equation}
where $\mu_n^{(b)}$ and $\sigma_n^{(b)}$ are the sample mean and standard deviation of each bootstrap series. $\mu_n$ is the sample mean of the original (non bootstrapped) data. We write $\tilde{t}_n^{b}$ for the statistics ordered such that $\tilde{t}_n^{(b)} \ge \tilde{t}_{n+1}^{(b)}$, so that, for each bootstrap sample $b$, $\tilde{t}_1^{(b)}$ is the largest statistic. We are then given a confidence level $l$, say $l=95$\%. The target threshold for the test is then the $l$ quantile of the vector $\tilde{t}_1^{(b)}$.

Now, instead of resorting to bootstrapped samples, we propose to use forking paths to generate alternative versions of the same problem. The rationale is that one path is subject to sampling bias, whereas many paths provide a richer characterization of possible states of the world (different sub-periods, various weighting schemes, etc.). Consequently, we propose the following adjustment to Equation \eqref{eq:boot}:

\begin{equation}
    t_n^{(p)}=\sqrt{T_p} \ \frac{\mu_n^{(p)}-\mu_n}{\sigma_n^{(p)}},
    \label{eq:emt}
\end{equation}
where bootstrapped samples $b$ are replaced by paths $p$ which have sample sizes $T_p$. In this case, $\mu_n$ can be the average over one representative path, or over any set of paths. Like for the bootstrapped statistics, the values are sorted to yield $\tilde{t}_1^{(p)}$.

The main difference between the two approaches is that, in the first case, $\mu_n^{(b)}$ is computed via the same data as $\mu_n$, so that the numerator is expected to be smaller in magnitude compared to $\mu_n^{(p)}-\mu_n$ because $\mu_n^{(p)}$ will be based on possibly very different samples. One important question is whether the denominators will mitigate these differences so that $\sigma_n^{(p)}$ will shrink the dispersion of path outcomes.

The consequences can further be theoretically illustrated. Say we consider $N$ Gaussian variables $t_n^{(\sigma)}$ which stand for anomalies' test statistics. For ease of exposition and analytical tractability, we assume that they are independent, with common zero mean and standard deviation $\sigma$. Then, we define the cdf of their maximum:
\begin{equation}
    F_\sigma(x)=P\left[\max_{n \le N} \ t_n^{(\sigma)} \le x\right]=\Phi_\sigma(x)^N,
\end{equation}
where $\Phi_\sigma(x)$ is the related Gaussian cdf. We are interested in the sensitivity of the quantile function (for the $p$-value, or significance threshold) with respect to $\sigma$, and for a fixed $x$, say $x=95\%$:
\begin{align*}
     \frac{\partial}{\partial \sigma}(F_\sigma^{-1})(x) &= \frac{\partial}{\partial \sigma} \left(\sigma \sqrt{2} \, \text{erf}^{-1}(2x^{1/N}-1) \right)= \sqrt{2} \, \text{erf}^{-1}(2x^{1/N}-1)>0 \quad \text{for} \ x > 2^{-N},
\end{align*}
where erf$^{-1}$ is the inverse error function. The above result intuitively proves that, as the dispersion of statistics increases, the threshold of their maximum will also increase. Given that we expect more dispersion from path-generated outputs, the resulting decision hurdles for significance should be larger. This will be investigated empirically and corroborated in Section \ref{sec:MT}.

The fact that path-generated decision thresholds will be more conservative implies that they will provide a stricter control over first type errors (false positives). However, this also means that they will be more prone to errors of the second kind (false negatives). The recent literature on multiple testing (\cite{romano2005stepwise}, \cite{harvey2020evaluation}, \cite{harvey2021lucky}) is very focused on this notion of \textbf{power}, i.e., the need to avoid as many false negatives as possible. This makes sense: controlling type 1 errors, while maximizing power amounts to seeking maximum accuracy and hence revealing a maximum number of \textit{true} factors.

Nevertheless, there are often asymmetries between the consequences of both types of errors. In the money management industry, false positives incur losses (at least relatively to a benchmark), whereas false negatives are simply missed opportunities. Hence, arguably, investors may be more sensitive to false positives. Exhaustive multiple testing provides a stringent filter: anomalies that pass the threshold are immune to a large amount of variation in protocols and sub-sampling. Seeking this kind of robustness should be a prerequisite in portfolio backtesting. However, a drawback is that our approach is not well suited for purely inferential purposes.

\subsection{Corroborating published results}

\label{sec:suspicious}

Published results can be subject to bias towards positive outcomes and consequently researchers often seek to reproduce studies. Most of the time, perfect replicability is impossible because of data availability, interpretation leeway, and coding choices (see \cite{perignon2022reproducibility} for more on the matter). Nevertheless, even if coefficients or $t$-statistics cannot exactly be reproduced, it is useful to evaluate if their magnitudes are reasonable - and forking paths are the appropriate tool for this task. 

Indeed, once paths have been generated, it is interesting to compare the distribution of their outcomes to results of contributions that have quantified the same effects. Let us illustrate this with an example from our study in Section \ref{sec:premium}. In Figure \ref{fig:bm} below, we plot the histogram of coefficients of predictive regressions when forecasting the market return with the aggregate book-to-market ratio, along 1,152 paths. From the distribution of the path-generated coefficients in the Figure, it is clear that the link between market returns and the valuation ratio is positive. Nevertheless, the value documented in \cite{goyal2021comprehensive} lies quite at the right of the distribution, meaning that it corresponds to some of the \textit{best} outcome from our 1,152 paths. 

\begin{figure}[!h]
\begin{center}
\includegraphics[width=14cm]{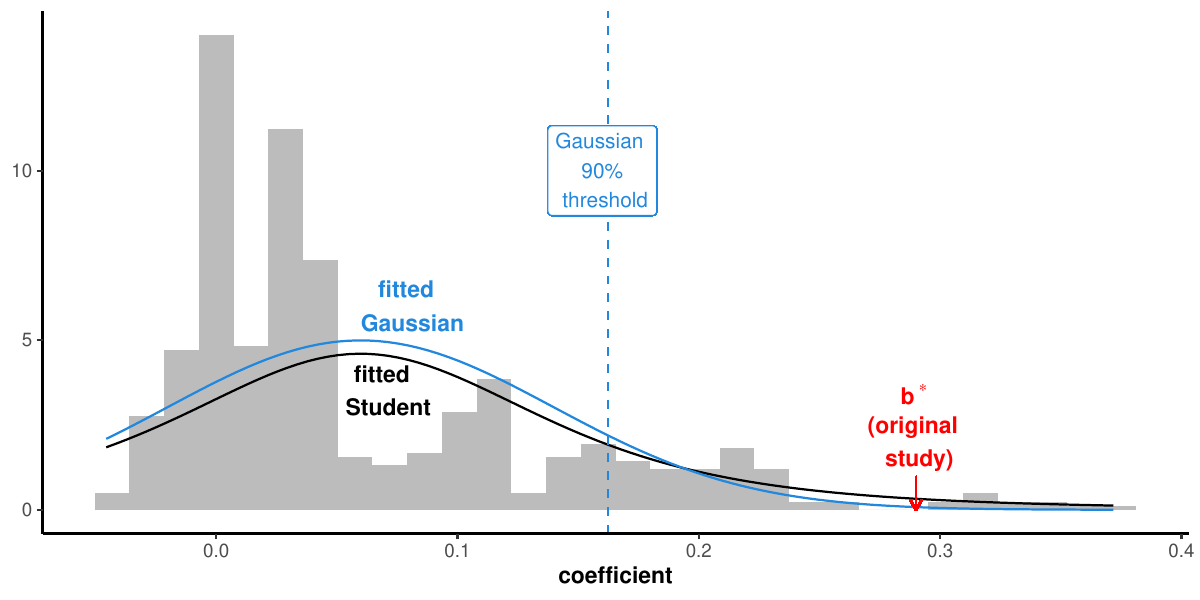}\vspace{-3mm}
\caption{\textbf{Comparison with prior work}. \small We report the distribution of coefficients in a predictive regression exercise. The independent variable is the \textit{aggregate book-to-market}. At the right of the histogram, we show the value ($b^*=0.29$) from the study of \cite{goyal2021comprehensive}. }
\label{fig:bm}
\end{center}
\end{figure}

In order to locate the position of outcomes from other studies within the distribution of path-generated results, we naturally propose to think in terms of quantiles. This way, the external results are assessed through the lens of their position within the paths' outcomes. If $\Phi_P$ is the empirical cdf of the path-generated outcome and if $b^*$ is the outcome of reference from a prior study, then we define the probability level of the prior effect as $\Phi_P(b^*)$. It is the probability that the documented effect size $b^*$ be larger - more extreme and favorable - than path values. Henceforth, we assume without loss of generality that the effect is positive and that we consider values to the right of the distribution to be favorable. For negative effects, we simply need to change the signs of results ex-ante.

For instance, if the probability $\Phi_P(b^*)$ is 0.6, then it means that the prior published result is close to the middle of the distribution obtained from the paths and thus there is little reason to suspect any cherrypicking because the reported value is so to speak \textit{representative} of the plurality of outcomes. However, if $\Phi_P(b^*)=0.99$, then the documented effect is more extraordinary. For example, in Figure \ref{fig:bm}, $\Phi_P(b^*)=0.98$.

Nevertheless, we see that this approach faces a potential limit when the external value lies outside the range of path-generated results, which will occur in Section \ref{sec:premium}. To address this issue, we propose to resort to parametric continuous cdfs, which we will write $\hat{\Phi}_P$, as their parameters will be estimated from paths' outcomes. This is illustrated in Figure \ref{fig:bm} with the \textcolor{bblue}{blue} curve which shows the fitted Gaussian density. Under this density, the result of \cite{goyal2021comprehensive} study is located at the 99.8\% level. If we want to allow for more diversity in outcomes and increase tails, it is possible to switch to other distributions, like the Student law (in black in Figure \ref{fig:bm} - with 3 degrees of freedom). In this case, the probability shrinks to 96.8\%. 

Based on these levels, our aim is to devise an indicator that evaluates the likelihood that a given outcome is \textit{outstandingly} favorable. To do so, we focus on the empirical distribution of $b$, conditional on $b$ being larger than some threshold effect $\theta$. For instance, we pick $\theta$ to be at the 90\% level of all paths-generated effects (or $t$-statistics) so that $\Phi(\theta)=0.9$. This is shown with the vertical \textcolor{bblue}{blue} line in Figure \ref{fig:bm}. Plainly, $\theta$ is a large yet realistic value for $b$. We then define the \textbf{odds of favorable outcome} (OFO) for $b^*$ as the probability with respect to the right-truncated distribution as follows

\begin{equation}
\mathbb{OFO}_P(b^*,\theta)=\mathbb{P}[b < b^*|b>\theta] := \frac{\Phi_P(b^*)-\Phi_P(\theta)}{1-\Phi_P(\theta)}\mathbb{1}_{\{b^*> \theta \}} \approx \frac{\hat{\Phi}_P(b^*)-q}{1-q}\mathbb{1}_{\{b^*> \theta \}} \in[0,1] ,
\label{eq:Phi}
\end{equation}
where, in the approximation, we have simply replaced $\Phi_P$ by its parametric proxy $\hat{\Phi}_P$ and $q=\hat{\Phi}_P(\theta)$ is a high benchmark level such as 0.9 for instance. Simply put, the OFO measures the likelihood of the reported effect to be above a value that is already considered large based on the path outcomes. Plainly, the OFO only makes sense for $\theta \le  b^*$. Moreover, it strongly relies on two key hypotheses. First, we assume that, as the number of paths increases, the average of the effects converges to the true value $\bar{b}$.\footnote{This is a very technical point which is the subject of a separate paper, essentially based on Theorem 1 of \cite{azriel2015empirical}.} Second, the OFO will depend, via $\hat{\Phi}_P(b^*)$, on the standard deviation of the effects across the paths. In order not be too conservative against published results, we hope to report values that can only understate the true OFO obtained with the actual distribution of effects, $\Phi$. This happens if the realized standard deviation of effects across paths is larger than $\sigma_b$, the true deviation.  

A favorable feature of the OFO metric is that it can be \textit{loosely} interpreted as a probability of original numbers being uncommonly favorable. Reversely, this can also be viewed as a measure of replicability: a small OFO signals a value that can be easily reproduced. For simplicity, we will henceforth report the EtC metric, which stands for ``\textit{Ease to Confirm}'' or ``\textit{Ease to Corroborate}'', and is equal to one minus the OFO: a value close to zero (\textit{resp.} one) signals a result that is hard (\textit{resp.} easy) to confirm. 

Let us exemplify the indicator with the values from Figure \ref{fig:bm}. In this case, $\hat{\Phi}(b^*)=0.98$ with a Gaussian proxy. We first fix $q=0.9$ as extreme benchmark quantile and, consequently, $\text{EtC}=1-\frac{0.998-0.9}{1-0.9}=0.02$ and the result from \cite{goyal2021comprehensive} stands out as clearly favorable and relatively hard to reproduce. If we impose an even more conservative threshold $q=0.95$, then $\text{EtC}=0.04$, a slightly higher score, but still underlining a figure that is hard to confirm. 

Importantly, effect sizes can be both highly statistically significant \textit{and} associated with EtC values that are very close to 100\%, i.e., easy to reproduce. This corresponds to situations in which it is not arduous to confirm positive findings. We will substantiate this claim with further examples in Subsections \ref{sec:premium_prior}, \ref{sec:sorts_prior} and \ref{sec:FM_prior}.

\section{Application: equity premium prediction}
\label{sec:premium}

\subsection{Data}

For the sake of reproducibility, the first illustration of the concepts of the paper rely on a public dataset as well as on a problem which is widely documented in the literature.\footnote{The code used to generate all results is available at \href{https://www.gcoqueret.com/files/misc/forking_paths.html}{https://www.gcoqueret.com/files/misc/forking\_paths.html}. The first version has been verified by the \textbf{cascad} certification service: \href{https://www.cascad.tech/certification/116-forking-paths-in-empirical-studies/}{https://www.cascad.tech/certification/116-forking-paths-in-empirical-studies/}} In financial economics, an old, still unresolved, question pertains to whether aggregate stock returns can be predicted by macro-economics indicators. The debate is likely impossible to settle, but recent results suggest a contingency on return horizon (\cite{bandi2019scale}), even if long-term predictability is biased by construction for simple estimators (\cite{boudoukh2008myth}, \cite{boudoukh2021biases}).

A critical view on the matter is the seminal article by \cite{welch2008comprehensive}, in which the authors document the poor forecasting ability of traditional macro-economic predictors. A favorable feature of the study is that the data is public, and has even been updated in the follow-up paper \cite{goyal2021comprehensive}. It is this material, updated until December 2021, that we use for our application.

For the sake of reproducibility, the first illustration of the concepts of the paper rely on a public dataset as well as on a problem which is widely documented in the literature.\footnote{The code used to generate all results is available at \href{https://www.gcoqueret.com/files/misc/forking_paths.html}{here}. The first version of the code has been verified by the \href{https://www.cascad.tech/certification/116-forking-paths-in-empirical-studies/}{\textbf{cascad} certification service}.} In financial economics, an old, still unresolved, question pertains to whether aggregate stock returns can be predicted by macro-economics indicators. The debate is likely impossible to settle, but recent results suggest a contingency on return horizon (\cite{bandi2019scale}), even if long-term predictability is biased by construction for simple estimators (\cite{boudoukh2008myth}, \cite{boudoukh2021biases}).

A critical view on the matter is the seminal article by \cite{welch2008comprehensive}, in which the authors document the poor forecasting ability of traditional macro-economic predictors. More recently, \cite{dichtl2021data} have confirmed the meager out-of-sample performance of most prediction methods. In addition, \cite{engelberg2021cross} also report weak results when using aggregated cross-sectional indicators. 

A favorable feature of the \cite{welch2008comprehensive} study is that the data is public, and has even been updated in the follow-up paper \cite{goyal2021comprehensive}. It is this material, updated until December 2021, that we use for our application. 

\subsection{Forking paths}

\begin{figure}[!h]
\begin{center}
\includegraphics[width=15.2cm]{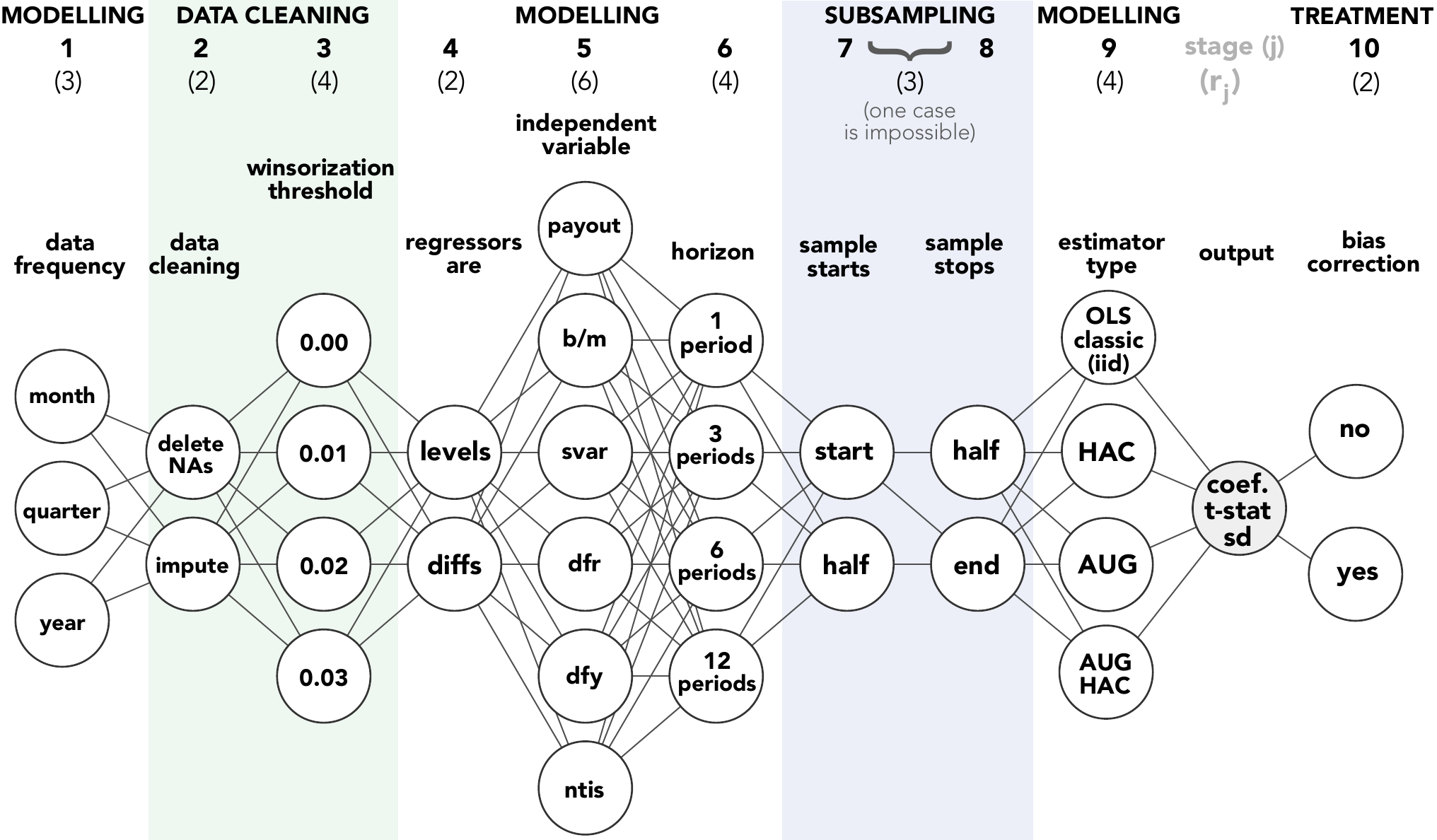}\vspace{-3mm}
\caption{\textbf{Diagram of empirical protocol} (forking paths). \small The graph, akin to a neural network structure, depicts the ten-step algorithm used to produce the $t$-statistic in the study. Each path has the same probability of realization (uniform distribution). The number of mapping options (circles) is reported between brackets below the stage number. There are 27,648 paths in total. }
\label{fig:scheme2}
\end{center}
\end{figure}

In order to generate enough metrics, we consider $J=10$ stages (layers) which are depicted in Figure \ref{fig:scheme2}. We briefly comment on each below: \vspace{-2mm}

\begin{enumerate}
   \setlength\itemsep{-0.2em}
    \item \textbf{data frequency} determines the horizon of returns, hence the left-hand side of the equation. In addition, this has a major effect on sample sizes, as annual samples are 12 times smaller, compared to monthly ones. 
    \item \textbf{handling missing points} boils to two options. The first is to remove rows of missing points, which means all regressors will start at the same point in time (1927). The second option (imputation of previous value) allows predictor-dependent sample sizes and some of them are available in 1871. Thus, this stage impacts sample depth but all the samples from the study have sizes above 40.
    \item \textbf{winsorization} defines the cutoff threshold for the taming of outliers, from none (0\%) to 3\%. The data is often verified, thus all values are trustworthy, so this step could theoretically be omitted. But it participates to increase the number of outputs, hence we keep it for the sake of exhaustiveness.
    \item \textbf{variable engineering} decides whether or not to use levels or differences in the regressions. 
    \item \textbf{independent variable} sets the predictor. Six options are possible and all are available across the three frequencies (monthly to annually).\footnote{\textbf{payout} is the difference between the log of dividends and the log of earnings, \textbf{b/m} is the the ratio of book value to market value for the Dow Jones Industrial Average, \textbf{svar} is the sum of squared daily returns on S\&P 500, \textbf{dfr} is the difference between the return on long-term corporate bonds and returns on the long-term government bonds, \textbf{dfy} is is the difference between BAA- and AAA- rated corporate bond yields, and \textbf{ntis} is the ratio of twelve-month moving sums of net issues by NYSE listed stocks divided by the total market capitalization of NYSE stocks.} 
    \item \textbf{horizon} fixes the number of periods that are used to compute the future return (dependent variable). We underline that three periods have different meanings depending on the original data frequency (chosen in step 1). 
    \item \textbf{starting point} determines if the sample commences at its first point, or at its middle point. This option leaves room for sub-sampling (on the two halves of each original sample).
    \item \textbf{end point} is either the end of the sample, or its middle point. The latter option is not possible if it also corresponds to the starting point.
    \item \textbf{estimation method} chooses between three specifications. First, the simple OLS with iid errors. Second, the improved HAC variance estimator of \cite{newey1987simple}. The regression model for these two variations is simply\footnote{To keep scales comparable across horizons and frequencies, we scale the dependent variable by the square root of horizon time frequency. If the frequency is quarterly and the horizon 12 periods, we divide $y$ by $\sqrt{36}$. Hence the baseline scale is the monthly return. }
    \begin{equation}y_{t+h}=a+bx_t+e_{t+h},\label{eq:remodel} \end{equation} where $y_{t+h}$ is the equity premium (at horizon $h$), $x_t$ the lagged predictor and $e_{t+h}$ the residual. The third version is the augmented regression suggested in \cite{hjalmarsson2011new}:
    \begin{equation}y_{t+h}=a+bx_t+\gamma \nu_{t+1}+e_{t+h},\label{eq:remodelaug} \end{equation}
    where $\nu_t$ is the innovation process stemming from the predictor. More precisely, $\nu_t=x_t-\hat{\delta}x_{t-1}$, where $\hat{\delta}$ is the estimated coefficient from $x_t=\alpha+\delta x_{t-1}+\varepsilon_t$ (the constant term does not matter). The rationale for this is to treat potential endogeneity upfront, in addition to being very close to the method proposed by \cite{amihud2004predictive} to solve the bias raised by \cite{stambaugh1999predictive}.\footnote{The only difference is that instead of using $\hat{\delta}$, \cite{amihud2004predictive} recommend to take $\hat{\delta}+(1+3\hat{\delta})/n+3(1+3\hat{\delta})/n^2$, which is close to $\hat{\delta}$ whenever $n$ is large enough, say $n>40$.} Finally, the fourth method combines the second and third (augmented regression with HAC variance estimation). The first two and last two have equal coefficients, but different $t$-statistics.  
    \item \textbf{post-treatment} seeks to correct potential error-in-variables bias. Once all paths have been generated, we have the choice follow or not Proposition 2 from \cite{barras2022skill} to adjust the distribution of $t$-statistics. 
\end{enumerate}

There are $\prod_{j=1}^{10}r_j=27,648$ possible paths from the data to the output. For simplicity, each is equi-probable so that we only need to consider each combination once.

\subsection{Model averaging}

\label{sec:prem_baseline}

\begin{figure}[!h]
\begin{center}
\includegraphics[width=13.5cm]{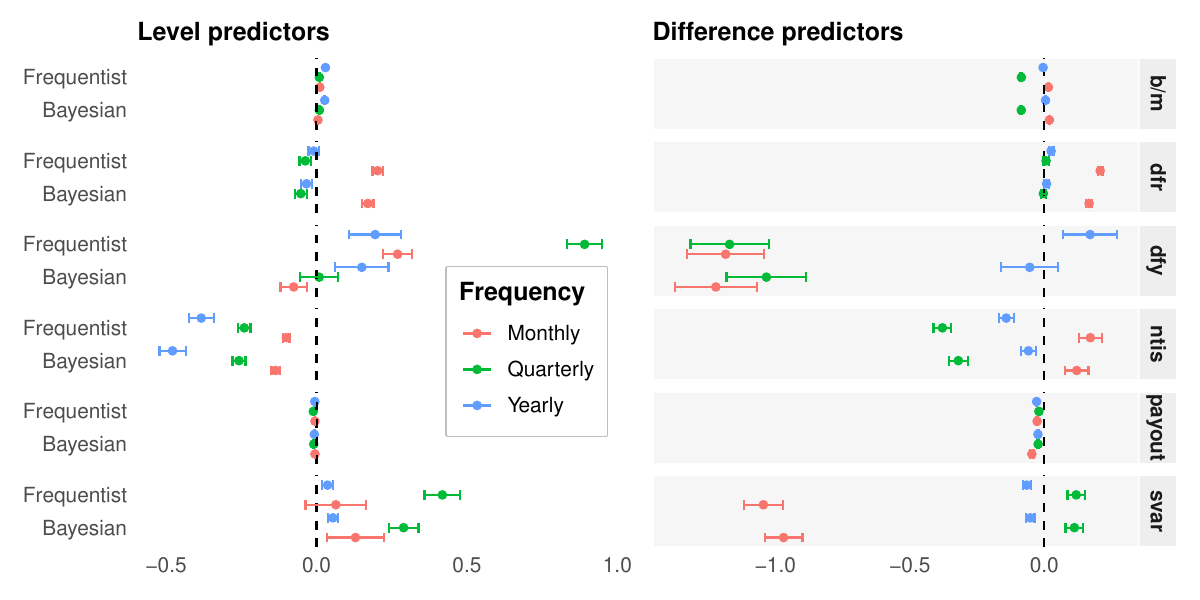}\vspace{-3mm}
\caption{\textbf{Frequentist model averaging}. \small We display average coefficients within their 95\% confidence interval. Coefficients stem from Equation \eqref{eq:coef} and predictors are scaled by their standard deviations prior to estimation to ease comparison of scales. Confidence intervals are defined by $[\hat{b}_*-1.96\sigma_*^2/\sqrt{T_*},\hat{b}_*+1.96\sigma_*^2/\sqrt{T_*}]$, where $T_*=\sum_{j=1}^Jw_jT_j$, with $T_j$ being the sample size of model $j$. The left panel displays results when predictors are levels, while the right one focuses on differences of variables. To allow comparisons, all predictors are scaled to have unit variance before estimation. Only the paths with no bias adjustment are considered.  }
\label{fig:MA}
\end{center}
\end{figure}

In Figure \ref{fig:MA}, we show the averaged coefficients within their 95\% confidence interval. We split the analysis along three axes: variable, level versus difference, and sampling frequency. The latter is important because it is determinant in the sample sizes which are used to compute the width of the intervals. For a given frequency and variable, they are homogeneous, though not exactly equal, and we use their weighted average $T_*=\sum_{j=1}^Jw_jT_j$. In the figure, the impact of sampling frequency on the width of intervals is obvious. Intervals pertaining to monthly data coincide with the average estimators, whereas intervals linked to annual samples are fairly large. By definition, large sample shrink the interval ranges.

We observe that \textit{ntis} yields only negative coefficients, and the intervals do not overlap with zero. The \textit{b/m} variable has mostly positive estimates, with one exception of the quarterly data in the right panel. Surprisingly, the \textit{dfy} variable also stands out with coefficients which are large in magnitude for quarterly and annual samples. For quarterly variables, the effect cancels out between levels (positive coefficients) and differences (negative ones). This partly explains why the variable was not previously identified as a potent driver of the equity premium.

\begin{figure}[!h]
\begin{center}
\includegraphics[width=14cm]{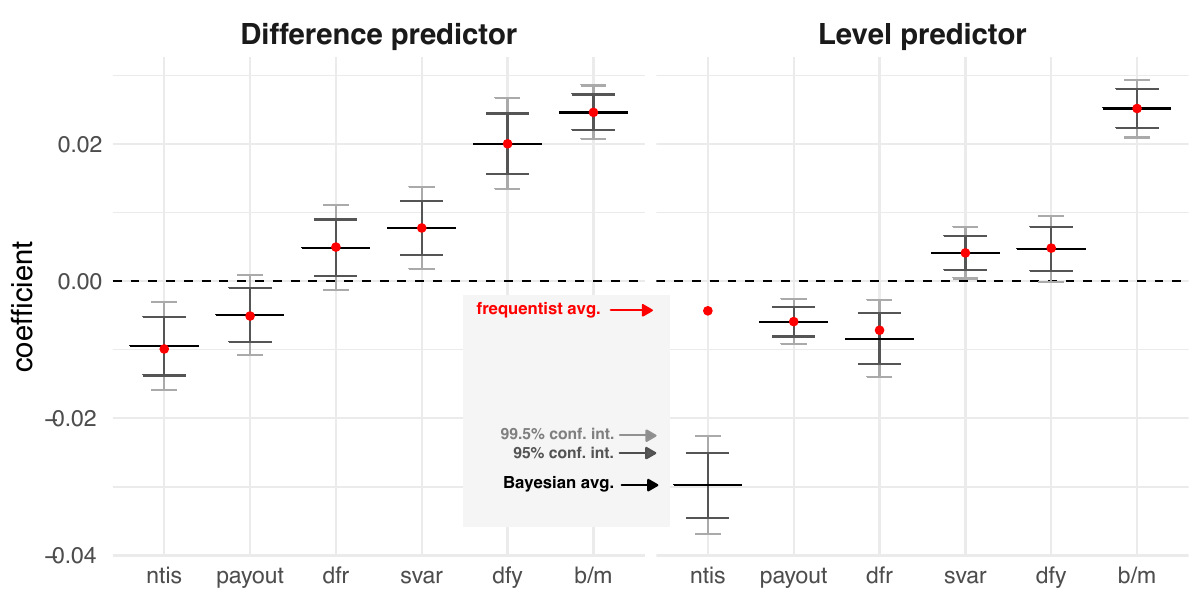}\vspace{-3mm}
\caption{\textbf{Bayesian model averaging} (annual data). \small We display average coefficients within their 95\% and 99.5\% confidence intervals. Averages stem from Equation \eqref{eq:mean} and predictors are scaled by their standard deviations prior to estimation to ease comparison of scales. . The bounds of the confidence intervals are defined by $\E[b|D]\pm \alpha \sqrt{\mathbb{V}[b|D]/T_*}$, where $T_*=\sum_{j=1}^Jw_jT_j$, with $T_j$ being the sample size of model $j$ and $w_j$ the posterior model probabilities. $\alpha$ relates to the confidence level. The left panel displays results when predictors are differences, while the right one focuses on levels. To allow comparisons, all predictors are scaled to have unit variance before estimation. Only the paths with no bias adjustment are considered. }
\label{fig:BMA} \vspace{-6mm}
\end{center}
\end{figure}

In Figure \ref{fig:BMA}, we plot the average coefficients computed according to Equation \eqref{eq:mean}, along with ad-hoc confidence intervals. We only report results for annually sampled variables in order to reduce sample sizes. Indeed, their exponentiation in some term of Equation \eqref{eq:odds} are problematic when two models have significantly contrasting sample sizes. This issue is circumvented with annual samples.

The intervals in Figure \ref{fig:BMA} tend to confirm those obtained for the frequentist averages. Both \textit{ntis} and \textit{b/m} are associated with intervals that do not overlap over zero. In fact, for all predictors except one, the Bayesian averages are mostly indistinguishable from their frequentist counterparts.

%There are several takeaways from this figure. First, naturally, coefficients depend strongly on the nature of predictors. This is true both for signs and magnitudes. Because intervals are computed according to paths, narrower intervals stem from smaller standard errors and not from sample sizes. 

%With respect to the the signs of effects \textit{across} frequencies, we find that results are consistent only for a few predictors. For raw predictors (levels in the left panel), we single out \textit{ntis}, \textit{b/m} and, to a lesser extent, \textit{svar}. With regard to differentiated variables (right panel), only \textit{payout} has coefficients that all have the same signs, albeit with small magnitudes. 

%Lastly, an interesting feature of the figure relates to the predominant agreement between frequentist and Bayesian means. In a large majority of case, both types of averages yield similar results, and inferential conclusions would not be impacted when switching from one to the other, which is reassuring.  

%Out of curiosity, we also tested the agnostic average for which weights are uniform $w_p=P^{-1}$. Unfortunately, the results were in some cases quite different from the former two and polluted the graphical output by generating outliers. This underlines the usefulness to resort to weights that depend on a quantification of the reliability of the models (goodness-of-fit or information criteria). All models are clearly not equal and we further illustrate this phenomenon in Subsection \ref{sec:case} in the Appendix.

\subsection{Conditional averages}
\label{sec:prem_cond_avg}

The above result show averages that are only conditioned by their sampling frequency. Below, we investigate the sensitivity of averages depending on other design choices, namely regression specification, subsampling, and estimated standard errors. 

We start with the first two, jointly, to highlight the differences in the results. The two mappings, or layers we focus are the following. First, the estimation model, standard versus augmented, which corresponds to Equations \eqref{eq:remodel} versus \eqref{eq:remodelaug} in the definition of the layers. Second, we shed some light on the period-dependence of our results. Our protocol allows to look at subsamples and we are able to discriminate between the first half of samples ($x$-axis) versus their second half ($y$-axis). 

\begin{figure}[!h]
\begin{center}
\includegraphics[width=15.5cm]{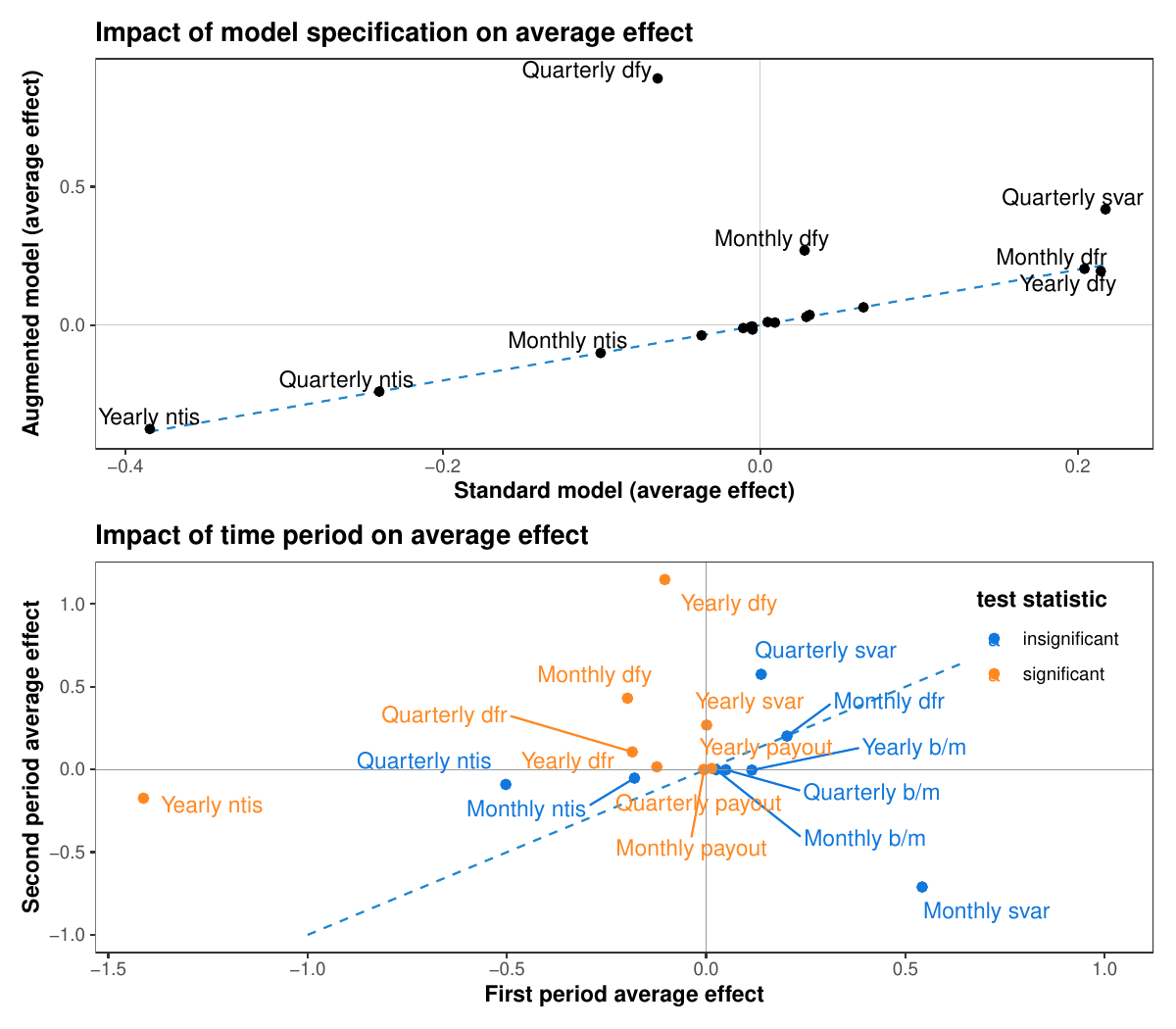}\vspace{-5mm}
\caption{\textbf{Conditional differences}. \small We plot the conditional averages of one layer option ($x$-axis) version the other option ($y$-axis). In the top panel, the layer is the regression model (standard versus augmented) whereas in the bottom panel, the layer is the subsample (first versus second period). For both plots, we focus on level predictors only and the \textit{frequentist} averages are obtained over 96 paths (top) and 128 paths (bottom). The dashed blue line marks the quadrant bisector. In the bottom figure, colors code for the significance of the simple $t$-test on the sequences \eqref{eq:test} at the 5\% confidence level.  }
\label{fig:cond_mean}
\end{center}
\end{figure}

In the top panel of Figure \ref{fig:cond_mean}, we plot conditional means of standard models (Equation \eqref{eq:remodel}) on the $x$-axis versus, on the $y$-axis the conditional means of coefficients obtained via the augmented models (Equation \eqref{eq:remodelaug}). In addition, we draw the $y=x$ line to see where the points lie compared to the bisector. Points show the frequentist averages for the level predictors.

Plainly, switching from one to the other model specification has very limited impact, as almost all points are located very closely to the bisector. There is only one clear outlier, the quarterly \textit{dfy} variable. When testing for the mean difference between the weighted values of the points (Equation \eqref{eq:test}), the null of zero mean was rejected only once at the 5\% confidence level (for \textit{dfr} on annual samples) with a $t$-statistic of 2.0. All other statistics were below 1.9, even for the outlier point. In short, the model type does not affect results very much. 

In the bottom panel of Figure \ref{fig:cond_mean}, we proceed with the same analysis, but this time for period comparison. Given the strong time-variation in predictive coefficients (see \cite{farmer2023pockets}), we expect to see some difference in this case. And indeed, our results contrast with the upper panel because points no longer perfectly align with the bisector. The test statistics of the $t$-test defined in \eqref{eq:test} and applied to the sequences of coefficients are split in two color categories depending on whether they are deemed significant at the 5\% level. Surprisingly, there is no clear pattern for colors: the points closest to the bisector are not necessarily blue. Upon verification, this comes from large standard deviations which shrink the average effect in the denominator of the statistic in some cases.  

\vspace{3mm}

Lastly, out of curiosity, we evaluate the impact of the standard deviation specification on the weighted average of $t$-statistics. In the above results, $t$-statistics are computed at the very end by taking the ratio between average effects and their standard deviations \eqref{eq:sigma}. This does not account for the model-specific estimates of standard deviations. In order to embed the latter in an aggregate measure, it is possible to average the $t$-statistics instead of raw effects. 

\begin{figure}[!h]
\begin{center}
\includegraphics[width=15.5cm]{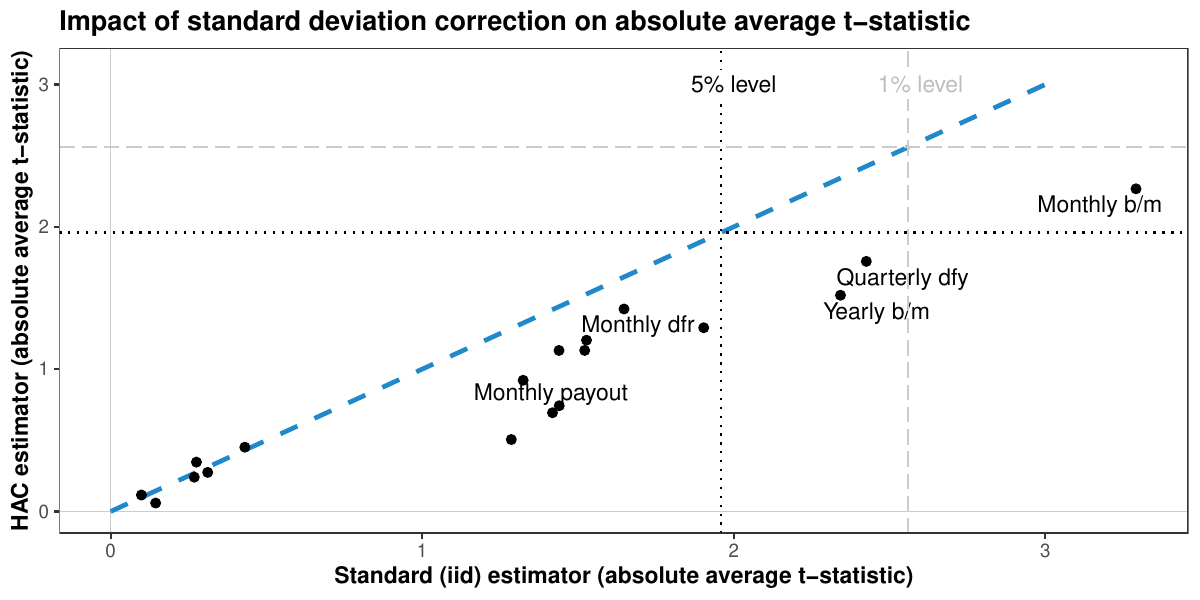}\vspace{-5mm}
\caption{\textbf{Standard deviations and \textit{t}-statistics}. \small We plot the absolute weighted averages of $t$-statistics (the absolute value is applied after the averaging). The $x$-axis relates to the average statistics computed with uncorrected standard deviations, whereas the $y$-axis pertain to statistics that adjust for auto-correlation and heteroskedasticity. Vertical and horizontal dashed lines mark the decision thresholds at the 5\% and 1\% levels.   }
\label{fig:hac}
\end{center}
\end{figure}

In Figure \ref{fig:hac}, we produce the absolute value of weighted averages in test statistics. The $y$-axis features the HAC-corrected statistics. Therefore, because HAC standard errors are usually more conservative (i.e., larger), we would expect that the corresponding statistics be \textit{smaller} than those from the i.i.d. estimator. And indeed, this is what we see in the plot, as almost all points lie below the dashed bisector. It is noteworthy to underline that the three points farthest to the right pass decision thresholds with the iid estimator and not with the HAC estimator. For instance, the \textit{Yearly b/m} (\textit{resp., Monthly b/m}) predictor is significant at the 5\% (\textit{resp.} 1\%) level with the iid estimator, but not with the HAC estimator. 

\subsection{Comparison with other studies}

\label{sec:premium_prior}

In \cite{goyal2021comprehensive}, the authors report the coefficient estimates from their regressions, both on their full sample and on the two halves of their initial sample. This is in contrast with \cite{dichtl2021data} and \cite{engelberg2021cross} who do not provide the raw estimates, but only out-of-sample fit. Henceforth, we focus on the monthly values provided by \cite{goyal2021comprehensive}, which also correspond to level predictors, and not difference variables (increments). 

\begin{figure}[!h]
\begin{center}
\includegraphics[width=15.5cm]{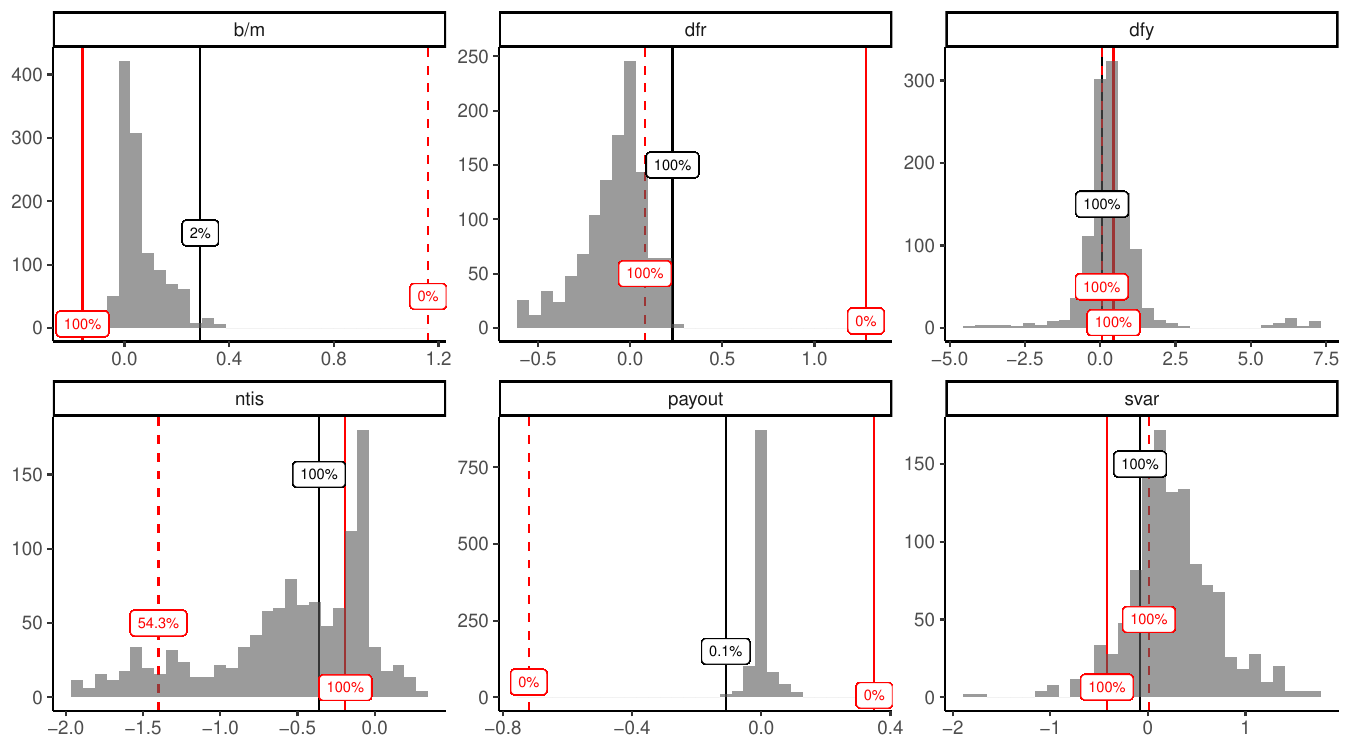}\vspace{-4mm}
\caption{\textbf{Comparison with \cite{goyal2021comprehensive}}. \small We display the distribution of effects across paths with grey histograms. The black line shows the value obtained by \cite{goyal2021comprehensive} over the full sample. The red lines represent the values for the two halves of the sample (dotted for the first half and full for the second half). The rounded rectangles provide the the \textit{ease to confirm} indicator (EtC) with respect to the Gaussian distribution (with mean and standard deviation computed from the histograms). It is equal to one minus \eqref{eq:Phi} with $q=0.9$.   }
\label{fig:premium_prior} 
\end{center}
\end{figure}

In Figure \ref{fig:premium_prior}, we depict the distribution of the path-generated values, alongside those from \cite{goyal2021comprehensive}. There are three of them: the full sample in black and the two halves in red. For four predictors (all but \textit{b/m} and \textit{payout}), the premia over the full sample are all well within the values obtained from the paths and they can hence be reproduced effortlessly. For the \textit{b/m} predictor, it seems the valued reported by \cite{goyal2021comprehensive} is slightly optimistic. However, for \textit{payout}, there appears to be a scale problem - though we stuck to the variable definition in the original paper. 

For the premia related to the sample halves, the ease to corroborate depends on variables, but paths are mostly compatible with at most one of the halves (e.g., having a EtC score above 50\%). Our results also underline, somewhat as expected, that large sample sizes are clearly associated with the highest generalization ability. Effect sizes computed on deep chronological samples are less prone to historical idiosyncrasies.

\subsection{Expansion rates of hacking intervals}
\label{sec:expanding}

\begin{figure}[!h]
\begin{center}
\includegraphics[width=15cm]{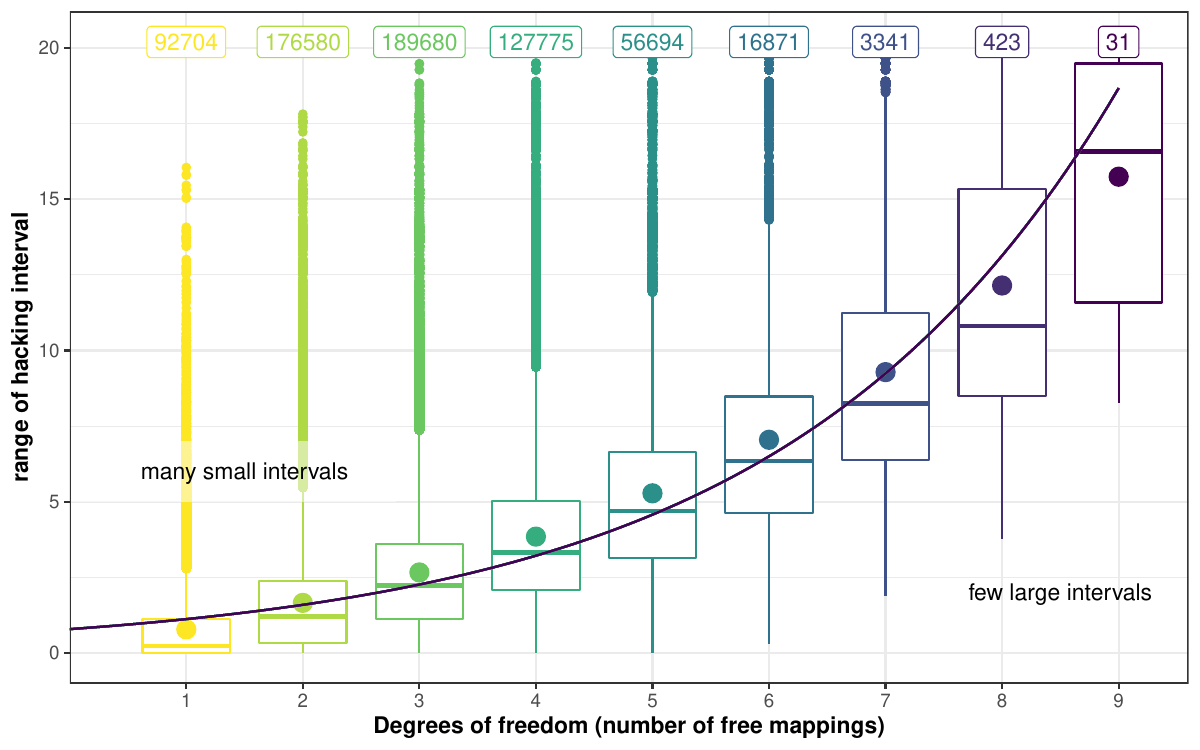}\vspace{-3mm}
\caption{\textbf{Expanding intervals}. \small We show the boxplots of hacking interval ranges defined in Equation \eqref{eq:hackint}). The small vertical points stand for outliers. The bigger circles represent the mean of ranges (ARI), introduced in Equation \eqref{eq:ari}. The curve is parametrized by: $ARI(n) = 0.78 \times 1.42^n$, where the constant were estimated by least square minimization to feat the mean points in a log-regression ($\log(ARI)=\log(a)+\log(b)\times n$). The numbers at the top provide the number of intervals available for each number of free mappings. }
\label{fig:hackint}
\end{center}
\end{figure}

The large number of paths generated in the study allows to compute the quantities defined in Equations \eqref{eq:hackint}, \eqref{eq:ari} and \eqref{eq:rate}, when the outcome is the $t$-statistic. In Figure \ref{fig:hackint}, we show the boxplots of interval ranges proposed in Equation \eqref{eq:hackint}. In the case when the $x$-axis $J-k=1$, to the left of the plot, there is only one free mapping, which generates small dispersion in $t$-statistics. This is why both the median and average of the ranges is small (below one). As one shifts to the right of the plot, more and more leeway is given to the researcher and the breadth of output increases. If we assume a constant rate of increase, we obtain (via log least-square optimization) that each mapping expands average intervals by 42\% (which we see with the black power line).

However, it is clear that the effect is not uniform: the line is below in the middle, but above towards the end. This signals the intuitive pattern that as $J-k$ increases, the speed of expansion slows down because intervals are already very large. This is confirmed in Table \ref{tab:rate} below, which provides speed at which the average intervals increase. As expected, the increase rate, $\rho_{J-k}$ defined in Equation \eqref{eq:rate}, decreases with $J-k$. Nonetheless, the final value (after 9 successive increments) remains close to 30\%.

\begin{table}[!h]
\centering
\begin{tabular}{rrrrrrrrrr}
  \hline
$J-k$ & 1 & 2 & 3 & 4 & 5 & 6 & 7 & 8 & 9 \\
  ARI & 0.779 & 1.656 & 2.662 & 3.847 & 5.278 & 7.048 & 9.285 & 12.149 & 15.745 \\
  rate ($\rho_{J-k}$)  &  & 1.124 & 0.608 & 0.445 & 0.372 & 0.335 & 0.317 & 0.308 & 0.296 \\
   \hline
\end{tabular}
\caption{\textbf{Interval expansion}. We report the average range of intervals (ARI, from Equation \eqref{eq:ari}, along with its growth with the number of free mappings ($J-k$). \label{tab:rate}}
\end{table}

The extreme case ($J-k=9$) occurs when only one mapping is fixed (each option being considered separately) and this is highlighted in Figure \ref{fig:intervals} in Appendix \ref{sec:figs}. There are 31 alternatives and the narrowest interval is the fourth one (\textit{dfy}), with a width close to 8, which is the lowest point in the rightmost candlestick in Figure \ref{fig:hackint}. Several choices yield intervals with range close to 20 (from -9 to +11), which is the upper limit of the candlestick.

All in all, our results confirm the self-evident idea that, as researchers consider more ways to run their protocol, they should expect a wider range of outcomes. Ideally, this would result in a better characterization of the effect they study.

%\clearpage

\section{Application: anomalies from portfolio sorts}
\label{sec:sorts}

The literature in financial economics has seen a surge in \textit{factors} (see \cite{harvey2016and}). The latter are also called \textit{anomalies} because they contradict the cornerstone result that is the capital asset pricing model (CAPM). The multiplication of publications in the field has even led researchers to devise new tests and approaches to detect or analyze when a factor is truly a factor (e.g., \cite{feng2020taming}, \cite{chinco2021estimating} and \cite{harvey2021lucky}).

The simplest way to proceed, since the seminal work of \cite{fama1992cross}, is to periodically sort stocks according to some characteristic and test if extreme quantile portfolios have significantly different means. This again gives rise to implementation leeway, such as holding periods, quantile thresholds, and portfolio weighting for instance. This section is therefore dedicated to the impact of these modelling choices on the significance of asset pricing anomalies. Its main goal is to illustrate the concept of exhaustive multiple testing presented in Section \ref{sec:multiple}. Our conclusions on the sensitivity of design choices corroborate some findings in the recent similar studies of \cite{bessembinder2022factor}, \cite{soebhag2022mind} and \cite{walter2022non}.

\subsection{Data and paths}

We rely on the dataset used in \cite{gu2020empirical} updated by the authors until the end of 2021. In the original study, 94 characteristics are used to predict returns, but some of them have limited support (e.g., they are binary), and are not suitable for sorting purposes. We remove these variables and are left with 82 characteristics.\footnote{The full list of abbreviated names is: absacc, acc, aeavol, agr, baspread, beta, betasq, bm, bm\_ia, cash, cashdebt, cashpr, cfp, cfp\_ia, chatoia, chcsho, chempia, chinv, chmom, chpmia, chtx, cinvest, currat, depr, dolvol, dy, ear, egr, ep, gma, grcapx, grltnoa, herf, hire, idiovol, ill, indmom, invest, lev, lgr, maxret, mom12m, mom1m, mom36m, mom6m, mvel1, mve\_ia, operprof, orgcap, pchcapx\_ia, pchcurrat, pchdepr, pchgm\_pchsale, pchquick, pchsale\_pchinvt, pchsale\_pchrect, pchsale\_pchxsga, pchsaleinv, pctacc, pricedelay, quick, rd\_mve, retvol, roaq, roavol, roeq, roic, rsup, salecash, saleinv, salerec, secured, sgr, sp, std\_dolvol, std\_turn, stdacc, stdcf, tang, tb, turn, zerotrade.} After this filter, some changes in data availability occur prior to 1983, hence we focus on the period from January 1983 to December 2021.\footnote{The code and data used for this part can be accessed \href{https://www.gcoqueret.com/files/misc/forking_paths2.html}{online}.}

\begin{figure}[!h]
\begin{center}
\includegraphics[width=15cm]{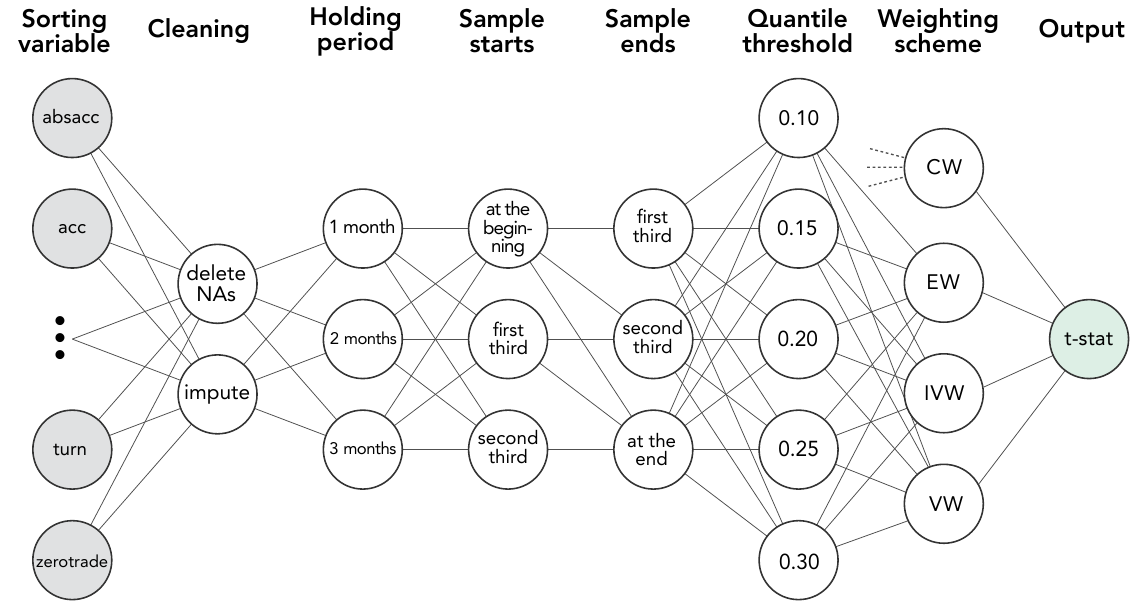}\vspace{-3mm}
\caption{\textbf{Forking paths for anomalies}. \small The graph, akin to a neural network structure, depicts the seven-step process used to produce the $t$-statistic (last layer) in the study. Each path has the same probability of realization (uniform distribution). }
\label{fig:nn_3}
\end{center}
\end{figure}

In Figure \ref{fig:nn_3}, we depict the steps (modelling choices) that we consider to qualify asset pricing anomalies. In total, there are 576 paths for each sorting variable, which makes 82$\times$576=47,232 paths in total.

More precisely, the seven modules are:
\begin{enumerate}
\setlength\itemsep{-0.2em}
    \item the \textbf{sorting variable};
    \item the \textbf{data cleaning} choice (imputation of prior value or removal);
    \item the \textbf{holding period} posterior to sorting (1, 2 or 3 months - this corresponds to reasonable choices for rebalancing frequency);
    \item the \textbf{starting point} of the study (minimum, first third or second third of the full sample's dates)
    \item the \textbf{ending point} of the study (first third, second third, or maximum of the full sample's dates). This means that the smallest samples have a length that is one third of the total sample ($\sim$39 years), hence encompassing several macro-economic environments.
    \item the \textbf{quantile threshold} ($q$) used to compute the long-short portfolios (long the upper $1-q$ stocks and short the lower $q$ stocks. The factors' sensitivity to breakpoints is thoroughly investigated in \cite{hollstein2021robust}.
    \item the portfolio \textbf{weighting scheme}: equally-weighted (EW), inverse volatilty-weighted (IVW) and value-weighted (VW). With IVW, weights are proportional to the inverse of the \textit{retvol} characteristic and with VW, they are proportional to the \textit{mvel1} indicator. Finally, we also consider an alternative scheme, CW (characteristics-weighting), for which the weight in the long-short portfolio is the scaled value of the characteristic such that each leg has weights that sum to one. In this case, weighting is smoother in the cross-section, as there is no threshold.  
\end{enumerate}

In addition, we define a default, or baseline, path:
 \vspace{1mm}  \\
$\blacktriangleright \,$ \textbf{default-representative path} (parametrization of the layer options): imputation of missing data (and not deletion), a one month holding period, 0.2 sorting threshold (quintiles), equally-weighted portfolios, and the average returns and statistics are computed on the full sample.  \vspace{3mm}

Each path produces a series of portfolio returns. The output corresponds to a simple $t$-test on the mean of the average return: $\sqrt{T}\mu/\sigma$, where $\mu$ is the average and $\sigma$ the standard deviation of returns. Up to the size scaling and assuming zero risk-free rate, the output is the Sharpe ratio of the portfolio.

\begin{figure}[!h]
\begin{center}
\includegraphics[width=15.2cm]{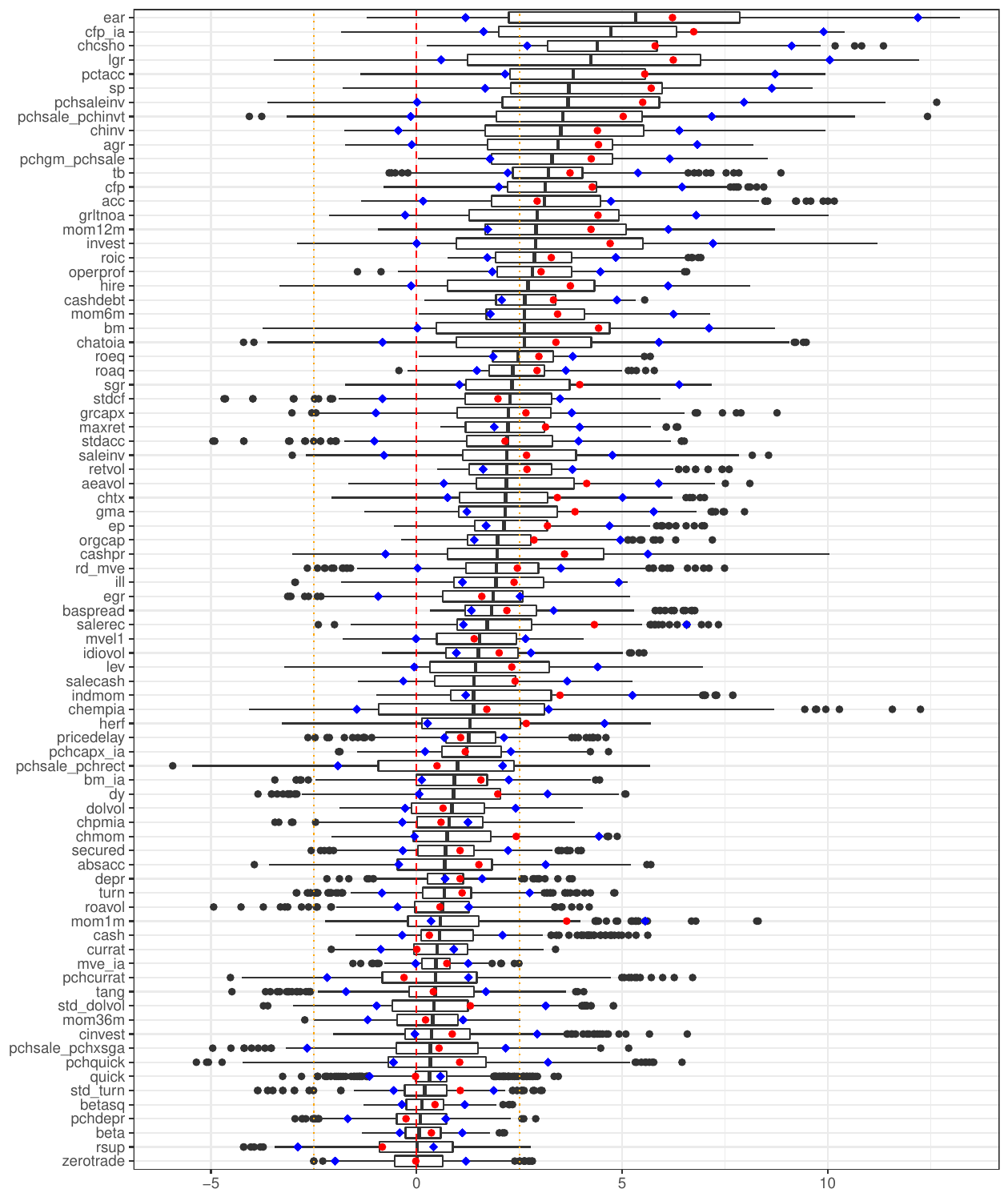}\vspace{-4mm}
\caption{\textbf{Distribution of output}. \footnotesize We produce the boxplots (over 576 paths) for the $t$-statistics for the simple test of significance of average returns of each of the 82 anomalies (shown on the $y$-axis). Variables are ordered by median $t$-statistic and for each variable the sign (long versus short) is adjusted so that this median be positive. The red point marks the statistic for the baseline path. The blue diamonds mark the range of statistics when paths are restricted to robustness checks (14 paths).  }
\label{fig:t_stats_anom}
\end{center} \vspace{6mm}
\end{figure}

These outputs are summarized in Figure \ref{fig:t_stats_anom}. Notably, the width between the extreme points of the boxplots correspond to the range of hacking intervals for each factor. An anomaly may be considered to be strong if these intervals are not centered around zero. A remarkable pattern is the contrast between the short-term reversal (negative returns stemming from one month momentum) and the one year trend-following strategy which is associated to positive returns. The variety of results suggests that some anomalies are definitely more robust than others.

Half a dozen factors are able to sustain positive average returns over the full wide scope of implementations: \textit{roic}, \textit{roeq}, \textit{retvol}, \textit{mom6m}, \textit{maxret} and \textit{cashdebt}. Because of the multiple environments in which they have been tested,\footnote{Formally, we cannot consider that each path represents an environment of its own. However, two paths (for a fixed characteristic) with no overlap in layer options may be considered as two complementary facets of the same factor. } they emerge as robust strategies within the zoo of factors. Our results are yet another confirmation of the momentum factor, which is widely documented as a persistent one (see \cite{smith2022have} for a recent appraisal).  

In addition, in Figure \ref{fig:t_stats_anom}, we also show with blue diamonds the extreme points of what we refer to as \textit{robustness checks}. These are all the paths which have exactly one deviation from the default path. Given the path structure depicted in Figure \ref{fig:nn_3}, this makes 14 paths. The average range, across anomalies, for intervals of $t$-statistics is 4.1 for robustness checks, while it is 9.1 if we span all the paths. This forcefully shows that simple robustness checks only provide a narrow picture of the diversity in outcomes, whereas paths are much more exhaustive.

\subsection{Multiple testing}
\label{sec:MT}

Asset pricing anomalies are a fertile ground for multiple testing. The aim here is to test if factors are genuine, or simple flukes (see, e.g., \cite{chen2021open}, \cite{harvey2021lucky}, \cite{jensen2021there} and \cite{chen2022peer}). We thus want to illustrate the exhaustive approach we advocate in Section \ref{sec:multiple}. In this setting, the observations $x_{t,n}$ are the time-$t$ returns of factor $n$.

We wish to compare the bootstrap reality check (BRC) approach at computing $t$-statistics (Equation \eqref{eq:boot}) to the exhaustive one that relies on paths (Equation \eqref{eq:emt}). To ensure comparability of the two methods, we rely on a number of bootstrap samples that is equal to the number of paths in the study (576). Moreover, we resort to block bootstrapping with blocks of size 12, i.e., coinciding to annual series.

\begin{figure}[!h]
\begin{center}
\includegraphics[width=13cm]{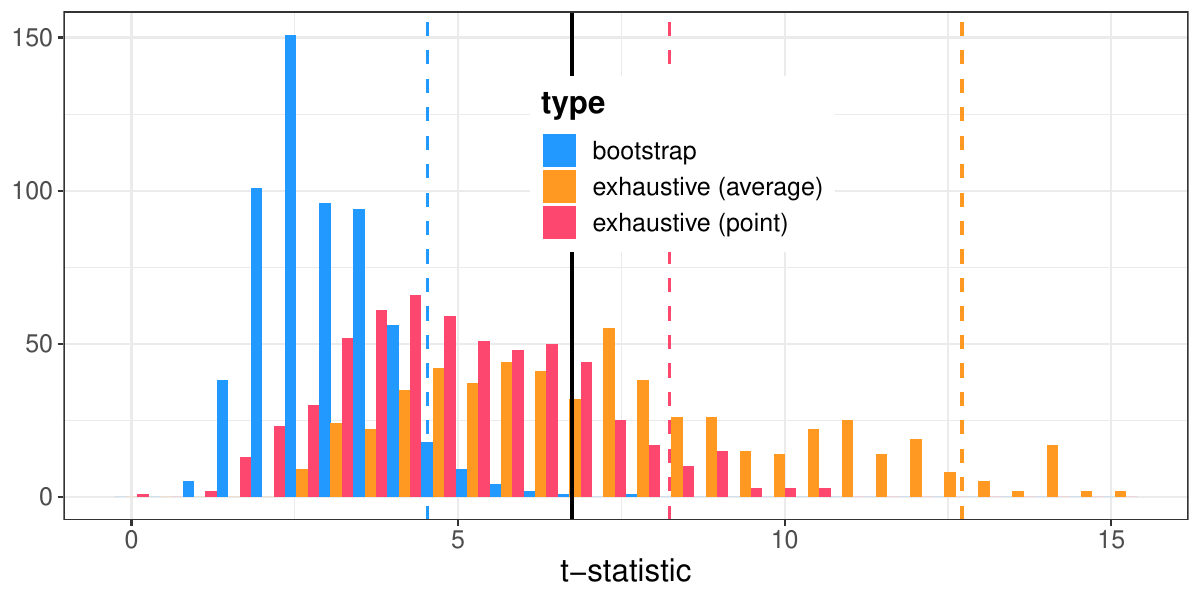}\vspace{-5mm}
\caption{\textbf{Distribution of bootstrapped and paths-related maximum statistics}. \small We produce the histogram of the maximum statistics stemming from bootstrapping ($\tilde{t}_1^{(b)}$, in blue) and forking paths ($\tilde{t}_1^{(p)}$, in orange and red). The sequences are derived from equations \eqref{eq:boot} and \eqref{eq:emt}, respectively. The vertical black line is the benchmark $t$-statistic of the \textit{best} anomaly (\textit{cfp\_ia} in this case) for the default path. The vertical dotted lines correspond to the 95\% quantile of the maximum statistics of each type. The difference from the two exhaustive distributions comes from the benchmark $\mu_n$ used to compute the statistics. The point-wise values are obtained when $\mu_n$ is the average anomaly return of the default path described above. The average values correspond to the case when $\mu_n$ is the average of factors' returns over all paths.  }  
\label{fig:EMT}
\end{center}
\end{figure}

In Figure \ref{fig:EMT}, we plot the distribution of maximum statistics obtained from bootstrapping returns versus forking paths. Plainly, the distribution of bootstrapped statistics lies to the left of those stemming from forking paths. The main reason for this is that alternative paths generate average returns that vary considerably, compared to the baseline case. One particularly important layer is the subsampling one because anomalies can evaluated over different time-frames. While this may seem odd, it makes sense from an investment standpoint: a reliable long-short strategy should perform similarly over various periods, as long as these periods are long enough to encompass a variety of market conditions (e.g., bull and bear markets).

The main takeaway from this exercise is that by examining a large spectrum of outcomes instead of bootstrapped returns, extreme cases become more likely, which raises the bar for significance. With the default path, the maximum statistic for anomalies is 6.74 (for the \textit{cpf\_ia} characteristic, see rightmost red point in Figure \ref{fig:t_stats_anom}), which is quite high. The hurdle at the 95\% level from bootstrapping is 4.5, so that the best original anomaly passes the test handily. For the sake of completeness, we have tested the case with 10 times more bootstrap samples (5,760) and the threshold remains the same, at 4.5.

However, if we consider thresholds generated by paths, we obtain 8.2 or 12.7, depending on the configuration. They are the vertical dotted lines in the figure. In both cases, the baseline path of \textit{cpf\_ia} is not longer significant. In \cite{chen2022peer}, the anomalies that cannot be matched by data mining techniques all have $t$-statistics above 5, and a handful of them are even above 12. When a baseline statistic is very high, it is likely that paths in its vicinity will also have large values, often above 2 or 2.5. In Figure \ref{fig:EMT}, we see that the variables for which the red point is far to the right (at the top) all have inter-quartile ranges do not include zero. For the corresponding accounting or risk characteristics, this means that 75\% of the paths at least lead to profitable strategies. Hence, raising the threshold helps immunize anomalies against implementation sensitivity and false positives.

Naturally, the symmetric cost of this is a substantial increase of the odds of false negatives. A higher decision threshold means that most anomalies will not be able to reject the null of zero return. Therefore, it is inevitable that genuine factors be missed. From an inferential standpoint (e.g., for the researcher in asset pricing), this is a severe limitation. In contrast, for a portfolio manager false positives correspond to where the money goes, which is arguably what matters most.

\subsection{Comparison with prior work}
\label{sec:sorts_prior}

The recent paper \cite{chen2021open} provides an interesting basis to work with, as it lists and reproduces a large array of prominent asset pricing anomalies. Given the authors' work,\footnote{They provide a \href{https://github.com/OpenSourceAP/CrossSection/blob/master/SignalDoc.csv}{CSV file} that compiles the information from the literature.} it is possible to compare the average returns (and the corresponding $t$-statistics) of sorted portfolios tested in prior studies. When identifying the academic articles from \cite{chen2021open} with the sorting variables we worked with, we are able to determine an overlap of 40 predictors for which both returns and $t$-statistics are documented. For a few of them, several results are reported and we only keep the one that corresponds to the smallest average return, i.e., the one that will maximise the EtC indicator (ease to corroborate).

\begin{figure}[!h]
\begin{center}
\includegraphics[width=15.5cm]{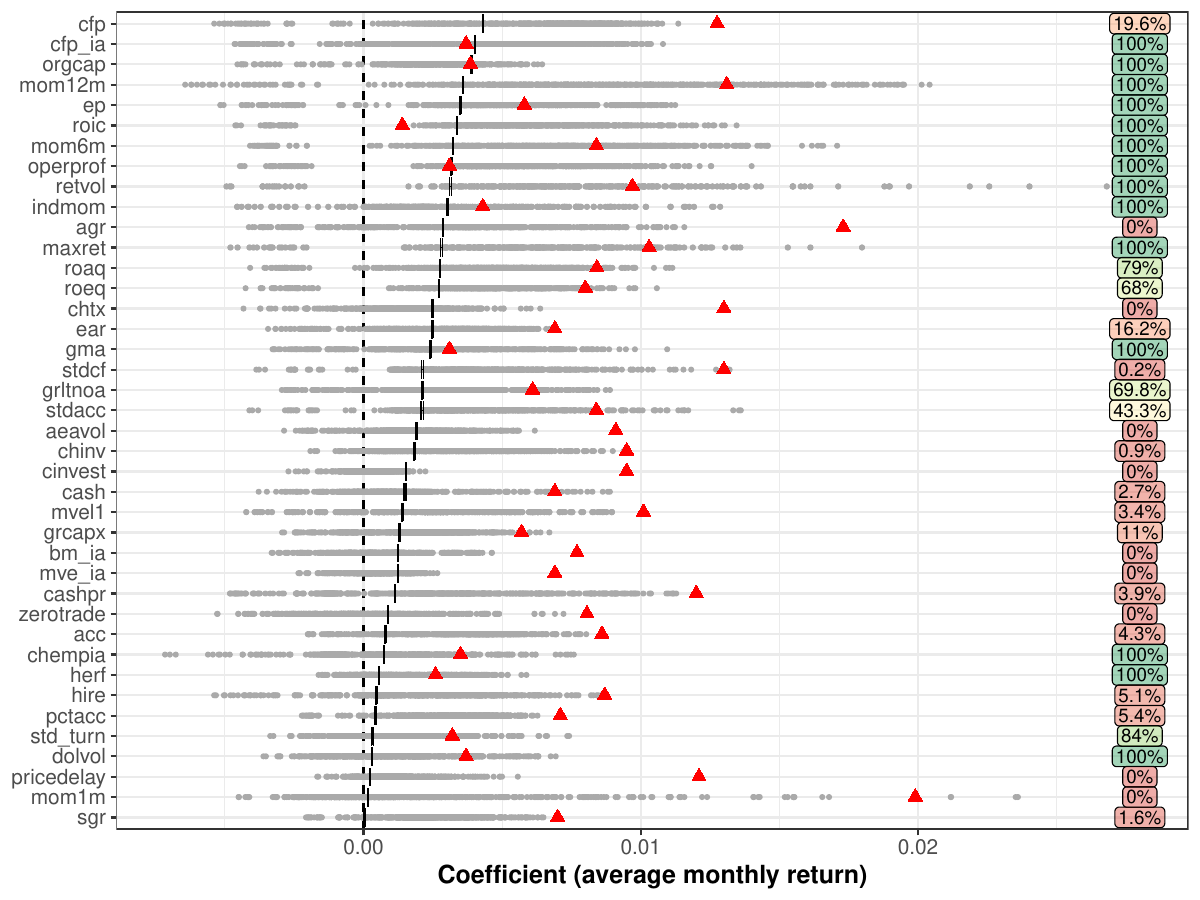}\vspace{-4mm}
\caption{\textbf{Comparison with prior studies}. \footnotesize We display the average returns obtained by the 576 paths with small grey circles for each predictor ($y$-axis). The very narrow confidence intervals defined in Equation \eqref{eq:IC} are shown in black. Anomalies are ranked according to the average with weights \eqref{eq:weights}. The average returns reported in the original studies are depicted with a \textcolor{red}{\textbf{red}} triangle. The values come from \cite{chen2021open}. At the very right of the plot in rounded rectangles, we provide the ease to confirm (EtC) of the triangle with respect to the Gaussian distribution fitted on the grey circles (mean and standard deviation) - equal to one minus \eqref{eq:Phi} with $q=0.9$. The colors code the facility to reproduce the original outcome.  }
\label{fig:comparison}
\end{center} \vspace{0mm}
\end{figure}

For each predictor, we can compute, given all average returns from the paths, the mean and standard deviation of these returns. It is then possible, as discussed in Subsection \ref{sec:suspicious}, to evaluate the position of the results from the former studies, if we assume a Gaussian distribution of outcomes from the paths. This is shown in Figure \ref{fig:comparison}. Therein, the average returns of paths are shown as grey points and the confidence intervals \eqref{eq:IC} are so narrow that they are in fact represented with black vertical segments. The figures from original values reported by \cite{chen2021open} are represented with \textcolor{red}{red} triangles. 

Then, we compute the sample mean and variance of the grey points for each anomaly and report the EtC defined as one minus the indicator \eqref{eq:Phi}, where $b^*$ is the red triangle value and $\Phi_P$ is the Gaussian cdf with mean and variance equal to their sample estimations. The EtC is reported at the right of the plot. Of the 25 anomalies we consider, 10 have an EtC above 50\%. For these studies, the reported effects may be large, but they are not suspiciously so.

In all of these 10 studies, the reported values are statistically significant, meaning that, based on the paths, these results are strong and \textit{likely} so. For instance, if we take the case of the \textit{maxret} factor, in the original paper by \cite{bali2011maxing}, the 1.03\% monthly return is associated with an absolute $t$-statistic of 2.83, which makes it significant even at the 1\% level. Therefore, this effect size is both significant and plausible. 

On the other side of the spectrum, we report that 11 anomalies have an EtC below 10\%, meaning that they correspond to values that were hard or even impossible to reach with the paths that we have spanned. This confirms that all anomalies are not equal. Some of them have returns that can sustain small alterations in the construction process, others do not.

\subsection{Conditional averages: stability through time}
\label{sec:sorts_stability}

Another way to test the potency of factors is to evaluate the shift of their performance through time. One convenient layer of the above protocol allows to split samples into three periods of equal sizes. This constitutes fertile ground to further illustrate the concept of conditional averaging introduced in Section \ref{sec:cond_avg}. The rationale for this is to evaluate if the average returns of sorted portfolios varies substantially across large non-overlapping periods. This is a major concern for money managers because a shift in one anomaly's profitability is likely to alter performance for agents who have invested in this particular anomaly. 

\begin{figure}[!h]
\begin{center}
\includegraphics[width=15.5cm]{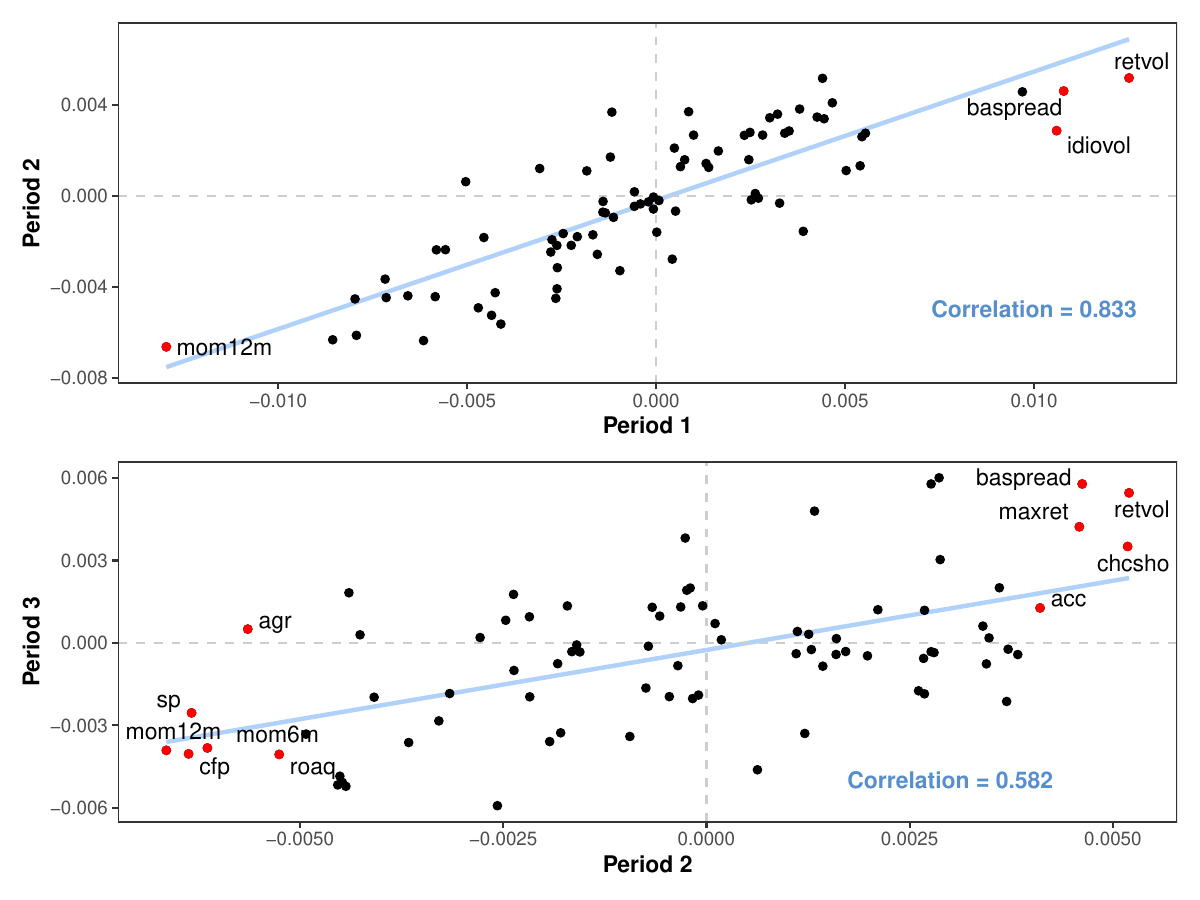}\vspace{-4mm}
\caption{\textbf{Consistency through time}. \footnotesize We depict the simple average returns of sorts from the first to the second period (upper plot) and from the second to the third (lower plot). Each point pertains to one sorting variable and corresponds to the average over 96 paths. The light blue line shows the linear relationship fitted on all points. Extreme points are shown in \textcolor{red}{red} and the corresponding anomalies are specified.  }
\label{fig:comp_sorts}
\end{center} \vspace{0mm}
\end{figure}

In Figure \ref{fig:comp_sorts}, we plot average returns of anomalies across two different periods, one for each axis. In the upper panel, we compare the first period (1983-1996) to the second one (1996-2008) and in the lower panel, the second period is linked to the third one (2008-2021). Clearly, the relationship between the periods is stronger in the upper panel: this means that an investor in 1996 would have not been too disappointed by the performance of anomalies until 2008. However, there seems to be some decoupling posterior to 2008, as shown by the evidence in the lower panel: the correlation between the periods shrinks from 83\% to 58\%. Importantly, there are significantly more anomalies in the upper left and lower right quadrants, meaning that performance has reversed between the two periods. Nonetheless, there are some similarities and a few extreme anomalies remain in the same zones of the plots: \textit{mom12m} (lower left) and \textit{retvol} and \textit{baspread} (upper right) emerge as stable factors. Only time will tell if this holds in the decades to come. 

\subsection{\textit{p}-hacking in the cross-section of characteristics}
\label{sec:crosshack}

\cite{elliott2021detecting} provide criteria on the distribution of $p$-values to determine if a sequence of outcomes was generated by $p$-hacking. Surprisingly, as discussed in Appendix \ref{sec:tests}, it may occur that even if $p$-values were not hacked, they fail to pass a test of no $p$-hacking. Therefore, after generating paths, it is interesting to determine how much falsification would be required to pass the most elementary $p$-hacking detection test. Because we only have 576 paths for each sorting variable, histograms may be somewhat noisy, which makes precise tests troublesome.

\begin{figure}[!h]
\begin{center}
\includegraphics[width=15.2cm]{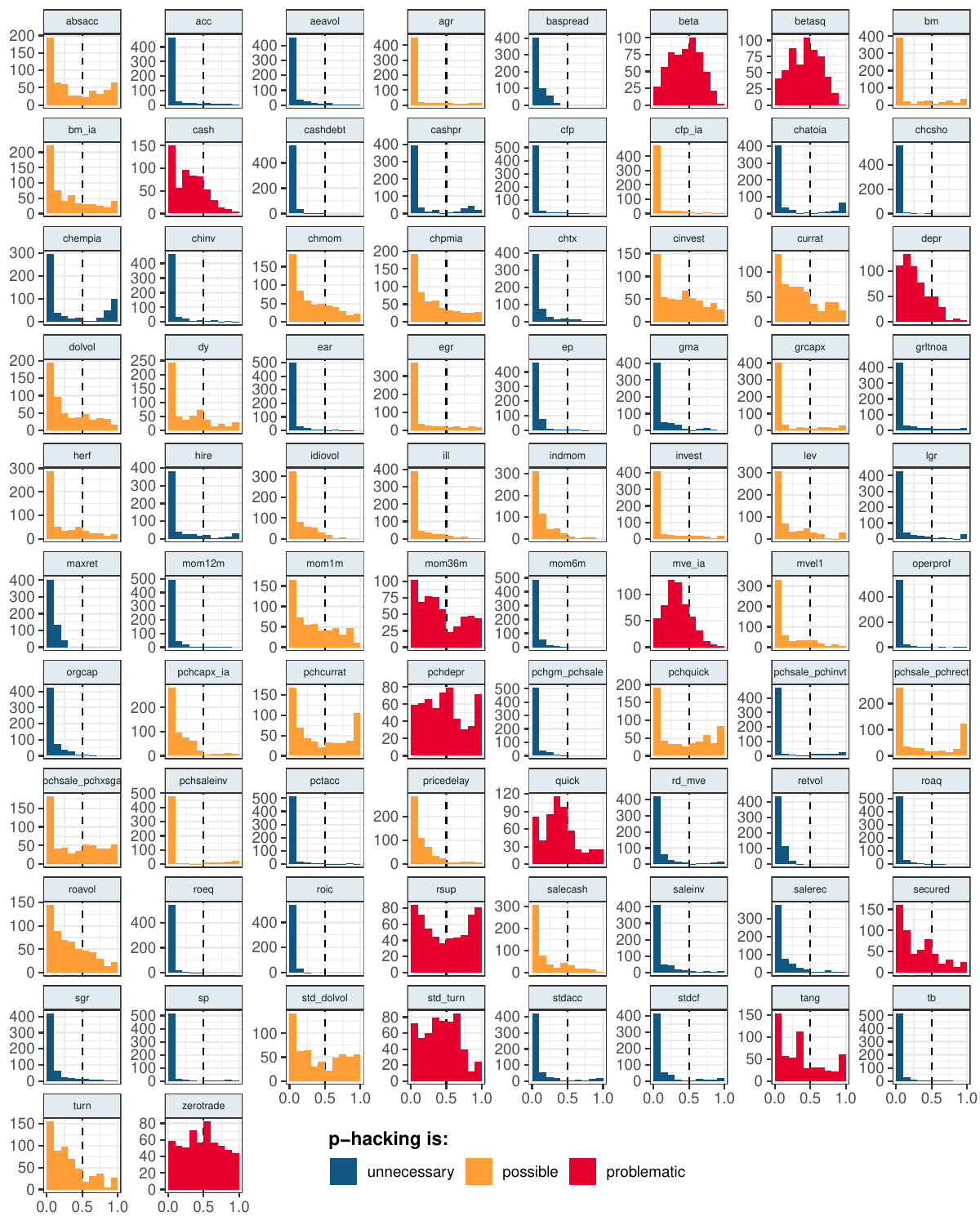}\vspace{-4mm}
\caption{\textbf{Distributions of $p$-values}. \small We produce the histogram of the $p$-values, for each sorting variable. There are $I=10$ breaks, which mark the deciles of the distribution. The vertical lines mark the limit of the critical zone ($x=1/2$). Colors code the level of $\kappa$ in Equation \eqref{eq:kappa}. }  
\label{fig:pval_dist}
\end{center}
\end{figure}

Instead, we resort to an ad-hoc classification of anomalies which is based on a weighted average of the decreasing rate of the $p$-curve. Namely, for each sorting characteristic, we compute
\begin{equation}
\kappa = \sum_{i=1}^{I/2}\frac{n_{i+1}}{n_i} \times \frac{1}{i}.
\label{eq:kappa}
\end{equation}
Without much loss of generality, we are assuming that the number of intervals in the histogram over the unit interval is even, hence all relevant frequency counts $n_i$ are spanned for $i$ between 1 and $I/2$ (the critical zone is $[0,1/2]$). The discounting factor $i^{-1}$ gives more weight to the first bars of the histograms because after a few rounds of decrease, the values of bar counts is very noisy and ratios are less trustworthy. Our classification is as follows:
\begin{itemize}
\setlength\itemsep{-0.1em}
    \item $p$-hacking is \textbf{unnecessary} if $\kappa < 0.25$.  
    \item $p$-hacking is \textbf{possible} if $\kappa \in (0.25, 0.4]$.
    \item $p$-hacking is \textbf{problematic}, otherwise.
\end{itemize}
Of course, these choices are arbitrary and determined ex-post, but, as is shown in Figure \ref{fig:pval_dist}, they yield coherent groups. By \textit{problematic}, we mean that the obtained vector of $p$-values would require a substantial amount of trafficking in order to pass a detection test and this may violate the requirement of a coherent structure for the paths. Typically what is possible is to remove, add, or change particular options or layers, but this is unlikely to alter the distribution exactly in the sought direction. Therefore, if the original $p$-curve is far from decreasing in the first place, minor adjustments will not suffice and hacking will be complicated.

\section{Application: Fama-MacBeth regressions}
\label{sec:FM}

\subsection{Method and paths}

A cornerstone question in financial economics pertains to the evaluation of the risk premium of asset pricing factors. Arguably, the most popular approach to answer this question is the double-pass regression of \cite{fama1973risk}. For the sake of completeness, we recall the two passes briefly below. First, asset returns are regressed on the target $F$ factors as follows:
\begin{equation}
    r_{t,n} = a_n + \sum_{f=1}^Fb^f_nf_t +e_{t,n}, \quad \text{estimated for all assets, }n. 
    \label{eq:pass1}
\end{equation}
The estimated loadings $\hat{b}_n^f$ are then recycled as independent variable in the second pass in which is run on a date-by-date basis:
\begin{equation}
r_{t,n} = \gamma_{t,0} + \sum_{f=1}^F \gamma_{t}^f\hat{b}_n^f+\epsilon_{t,n}, \quad \text{estimated for all dates, }t.
    \label{eq:pass2}
\end{equation}

Nevertheless, the methodology leaves several open choices. One important option is how to perform the first pass. For instance, it is possible to estimate the loadings on the full sample, an approach which we henceforth qualify as ``\textit{in-sample}'', or to estimate them on rolling samples (``\textit{out-of-sample}''). Another important degree of freedom is the set of assets from which these loadings are estimated. For instance, \cite{ang2020using} argue that, contrary to conventional wisdom, using portfolios instead of individual assets ``\textit{destroys information by shrinking the dispersion of betas, leading to larger standard errors}''.

\begin{figure}[!h]
\begin{center}
\includegraphics[width=13cm]{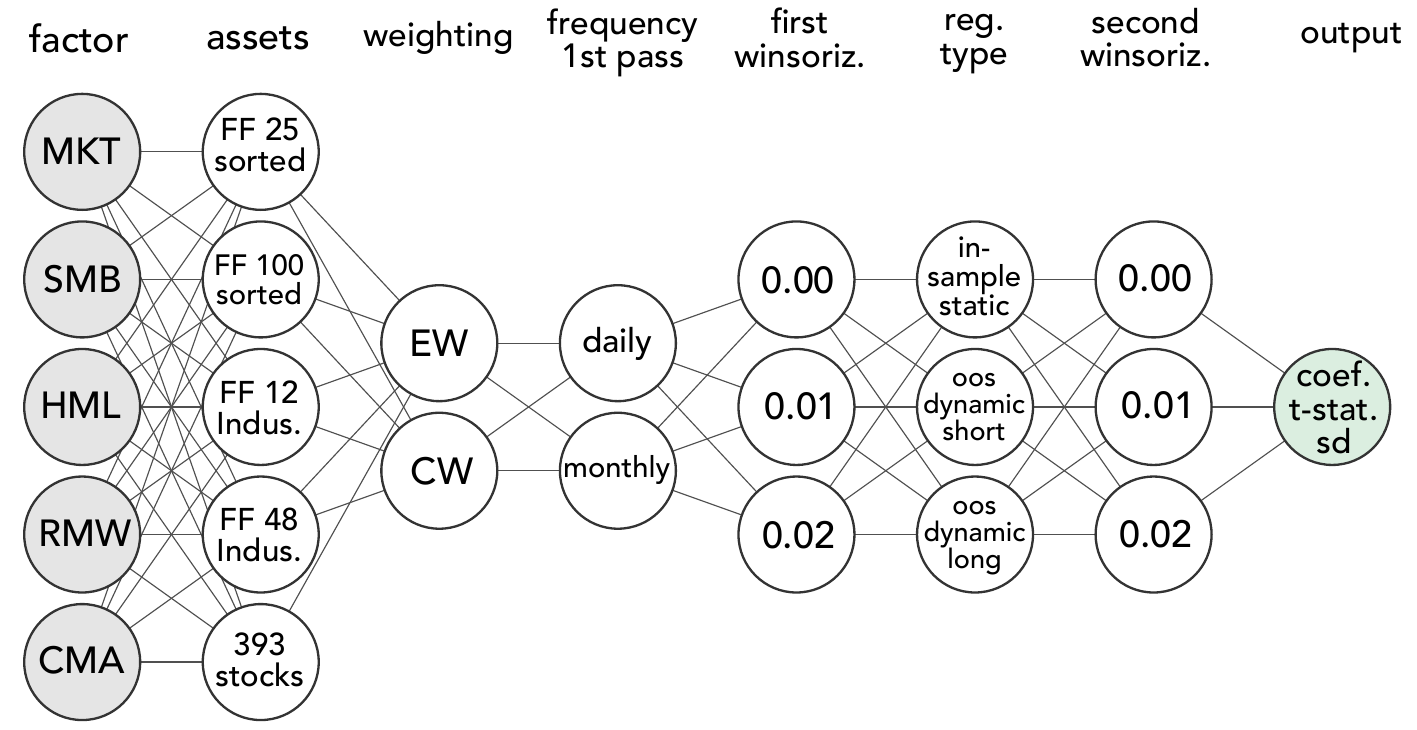}\vspace{-3mm}
\caption{\textbf{Forking paths for Fama-MacBeth regressions}. \small We depict the seven steps of the protocol. Note: for individual stocks, weighting does not apply - it is fixed to EW by default. In total, there are $9\times2\times3\times3\times3=486$ paths for each factor.}
\label{fig:nn_FM} 
\end{center}
\end{figure}

In Figure \ref{fig:nn_FM}, we show the diversity of paths for this study, with the following choices and layers:

\begin{enumerate}
\setlength\itemsep{-0.2em}
    \item the \textbf{factor} for which the risk premium is computed, among the five \cite{fama2015five} factors;
    \item the base \textbf{assets} that are used for the estimation. There are five alternatives: two Fama-French sorted portfolios on book-to-market (25 or 100 portfolios), two Fama-French industry portfolios and 393 individual stocks, available in open source for reproducibility;
    \item the \textbf{weighting scheme} of portfolios in base assets, either equally-weighted or value-weighted (does not apply to individual stocks);
    \item the data \textbf{frequency} for the first pass (initial loading estimation), whether it is daily or monthly;
    \item the \textbf{winsorization level} for returns before the first pass;
    \item the \textbf{regression type}, i.e., whether the first pass is run on the full sample, or on rolling windows of short or long samples (24 versus 60 months or 120 versus 300 days depending on data frequency).
    \item the \textbf{winsorization level} for returns before the second pass, which is applied to treat outliers in estimated betas.
\end{enumerate}

This procedure generates, for each date $t$ and factor $f$, the $\hat{\gamma}_t^f$, along with the related AIC criteria and standard errors which can be used to compute weights \eqref{eq:weights} and confidence intervals \eqref{eq:IC}. The results are obtained for the five factors from \cite{fama2015five}.

\subsection{Baseline results}
\label{sec:FM_baseline}

\begin{figure}[!h]
\begin{center}
\includegraphics[width=15.5cm]{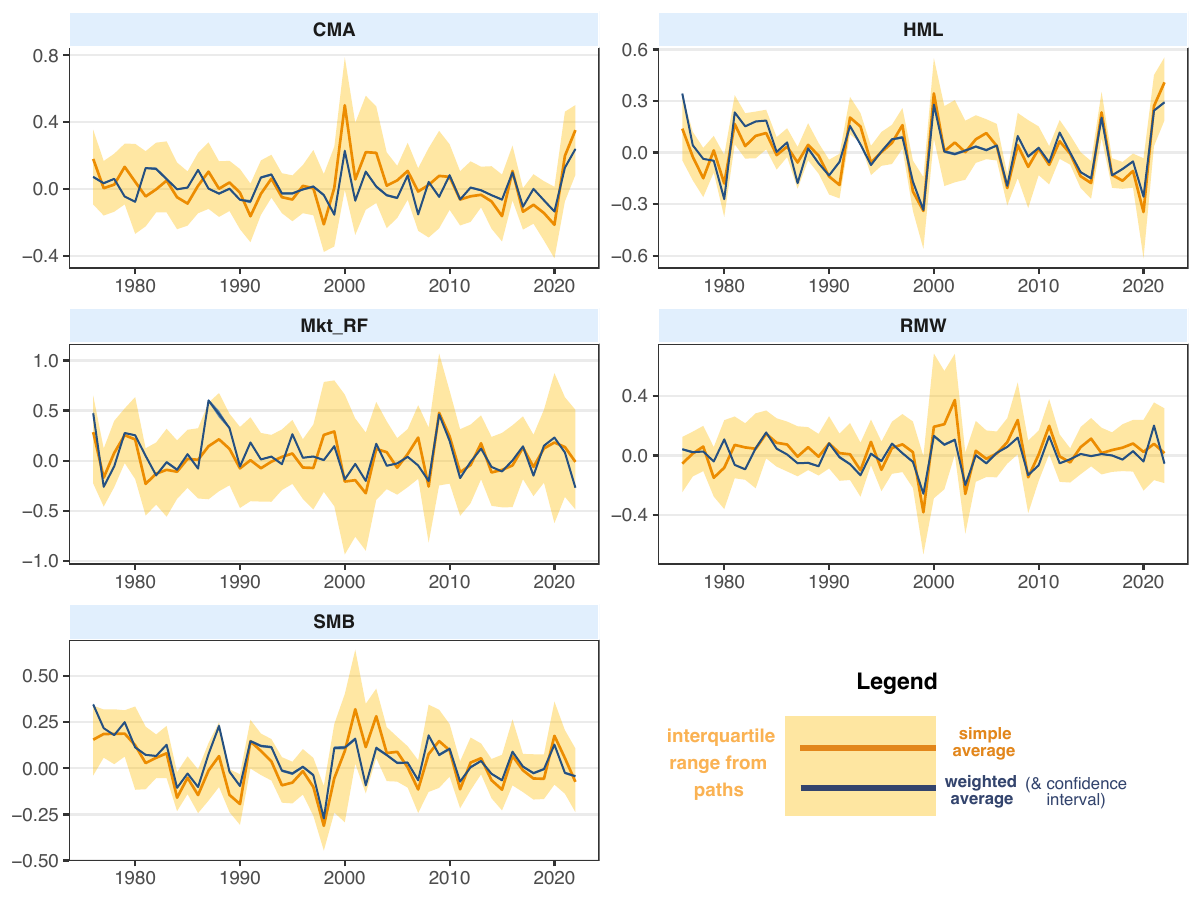}\vspace{-4mm}
\caption{\textbf{Fama-MacBeth premia: distribution across time}. \footnotesize We represent some properties of loadings from the second pass of \cite{fama1973risk} regressions. Monthly values are averaged at the annual scale to ease readability. The \textbf{\textcolor{orange}{orange}} line shows the simple average of loadings across all 486 paths. The \textbf{\textcolor{dblue}{dark blue}} line depicts the Bayesian weighted average from Equation \eqref{eq:bayes_avg}. Around these latter averages, the 99\% confidence intervals from Equation \eqref{eq:IC} are indistinguishably thin. Finally, the light yellow shaded areas represent the inter-quartile range across all paths.     }
\label{fig:FM_baseline}
\end{center} \vspace{0mm}
\end{figure}

In Figure \ref{fig:FM_baseline}, we plot the average premia for calendar years.\footnote{The code for this section can be accessed \href{https://www.gcoqueret.com/files/misc/forking_paths3.html}{here}.} First, we compute simple and weighted averages for each month and these values are then averaged (simply) over the year (orange curve). We also report the Bayesian confidence intervals in light blue, but they are so narrow that they are indistinguishable from the corresponding weighted average curves (Equation \eqref{eq:bayes_avg}, in dark blue). Results for frequentist averages are qualitatively the same and the intervals remain too thin to discriminate from the means, as in Figure \ref{fig:MA} for monthly frequencies. In unreported results, we find that simple averages have larger standard errors and thus yield larger hence more conservative intervals. Overall, the different averaging schemes  produce quite consistent values, especially for the value factor (HML).

Lastly, in the background of the figures, we show the inter-quartile range across all paths. The latter is the widest for the market factor and illustrates the diversity of outcomes and the richness of paths. Notably, we find that the paths often yield values of contrasting signs, which is not surprising since premia oscillate around zero. There are however some instances where a large majority of path values lie either above or below zero: they correspond to cases when many paths agree on the sign of the premium (e.g., in 1992 or 2021 for HML or in 1990 for SMB).

\subsection{Which design options matter?}
\label{sec:FM_options}

To illustrate the conditional averages proposed in Section \ref{sec:cond_avg}, we focus on the layers for which there are 3 possible options in our protocol: they are the first pass regression type and the two winsorization levels (before and after the first pass). 

Our results are gathered in Table \ref{tab:FM_options}. The left part of the table reports the average premium of factors when fixing either of the three options $o_j$. For instance, for Panel B and C, $o_1$ corresponds to an absence of winsorization, while $o_2$ and $o_3$ winsorize at the 1\% and 2\% level, respectively. For panel A, the $o_j$ pertain to the type of first pass regression (sampling scheme). The right part of the table gathers the difference between the options and the statistical significance of the related simple $t$-test. We see that for panel C, differences are hardly significant, which means that the second winsorization, after the loadings have been estimated, has little impact on premia estimates.

\begin{table}[ht]
\centering
\small
\begin{tabular}{l | rrr | lll}
  \midrule
\multirow{2}{*}{Factor}  & \multicolumn{3}{c}{Means of options $o_j$} & \multicolumn{3}{c}{Difference in means} \\ \cline{2-7}
 & $o_1$ & $o_2$ & $o_3$ & $o_1-o_2$ & $o_1-o_3$ & $o_2-o_3$ \\ 
  \midrule
  \multicolumn{7}{l}{\textbf{Panel A: Regression type (first pass)}} \\
CMA & 0.149 & 0.187 & 0.264 & -0.038 * & -0.115 *** & -0.077 ** \\ 
  HML & 0.091 & 0.110 & -0.115 & -0.019  & 0.206 *** & 0.225 *** \\ 
  MKT & 0.141 & 0.147 & 0.267 & -0.006  & -0.126 ** & -0.12 ** \\ 
  RMW & 0.135 & 0.173 & 0.418 & -0.038 * & -0.283 *** & -0.245 *** \\ 
  SMB & 0.180 & 0.283 & 0.068 & -0.103 *** & 0.112 *** & 0.215 *** \\ \midrule
    \multicolumn{7}{l}{\textbf{Panel B: Winsorization threshold before the first pass}} \\
  CMA & 0.430 & 0.107 & 0.070 & 0.323 *** & 0.36 *** & 0.037  \\ 
  HML & -0.033 & 0.038 & 0.066 & -0.071 ** & -0.099 *** & -0.028  \\ 
  MKT & 0.074 & 0.189 & 0.298 & -0.115 ** & -0.224 *** & -0.109 * \\ 
  RMW & 0.194 & 0.267 & 0.283 & -0.073 ** & -0.089 *** & -0.016  \\ 
  SMB & 0.202 & 0.146 & 0.173 & 0.056 * & 0.029  & -0.027  \\ \midrule
      \multicolumn{7}{l}{\textbf{Panel C: Winsorization threshold after the first pass}} \\
  CMA & 0.210 & 0.204 & 0.193 & 0.006  & 0.017  & 0.011  \\ 
  HML & 0.021 & 0.023 & 0.027 & -0.002  & -0.006  & -0.004  \\ 
  MKT & 0.150 & 0.194 & 0.223 & -0.044  & -0.073  & -0.029  \\ 
  RMW & 0.235 & 0.253 & 0.256 & -0.018  & -0.021  & -0.003  \\ 
  SMB & 0.164 & 0.177 & 0.180 & -0.013  & -0.016  & -0.003  \\  
   \midrule
\end{tabular}
\caption{\textbf{Which options matter?} \small We compute the average premia for choices with three options $o_1$, $o_2$ and $o_3$. These choices are short rolling regression ($o_1$), long dynamic rolling regression ($o_2$) and static regression ($o_3$) for the first pass and 0.00, 0.01 and 0.02 respectively for the winsorization thresholds, before and after the first pass. The significance levels are: (***)$<$0.001$<$(**)$<$0.01$<$(*)$<$0.05.   \label{tab:FM_options}}
\end{table}

In contrast, in the first two panels, we report some substantial differences. From panel B, we infer that switching from no winsorization to 1\% or 2\% does alter the results, though not always in the same direction, depending on the factor. The figures from panel A imply that the choice of samples for the first pass also matters. The largest changes occur between full sample versus both types of rolling samples. Interestingly, switching from small to longer dynamic samples always has a negative impact on premia, though we cannot rationalize why that may be the case.

\subsection{Comparison with prior work}
\label{sec:FM_prior}

We wish to compare the distribution of the 486 paths we generated for each date with the values of prior studies, namely \cite{fama1973risk} and \cite{ang2020using}. For simplicity, we will only compare premia that are averaged over the longest periods in these articles. 

\begin{figure}[!h]
\begin{center}
\includegraphics[width=15cm]{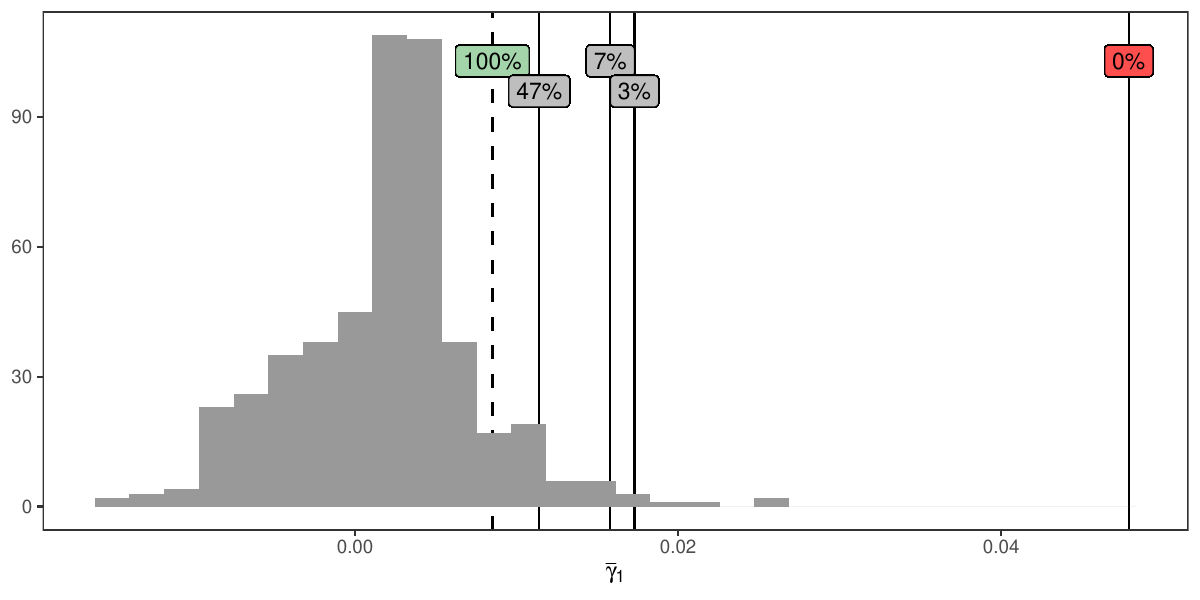}\vspace{-5mm}
\caption{\textbf{Comparison with prior work (market factor)}. \footnotesize We depict the distribution of average monthly premia of the market factor spanned by the 486 paths. In addition, we locate the corresponding values for the market factor from the studies of \cite{fama1973risk} and \cite{ang2020using} with dashed and full vertical lines, respectively. The colored rounded squares indicate the ease to confirm (EtC) of the reported value with respect to the Gaussian distribution fitted on the grey histogram (mean and standard deviation). The EtC is equal to one minus \eqref{eq:Phi} with $q=0.9$. The colors code the facility to reproduce the original outcome.  }
\label{fig:FM_prior}
\end{center} \vspace{0mm}
\end{figure}

In \cite{fama1973risk}, Table 4 in Section V-B, the average $\gamma$ value for the market premium is 0.0085 for the longest sample, from 1935 to 1968. \cite{ang2020using}, in Table 5 in Section IV-A, report four values, depending on the set of assets used for the estimation: 0.0114, 0.0158, 0.0173 and 0.0479. We locate these values in comparisons to the ones we obtained with vertical lines in Figure \ref{fig:FM_prior}.

In addition, we specify the ease-to-replicate indicator at the top of the vertical lines. We find that the value reported by \cite{fama1973risk} is very reasonable, as it lies not too much to the right of the distribution obtained from the paths. With regard to the average premia found in \cite{ang2020using}, the ease to reproduce results is more mixed and lukewarm, with one value (4.8\% per month) which is found to be virtually impossible to reach with the paths that we were able to span. This large value correspond to the estimation with the individual stocks, which seem to suggest that granular test assets yield higher estimated premia. We are able to test this conjecture and do not find much support for it. In Figure \ref{fig:FM_assets} in the Appendix, we plot average absolute premia across all factors and test asset groups and do not find a strong relationship between the granularity of assets and the magnitude of premia. If anything, it is the least granular assets (the 12 industries) that are associated with the largest absolute premia. 

A last word on sample sizes. In \cite{fama1973risk}, the average premia for the market change substantially from period to period. For example, it is equal to 0.163 in 1935-1945 and to 0.143 in 1961-1968. These values are much larger than the estimate over the total period and would obtain an EtC score well below 50\%. Hence, again, considering chronologically deep samples increases the odds of confirmation.

%\clearpage

\section{Conclusion}
\label{sec:conc}

Amid debates in some scientific communities about the validity of empirical results, we make the case for an exhaustive approach. Formally, we suggest to report results for a large number of design choices, as if extensive robustness checks were in fact constituent of the baseline research protocol. Small variations in designs allow the generation of many estimates and  test statistics. The distribution of these statistics can help figure out if one configuration yielded a favorable outlier, or if the sought effect is indeed statistically strong. Moreover, having many coefficient or statistics at one's disposal allows to resort to aggregation so as to obtain more robust estimates and confidence intervals.

The application of these ideas to three exercises in financial economics confirms the intuition that by considering many design options, the range of outcomes can increase rapidly. Because of this, it is important to report which particular choices in the protocol may shift the distribution of outcomes. Moreover, spanning large numbers of paths allow to determine when prior results are plausible and robust - or not. We craft an indicator called the \textit{ease-to-corroborate} (EtC) which evaluates how realistic effect sizes published in prior studies can be. We find substantial heterogeneity in the asset pricing literature with respect to this criterion, with some contributions that report very reproducible outcomes and others that propose results that appear much harder to replicate. 

Forking paths can also be very useful to generate conclusions that are more trustworthy, especially for investment purposes. Averaging returns, premia or loadings across many configurations strengthens inference. We find high cross-period correlation between anomalies returns' once they have been averaged across many paths. This removes the risk of an outlier point from one specific set of implementation choices and increases the odds of generalization out-of-sample. 

Paths can also be applied to multiple testing. They allow to generate outcomes that are less homogeneous, compared to bootstrapped series. This produces distributions for maximum statistics which have heavier tails. Consequently, the corresponding significance thresholds are higher. For asset anomalies to hold, our framework requires that they remain statistically profitable under various weighting schemes and across several sub-periods. Our second empirical analysis finds that the bar for $t$-statistics in portfolio sorts should be raised to 8.2, an almost prohibitive level that is much higher than those typically used in the literature. However, higher decision hurdles also come at the cost of more false negatives, which may or may not matter, especially for investment purposes.

There are of course several limitations to our suggestions. First, it is possible to push the limits of data-snooping to the extreme by reporting only the combinations of design choices that fit a particular narrative, but this is arguably cumbersome. Second, given the amount of time required to generate comprehensive results, the research question must be inherently simple. Each path should not take more than a handful of minutes, so that hundreds, or thousands, of them can be generated in less than one day. The aim of the paper is clearly not to increase the carbon footprint of researchers. Long computation times may contribute to this footprint and we refer to \cite{mariette2021open} for a discussion on this matter. This is why a precise framing of the research question, as well as its relevant ramifications, is imperative to avoid superfluous digressions.

\clearpage

\appendix

\section{Examples of Lipschitz constants}
\label{sec:exlip}

This section relies heavily on norms. For vectors, we work with $L^p$ norms: $\|\bm{v}\|_p^p=\sum_{n=1}^N|v_n|^p$ and for $(N\times K)$ matrices we will consider the following:
\begin{itemize}
\setlength\itemsep{-0.2em}
    \item $\|\bm{M}\|_2^2=\sum_{n=1}^N\sum_{k=1}^KM_{n,k}^2$: Frobenius Norm;
    \item $\|\bm{M}\|_1=\underset{k}{\max}\sum_{n=1}^N|M_{n,k}|$: maximum absolute column sum;
    \item $\|\bm{M}\|_\infty=\underset{n}{\max}\sum_{k=1}^K|M_{n,k}|$: maximum absolute column row.
\end{itemize}

\subsection{Descriptive statistics}

Sample moments and other mainstream metrics play an important in empirical studies. We begin our journey of illustrations with the simplest of them all: the sample mean. In the sequel, we will use the notation $f$ as generic mapping. The case of the (biased) \textbf{sample variance} is more tricky:
\begin{align}
    \|f(\mathbb{d})-f(\bm{d}) \|_p &=   \left| \frac{1}{N} \sum_{n=1}^N \left(\mathbb{d}_n-\frac{1}{N} \sum_{n=1}^N \mathbb{d}_n\right)^2 - \frac{1}{N} \sum_{n=1}^N \left(\mathpzc{d}_n- \frac{1}{N} \sum_{n=1}^N \mathpzc{d}_n\right)^2   \right| \nonumber \\
    &=\frac{1}{N} \left|N(\bar{d}^2-\bar{\mathbb{d}}^2)+\sum_{n=1}^N\mathbb{d}_n^2-\mathpzc{d}_n^2 \right| \nonumber \\
    &= \frac{1}{N}\left|N(\bar{d}-\bar{\mathbb{d}})(\bar{d}+\bar{\mathbb{d}})+\sum_{n=1}^N(\mathbb{d}_n+\mathpzc{d}_n)(\mathbb{d}_n-\mathpzc{d}_n) \right| \nonumber \\
    &\le |\bar{\mathbb{d}}+\bar{d}|\times |\bar{\mathbb{d}}-\bar{d}| +\frac{d^*}{N}\left|\sum_{n=1}^N(\mathbb{d}_n-\mathpzc{d}_n) \right| \nonumber \\
    & \le c_p \|\mathbb{d}-\bm{d} \|_p, \label{eq:var}
\end{align}
where $c_p =N^{-1/p}( |\bar{\mathbb{d}}+\bar{d}|+d^*)$, with $d^*=\max_n|\mathbb{d}_n+d_n|$.  $\bar{\mathbb{d}}$ and $\bar{d}$ are the sample means. The last inequality comes from \eqref{eq:mean2}. In this case, and as will be recurrent, the constant depends on the magnitude of the series. To remove this dependence, it is imperative to specify some properties of the vectors (e.g., if they belong to the unit sphere, or if their range is restricted to particular intervals). This comment holds for the remainder of the paper, as many constants will be input-dependent below.

Typically, for the sample \textbf{covariance}, we have that
\begin{align*}
|f(\mathbb{d}_1,\bm{d}_1)-f(\mathbb{d}_2,\bm{d}_2)|&=
    |(\mathbb{d}_1-\bar{\mathbb{d}}_1)'(\bm{d}_1-\bar{\bm{d}}_1)-(\mathbb{d}_2-\bar{\mathbb{d}}_2)'(\bm{d}_2-\bar{\bm{d}}_2)| \\
    &= |(\mathbb{d}_1-\bar{\mathbb{d}}_1 - (\mathbb{d}_2-\bar{\mathbb{d}}_2))'(\bm{d}_1-\bar{\bm{d}}_1)-(\mathbb{d}_2-\bar{\mathbb{d}}_2)'(\bm{d}_2-\bar{\bm{d}}_2-(\bm{d}_1-\bar{\bm{d}}_1))| \\
    & \le \frac{\|\bm{d}_1-\bar{\bm{d}}_1 \|_1}{N} (|\mathbb{d}_1-\mathbb{d}_2|+|\bar{\mathbb{d}}_1-\bar{\mathbb{d}}_2|) + \frac{\| \mathbb{d}_2-\bar{\mathbb{d}}_2\|_1}{N}(|\bm{d}_1-\bm{d}_2|+|\bar{\bm{d}}_1-\bar{\bm{d}}_1|)
\end{align*}
which is again a similar form.
\vspace{3mm}

Let us now mention the \textbf{maximum} of vectors. We have that
\begin{align*}
    \max_n \mathbb{d}_n = \max_n  [ \mathbb{d}_n-\mathpzc{d}_n +\mathpzc{d}_n] \le  \max_n |\mathbb{d}_n-\mathpzc{d}_n| + \max_n \mathpzc{d}_n,
\end{align*}
so that
$$\|\max_n\mathbb{d}_n -\max_n d_n\|_\infty=|\max_n\mathbb{d}_n -\max_n d_n| \le \max_n   |\mathbb{d}_n-\mathpzc{d}_n|=\|\mathbb{d}-\bm{d}\|_\infty,$$
i.e, the Lipschitz constant in this case is one. Straightforwardly, the same applies to the minimum operator.

\subsection{Other examples of Lipschitz constants}

In empirical studies, the \textbf{data collection} stage is the hardest to model, because of its heterogeneity. It can be quite constrained if data comes from a provider (e.g., WRDS, Bloomberg, etc.), in which case the researcher has a few degrees of freedom: which variables to import, for which universe, at which frequency, over which time frame, etc. Providers often update their data so that downloading a sample at two different periods may generate discrepancies if series are not kept point-in-time. This has been recently shown for the Fama-French factors in \cite{akey2021noisy}. There is also relatively little room for initiative in economics or physics when working with official series, such as GDP output, inflation, unemployment, CO$_2$ concentration, temperatures, etc.

However, when the study is based on surveys, the researcher has more latitude, and we for instance refer to the guide of \cite{bergman2020survey} for an overview of the range of options in that case. In qualitative studies, there is also an important coding phase (see, e.g., the review by \cite{basit2003manual}), which is difficult to model neatly and efficiently.

For all these reasons, we commence this section with the step that comes right \textit{after} data collection, namely \textbf{data cleaning}. We underline that \cite{mitton2021methodological} reports that ``\textit{the methodological decisions that affect statistical significance the most are dependent variable selection, variable transformation, and outlier treatment}''. In this subsection, we tackle all of these elements. Lastly, the purpose of the section is to show that most classical operations on data can be represented as Lipschitz mappings. It is not to provide sharp constants for these mappings.

\subsubsection{Data cleaning}

The first issue that most, if not all, researchers encounter, is \textbf{missing data}. When working with time-series, a common practice is to impute with the most recent well-defined point prior to the missing value, if it exists. Another option is to resort to cross-sectional means or medians. Interpolation is usually avoided because it introduces a forward-looking bias. In this subsection, to ease the exposition, we make strong assumptions on the vectors on which the imputation mapping will operate.

Formally, we consider two vectors $\mathbb{d}$ and $\bm{d}$ such that $S$ is the \textit{common} set of indices for which a value is missing. The fact that $S$ is common to the two vectors comes from the fact that we need $\|\mathbb{d}-\bm{d}\|$ to be well defined. In the above norm, two points that are not defined are assumed to be equal, but if one value is defined and the other is not, there is no unambiguous way to proceed. For simplicity, we assume that 1 does not belong to the $S$ set, so that the imputation values will always be defined. In addition, we impose that the indices in $S$ are never consecutive numbers, though this assumption can be relaxed easily. We then have
\begin{align*}\|f(\mathbb{d})-f(\bm{d})\|_p^p= \sum_{n \in S} |\mathbb{d}_{n-1}-d_{n-1}|^p + \sum_{n \notin S} |\mathbb{d}_{n}-d_{n}|^p \le 2 \|\mathbb{d}-\bm{d}\|_p^p
\end{align*}
where the last inequality comes from the fact that the values that precede missing points get counted twice. The Lipschitz constant is not very sharp in this case.

Other examples of methods include cross-sectional imputation, whereby a missing value is replaced by the cross-sectional mean (or median) across other observations. In this case, via inequality \eqref{eq:mean} it is also possible to derive a Lipshitz constant for mean-driven imputation. Parametric imputation based on some distributional assumption follow the same logic, though their treatment is substantially more involved.

\vspace{3mm}

One extreme solution when facing missing data is simply the \textbf{removal of observations}. To illustrate this issue, we consider two matrices of numerical data $\mathbb{D}$, $\bm{D}$ with equal sizes, $N$ rows and $M$ columns. In line with the above assumptions, we write $S$ for the (common) indices of their rows which contain missing points. Naturally, we again assume that the cardinal of $S$ is much smaller than the total number of rows $N$. In this case, it is straightforward that for the usual matrix norms (Frobenius, $\|\cdot\|_1$ and $\|\cdot\|_\infty$), the Lipschitz constant is one at most, i.e., that
\begin{align*}
    \|f(\mathbb{D})-f(\bm{D})\| \le \|\mathbb{D}-\bm{D}\|.
\end{align*}
Similarly, the researcher may want to remove columns (instead of rows) of the data because of \textbf{co-linearity} issues. The Lipschitz constant of such an operation is also one at most.

\vspace{3mm}

Another important stage in data processing is \textbf{outlier management}. One of the most frequently used tools to this purpose is \textbf{winsorization}, whereby extreme values are replaced by given quantiles, often at the 1\% and 99\% levels.

Without loss of generality, let us assume that the vector $\mathbb{d}$ is ordered, i.e., that $\mathbb{d}_1 < \dots <\mathbb{d}_N$. For a given integer $k \ll N/2$, the winsorization operator is defined as:
\begin{equation}
    f(\mathbb{d}_n)=\left\{\begin{array}{ll}
         \mathbb{d}_n & \text{if } n \in [k+1,N-k] \\
         \mathbb{d}_{k+1} & \text{if } n \le k \\
         \mathbb{d}_{N-k} & \text{if } n > N-k
    \end{array} \right.,
\end{equation}
so that exactly $2k$ values are replaced: the most extreme $k$ values in both tails. If we abusively write $f(\mathbb{d})$ for the vector of $f(\mathbb{d}_n)$ values, it holds that
\begin{align*}
    \|f(\mathbb{d})-f(\bm{d}) \|_1&=\sum_{n=1}^k |\mathbb{d}_{k+1}-\mathpzc{d}_{k+1}| + \sum_{n=k+1}^{N-K}| \mathbb{d}_n-\mathpzc{d}_n| + \sum_{n=N-K+1}^N |\mathbb{d}_{N-k}-\mathpzc{d}_{N-k}| \\
    &=K(|\mathbb{d}_{k+1}-\mathpzc{d}_{k+1}| + |\mathbb{d}_{N-k}-\mathpzc{d}_{N-k}|)+ \sum_{n=k+1}^{N-K} |\mathbb{d}_n-\mathpzc{d}_n | \\
    &\le (1+K)\|\mathbb{d}-\bm{d}\|,
\end{align*}
where the constant is clearly sub-optimal. It could be improved but would then rely on the properties of the underlying vectors.

%trimmed mean \cite{caperaa1995variance} + winso \cite{olbricht1992robustification}

\subsubsection{Variable engineering}

Once the data has been cleaned, the researcher will often perform additional adjustments. We list a few below.

\textbf{Normalization} is a common step in data preparation: it ensures that all variables have roughly the same scales. This is convenient when one wants to compare effect sizes for example. There are several ways to proceed, such as standardization, or min-max rescaling. Let us analyze the former:
\begin{align}
    \| f(\mathbb{d})-f(\mathpzc{d})\|_2 &= \left\|\frac{\mathbb{d}-\bm{m}_\mathbb{d}}{\sigma_\mathbb{d}}-\frac{\mathpzc{d}-\bm{m}_\mathpzc{d}}{\sigma_\mathpzc{d}} \right\|_2=\sigma_\mathbb{d}^{-1}\sigma_\mathpzc{d}^{-1}\|\sigma_\mathpzc{d}(\mathbb{d}-\bm{m}_\mathbb{d})-\sigma_x(\mathpzc{d}-\bm{m}_\mathpzc{d})\|_2 \nonumber \\
    &=\sigma_\mathbb{d}^{-1}\sigma_\mathpzc{d}^{-1}\|(\sigma_\mathpzc{d}-\sigma_\mathbb{d}+\sigma_\mathbb{d})(\mathbb{d}-\bm{m}_\mathbb{d})-\sigma_\mathbb{d}(\mathpzc{d}-\mathbb{d}+\mathbb{d}-\bm{m}_)\|_2 \nonumber \\
    &\le \sigma_\mathbb{d}^{-1}\sigma_\mathpzc{d}^{-1}|\sigma_\mathpzc{d}-\sigma_\mathbb{d}| \times \|\mathbb{d}-\bm{m}_\mathbb{d}\|_2+  \sigma_\mathpzc{d}^{-1}\|\mathbb{d}-\mathpzc{d}\|_2+\sigma_\mathpzc{d}^{-1}\|\bm{m}_\mathpzc{d}-\bm{m}_\mathbb{d}\|_2  \label{eq:7}\\
    &\le \sigma_\mathpzc{d}^{-1}\sqrt{N}\frac{|\sigma_x^2-\sigma_\mathpzc{d}^2|}{\sigma_\mathbb{d}+\sigma_\mathpzc{d}}+\sigma_\mathpzc{d}^{-1}\|\mathbb{d}-\mathpzc{d}\|_2+\sigma_\mathpzc{d}^{-1}N^{-1/2}\|\mathbb{d}-\mathpzc{d}\|_2 \label{eq:8}\\
    & \le c \|\mathbb{d}-\mathpzc{d}\|_2  \nonumber
\end{align}
where $\bm{m}_\mathbb{d}$ is the constant mean vector of $\mathbb{d}$, $\sigma_\mathbb{d}$ its standard deviation and $$c=\sigma_\mathpzc{d}^{-1}(1+N^{-1/2}+\sqrt{N}c_1(\sigma_\mathbb{d}+\sigma_\mathpzc{d})^{-1}),$$
the constant $c_1$ being the one from Inequation \eqref{eq:var} for $p=1$. We have used that $\|\mathbb{d}-\bm{m}_\mathbb{d}\|_2=\sqrt{N}\sigma_\mathbb{d}$ and applied \eqref{eq:mean2} and \eqref{eq:var} in lines \eqref{eq:7} and \eqref{eq:8}, respectively.
 
 \vspace{3mm}

Sometimes, when working with time-series, the model requires stationary variables, but the collected data is integrated and has unit roots. In other contexts, the level of the independent variable may matter less than its variations from a predictive standpoints. Thus it is relevant to consider variable differences in these settings too. In any case, the solution is \textbf{differentiation}:
\begin{equation}
    f(\mathbb{d}_n)=\left\{\begin{array}{ll }
         \text{NA} & \text{if } n=1  \\
         \mathbb{d}_n-\mathbb{d}_{n-1}& \text{otherwise}
    \end{array} \right. .
    \label{eq:diff}
\end{equation}
In practice, the first missing point is often removed so that the resulting vector has length $N-1$. For two numerical vectors with no missing points $\mathbb{d}$ and $d$,
\begin{align*}
    \|f(\mathbb{d})-f(\mathpzc{d})\|_1&=\sum_{n=2}^N|\mathbb{d}_n-\mathbb{d}_{n-1}-\mathpzc{d}_n+\mathpzc{d}_{n-1}| \\
    &\le \sum_{n=2}^N|\mathbb{d}_n-\mathpzc{d}_n|+|\mathbb{d}_{n-1}-\mathpzc{d}_{n-1}|\le 2\|\mathbb{d}-\mathpzc{d}\|_1
\end{align*}

\vspace{3mm}

It may also happen that researchers seek to explain long term effects. For instance, in the predictability literature, there is a debate between short-term and long-term predictability. At a first order approximation, long term returns can be viewed as \textbf{cumulative sums} of shorter horizon returns, which is why we briefly mention the topic below. For a given well-defined numerical vector, we have in this case, for $n>0$,
$$f(\mathbb{d}_n)=\sum_{k=1}^n\mathbb{d}_k,$$
and
\begin{align*}
    \|f(\mathbb{d})-f(\bm{d})\|_1=\sum_{n=1}^N\left| \sum_{k=1}^n\mathbb{d}_k-\sum_{k=1}^n d_k\right|\le \sum_{n=1}^N \sum_{k=1}^n\left|\mathbb{d}_k-d_k\right|\le N\|\mathbb{d}-\bm{d}\|_1,
\end{align*}
where the bound may seem loose, but can be sharp if $\mathbb{d}_n=d_n=0$ for $n>1$ for instance.

\vspace{3mm}
 
To conclude this subsection, we acknowledge that many more operations exist in the data preparation phase and some would require a lengthy treatment. For instance, joining procedures that merge two tables according to a common key are a widespread practice. They are however more complex to handle and we leave their analysis to future work.

\subsubsection{Testing}

An ubiquitous tool in the researcher's arsenal is the \textbf{linear regression}. Given a matrix of independent variables $\bm{X}$ and the vector of dependent variable $\bm{y}$, the ordinary least square (OLS) estimator $\bm{b}$ that minimizes the quadratic error
\begin{equation}
e^2(\bm{X},\bm{y})=(\bm{y}-\bm{Xb})'(\bm{y}-\bm{Xb})
\label{eq:e2}
\end{equation}
is
\begin{equation}
    \bm{b}(\bm{X}, \bm{y})=(\bm{X}'\bm{X})^{-1}\bm{X}'\bm{y}
    \label{eq:b}
\end{equation}
where $\bm{v}'$ denotes the transpose of $\bm{v}$ (vector or matrix). In the sequel, we will always assume that the inverse matrix is well-defined. The issue is then that there are two inputs. Given one of them, it is possible to intuitively deduce data-specific Lipschitz constants. For instance, if $\bm{X}$ is fixed, then factorizing the $\bm{X}$-dependent matrices yields
\begin{equation}
    \| \bm{b}(\bm{X}, \bm{y})-\bm{b}(\bm{X}, \bm{z}) \| \le c_X \|\bm{y}-\bm{z} \|.
\end{equation}
The case when $\bm{y}$ is fixed is less straightforward but can be handled with suitable norms. The general case when both $\bm{X}$ and $\bm{y}$ are subject to perturbation is more intricate. It is reviewed in Section 5 of \cite{grcar2003optimal}. One foundational result (\cite{golub1966note}) is that if $\|\bm{X}\|_2=\|\bm{y}\|_2=1$, then
\begin{equation}
    \|\bm{b}(\bm{X}, \bm{y})-\bm{b}(\bm{Z}, \bm{v})\|_2 \le c (\|\bm{X}-\bm{Z}\|_2+\|\bm{y}-\bm{v}\|_2)+R,
    \label{eq:lm}
\end{equation}
where $c$ depends on the smallest singular value of $\bm{X}$, on $\|\bm{b}\|_2$, and on the quadratic error $e^2$ defined in Equation \eqref{eq:b}. The matrix norms are of Frobenius type: $\|\bm{X}\|_2^2=\text{tr}(\bm{XX}')$, where tr$(\cdot)$ is the trace operator. The above result holds in the case when the norms are arbitrarily small and the residual term $R$ is a second order term which is quadratic in the maximum of the two norms.

While the coefficients in linear regressions (or more general models) are undoubtedly analyzed by researchers, it is their \textbf{statistical significance} which often matters more because it will determine if the effect revealed by the study is strong enough.

In the case of a linear model, the expressions for the $t$-statistics are $t_k=b_k/\sqrt{s^2S_k}$, where $S_k$ is the $k^{th}$ diagonal element of $(\bm{X}'\bm{X})^{-1}$ and $s^2=e^2/(N-K)$, with $K$ being the number of columns of $\bm{X}$ (Equation 4-47 in \cite{greene2018econometric}). Lipschitz numbers can be obtained for $S_k$ (see e.g., \cite{demmel1992componentwise} and \cite{loh2018high}) and for $s^2=(N-K)^{-1}(\bm{y}'\bm{y}-\bm{y}'\bm{X}(\bm{X}'\bm{X})^{-1}\bm{X}'\bm{y})$ as well. From these numbers, it is possible to derive a Lipschitz constant similar to \eqref{eq:lm} for the statistic, based on norms that depend on the inputs.

Often, models are tested with several control variables and model permutations, in order to check the robustness of the initial specification. It is even possible to aggregate coefficients of multiple specifications via weighting, as in \cite{hansen2007least} and \cite{zhang2019inference}. In linear models, adding columns to the matrix $\bm{X}$ is often handled via the Frish-Waugh-Lovell theorem. The Lipschitz continuity of this operation with respect to estimates and $t$-statistics is currently out of the scope of the paper (it requires to replace the concatenation $[\bm{X}_1 \ \bm{V}_1]$ with $[\bm{X}_2 \ \bm{V}_2]$ as the independent matrix in Equation \eqref{eq:b}). In the field of randomized experiments, \cite{muralidharan2022factorial} analyze the change in significance for coefficients of short versus long models.

\vspace{3mm}
Naturally, modern studies rely on much more complex apparatus, including structural equations, difference-in-differences, dynamic panels, HAC estimators (\cite{white1980heteroskedasticity}, \cite{newey1987simple} and \cite{andrews1992improved}, to cite the most commonly used), etc. Given the exhaustiveness and complexity of the related methods, we cannot treat them comprehensively, but it is possible that many of them can be described as Lipschitz operators. Obviously, the variety of strata in the design of empirical studies is such that most intermediate steps cannot be listed in the present paper - one reason being that many of them can be discipline-specific.
\clearpage

\section{Handling the correlations between paths}
\label{app:CLT}

Intuitively, close paths are expected to yield a vicinity of outcomes. Given two paths $p=(r_{p,1},\dots, r_{p,J})$ and $q=(r_{q,1},\dots r_{q,J})$, we recall the assumption from Equation \eqref{eq:distass}: $ d(p,q)= \#\{j, r_{p,j}\neq r_{q,j} \}$. This assumes that all layers have the same importance, and the rationale for this simplification is that it will greatly help the analytical derivations below.

%First, for a fixed $J$, the correlation between two paths decreases with the distance between these two paths, which is an intuitive feature. Second, as the number of paths increases (with $J$), this correlation decreases even more steeply. This required multiplicative speed of decay is a strong assumption. It could be alleviated to $\E[b_pb_q]=\sigma_b^2\rho^{d(p,q)+J}$, but this would require additional hypotheses on $\rho$ and the $r_j$, as is shown at the end of the proof of the following Proposition.

\vspace{3mm}

Below, we prove Proposition \ref{prop:CLT}.

The main issue lies with the distribution of the distance measures $d(p,q)$. They are supported on the set of integers between zero (when $p=q$) and $J$, which occurs when the paths have no option in common. In order to characterize the number of paths which have exactly a distance of $d$, we introduce the notion of elementary symmetric polynomials:
\begin{equation}
    e_k(x_1,\dots,x_J) = \sum_{1 \le j_1 < \dots < j_k \le J}x_{j_1}\dots x_{j_k},
    \label{eq:esp}
\end{equation}
in which there are ${J \choose k}$ terms. Let us start with the simplest case, $d=1$ ($d=0$ being trivial). We are seeking all the pairs of paths with a distance of one. If we pick one path, say, the first one, then we first need to find which mapping will be the one where the difference occurs, and then count how many options there are. In this case, there are $J$ possible mappings and for each mapping $j$, the number of options is $r_j-1$, that is, all options, except the one from the first path. The number of possibilities is thus $\sum_{j=1}^J(r_j-1) $.

More generally, or any path $p$, the number of other paths which have an arbitrary distance of $d$ (with $p$) is
\begin{equation}
    \sum_{q=1}^{J \choose d}\prod_{s=1}^d(r_{r_{q,s}}-1)=e_d(r_1-1,\dots,r_J-1), \quad d\ge 1
    \label{eq:freq}
\end{equation}

where the sum is over all permutations of relevant mappings and the product counts the number of remaining options. Note that the number does not depend on $p$. If we then sum over $d=0,1,\dots, J$, we recover the total number of paths $\prod_{j=1}^Jr_j$ via Vieta's formula:
\begin{equation}
\prod_{i=1}^n(x-z_i)=x^n+\sum_{k=1}^n(-1)^ke_k(z_1,\dots,z_n)x^{n-1},
\label{eq:vieta}
\end{equation}
which we have combined, for $x=1$, with $\prod_{j=1}^Jr_j=\prod_{j=1}^J(1-(1-r_j))$ and
$$e_k(r_1-1,\dots,r_J-1)=(-1)^ke_k(1-r_1,\dots,1-r_k).$$
Importantly, one term in the r.h.s. of \eqref{eq:vieta} is isolated and corresponds to the case $d=0$, so that we do not forget to count the initial path $p$.

Coming back to the norm of the correlation matrix, the fact that the distribution of distances does not depend on the paths allows us to consider only one sum (multiplied $P$ times), as follows:
\begin{align*}
    \|\bm{\Sigma}_P \|_1&=P^{-1}\left( 1+\sum_{p=2}^P \rho^{d(1,p)} \right) = P^{-1}\left( 1+\sum_{d=1}^J\rho^{d} \times e_d(r_1-1,\dots,r_J-1) \right) .
\end{align*}
In the second equality, we switch from a given ordering of the paths to the sum of correlation values ($\rho^{d}$) multiplied by the number of times they appear in the sum - see Equation \eqref{eq:freq}. Given the fact that each term in the polynomial is of order $d$, we can factor in the $\rho$:

\begin{align*}
    \|\bm{\Sigma}_P \|_1 &=P^{-1}\left(1 + \sum_{d=1}^J e_d(\rho(r_1-1),\dots,\rho(r_J-1)) \right) \\
    &=P^{-1}\left(1 + \sum_{d=1}^J (-1)^{k} e_d(\rho(1-r_1),\dots,\rho(1-r_J)) \right) \\
    &=P^{-1}\left(\prod_{j=1}^J(1+\rho(r_j-1)) \right) \quad \text{via } \eqref{eq:vieta} \ \text{for } \ x=1   \\
    &= \prod_{j=1}^J\frac{1+\rho(r_j-1)}{r_j}, \quad \text{because }
 \ P=\prod_{j=1}^Jr_j.
\end{align*}
We recall that $r_j \ge 2$, so that $\frac{1+\rho(r_j-1)}{r_j}<1$ for $\rho <1$. Hence, as $J\rightarrow \infty$, $\|\bm{\Sigma}_P \|_1\rightarrow 0$. However, if $J$ is fixed and $r_j\rightarrow\infty$, then $\|\bm{\Sigma}_P \|_1$ does not shrink to zero - unless $\rho=0$, which is a trivial case we exclude.

%\clearpage

\section{Discussion on $p$-hacking tests}
\label{sec:tests}

With forking paths, we should not be interested in $p$-hacking tests such as the ones in \cite{elliott2021detecting} because the purpose of exhaustive paths is precisely to \textit{avoid} $p$-hacking in the first place. However, out of curiosity and in full disclosure, we have used these tests on the series of paths obtained from our two empirical studies and got disappointing results: while we did not hack, the tests in some cases concluded that we did. For instance, testing for $p$-hacking for the $p$-values with red histograms in Figure \ref{fig:pval_dist} will clearly lead to reject the null of no $p$-hacking.  

Let us understand why that may be the case. In \cite{elliott2021detecting}, there are 3 important distributions: $F_h$, the one of the test statistic, $F$ the distribution that determines the critical values under the null, and $\Pi$, the true distribution of the effect under scrutiny. The latter does not matter, as all tests in \cite{elliott2021detecting} hold irrespective of $\Pi$. Critical values are also rarely an issue, as they are often determined based on standard asymptotic results.

In our case, the problem comes from the empirical distribution of test statistics. Indeed, the quantities we compute are supposed to be distributed according to the Student or Gaussian distribution. However, because paths share similarities, correlation issues arise, as statistics are clearly not independent. This is a possible explanation to why distributions may be distorted.

\clearpage
\section{Additional figures}
\label{sec:figs}

\subsection{Baseline results}

\vspace{-6mm}

\begin{figure}[!h]
\begin{center}
\includegraphics[width=14.8cm]{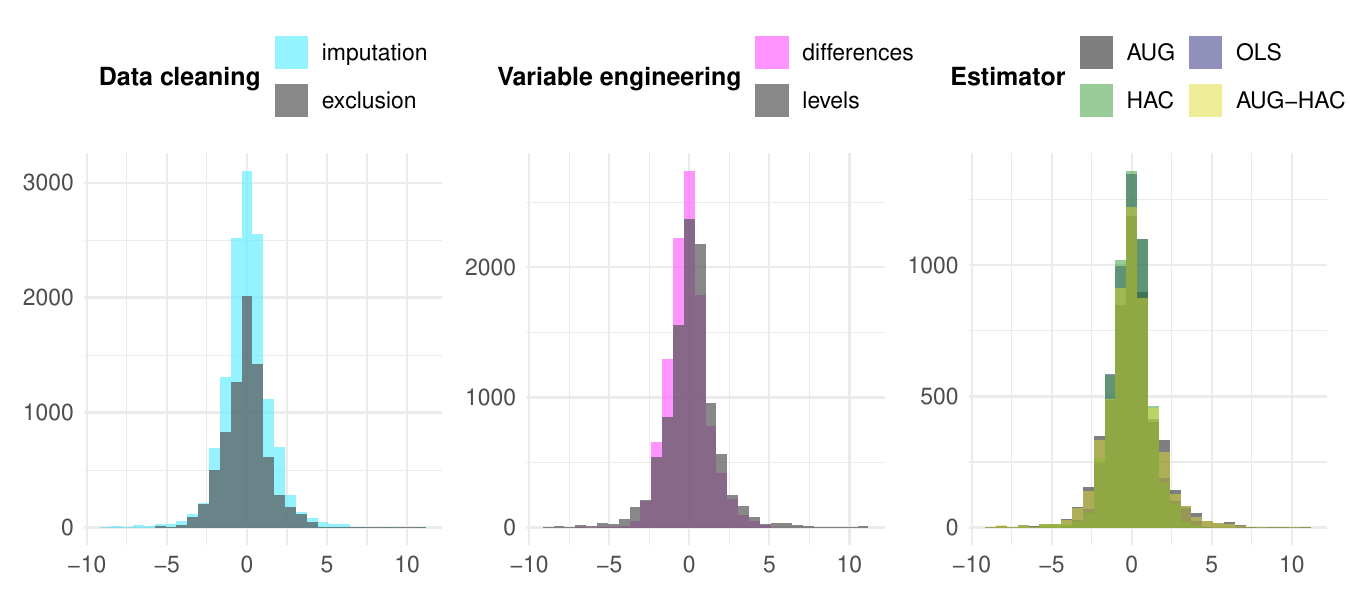}\vspace{-3mm}
\caption{\textbf{Impact of mappings: robustness checks}. \small We report the distribution of $t$-statistics for two binary choices in mappings, plus the final estimator type. Results for regressions with fewer than 30 observations are discarded.  }
\label{fig:histos}
\end{center}
\end{figure}

\vspace{-8mm}

\begin{figure}[!h]
\begin{center}
\includegraphics[width=14.8cm]{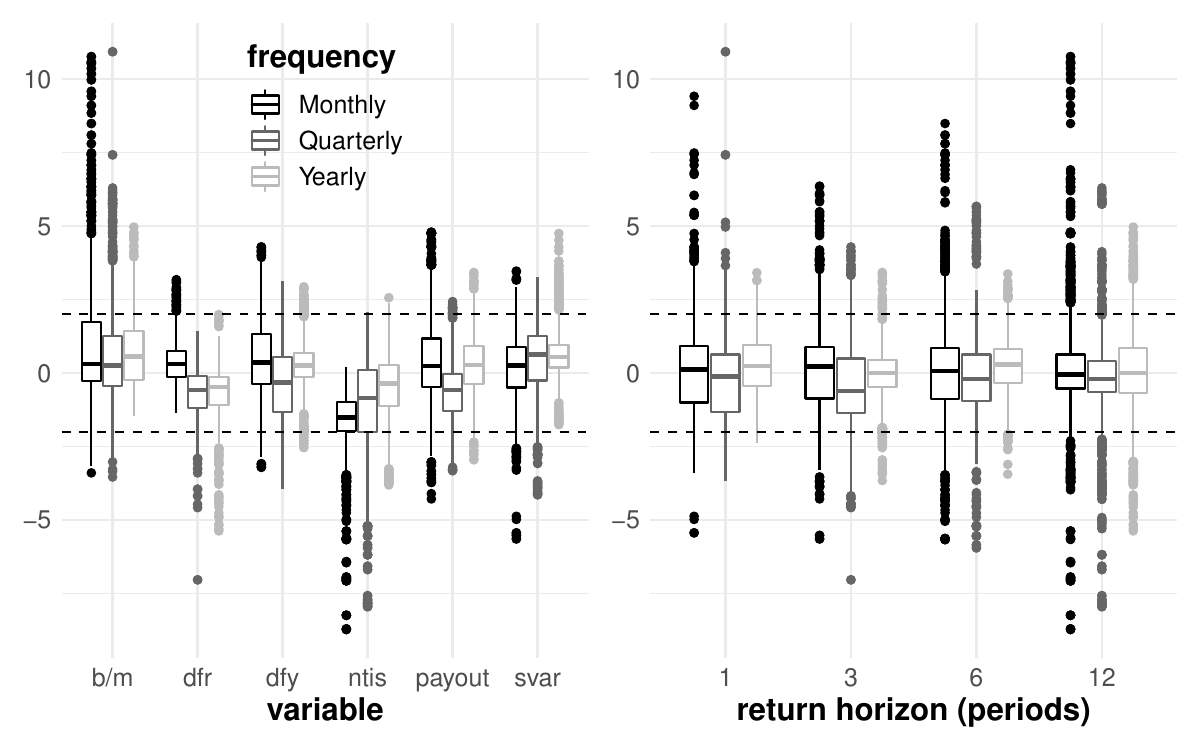}\vspace{-4mm}
\caption{\textbf{Drivers of scalar output: modelling assumptions}. \small We report the distribution of $t$-statistics for two important modelling choices: the independent variable (left panel), and the return horizon of the dependent variable. Results for regressions with fewer than 30 observations are discarded.  }
\label{fig:map}
\end{center}
\end{figure}

\vspace{-6mm}

\clearpage
\subsection{Hacking intervals}

\vspace{-5mm}

\begin{figure}[!h]
\begin{center}
\includegraphics[width=14.5cm]{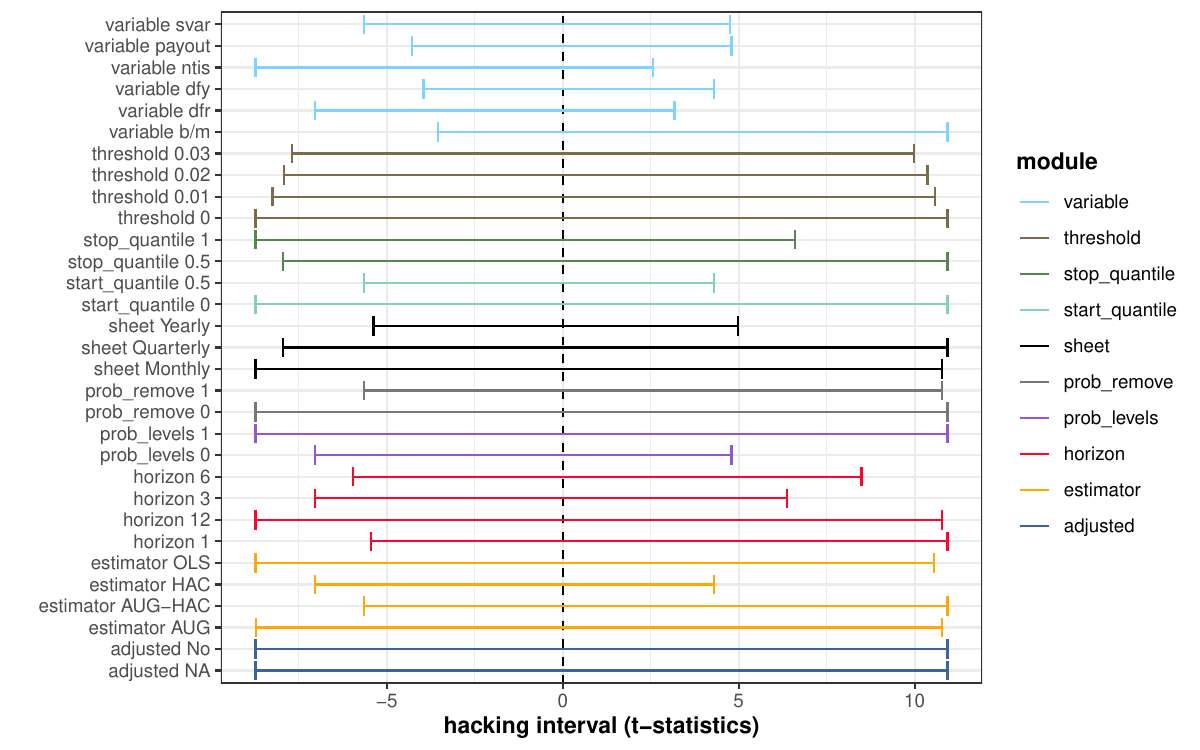}\vspace{-5mm}
\caption{\textbf{Hacking intervals with one fixed mapping}. \small We show the intervals of $t$-statistics obtained when fixing one mapping. Each option of the mapping is tested and all combinations of all other mappings are spanned to generate the hacking intervals. The ten modules (i.e., mappings) are shown with colors.  }
\label{fig:intervals}
\end{center}
\end{figure}

\vspace{-7mm}

\subsection{Test assets and the magnitude of risk premia}

\vspace{-5mm}

\begin{figure}[!h]
\begin{center}
\includegraphics[width=15cm]{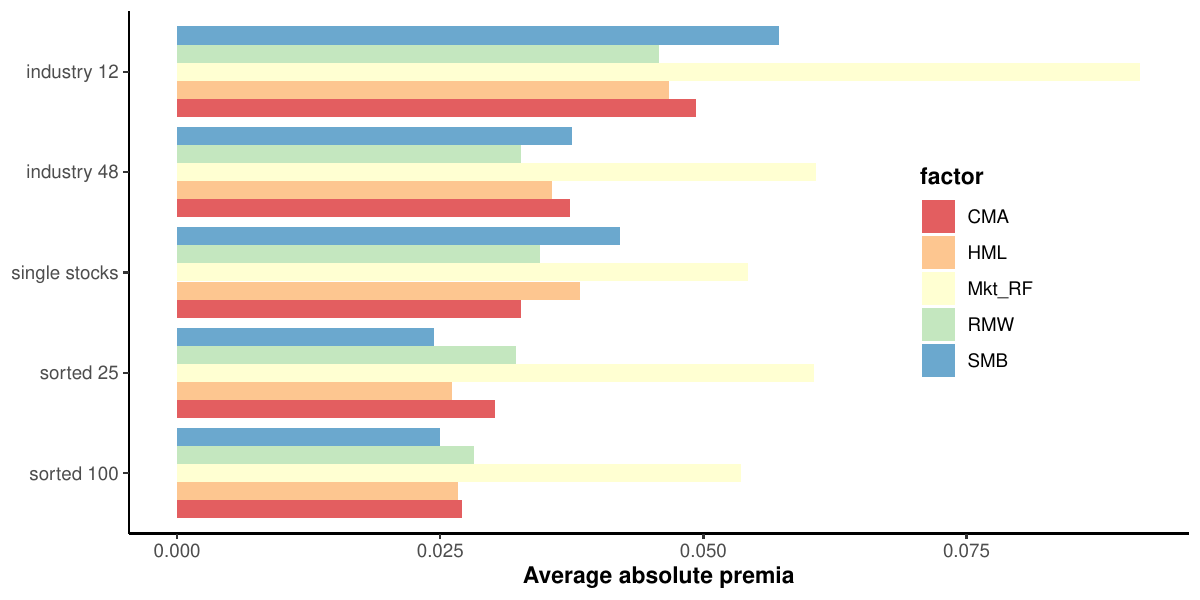}\vspace{-4mm}
\caption{\textbf{Test asset and average abolute premia}. \small We plot the mean absolute premium ($x$-axis) of factors (shown with colors) for each group of test assets ($y$-axis).}
\label{fig:FM_assets}
\end{center}
\end{figure}

\clearpage

\setstretch{1}

\bibliographystyle{chicago}
\bibliography{bib}

\end{document}